\documentclass[12pt, reqno, oneside,amsart]{article}
\usepackage[
left=1.2in,
right=1.2in,
top=1.in,
bottom=1.in,
includeheadfoot,
footskip=0.7in
						]{geometry} 
\usepackage{amsmath}

\usepackage{cite}
\usepackage{graphicx}
\usepackage{amsfonts}
\usepackage{amssymb}
\usepackage{dsfont}		

\input epsf

\usepackage[shortlabels]{enumitem}

\usepackage{framed}

\usepackage{braket}

\makeatletter
\newenvironment{sqcases}{%
  \matrix@check\sqcases\env@sqcases
}{%
  \endarray\right.%
}
\def\env@sqcases{%
  \let\@ifnextchar\new@ifnextchar
  \left\lbrack
  \def\arraystretch{1.2}%
  \array{@{}l@{\quad}l@{}}%
}
\makeatother

\usepackage{lipsum}

\usepackage[titletoc]{appendix}

\usepackage{enumitem}

\usepackage[dvipsnames]{xcolor}

\usepackage{hyperref}
\hypersetup{
    colorlinks=true,
    anchorcolor=red,
    urlcolor=magenta,
    citecolor=red,
    linkcolor=cyan}
        
\usepackage{simplewick}    

\usepackage{empheq}

\usepackage{mathrsfs}

\usepackage{stmaryrd}

\usepackage{subfigure}

\usepackage{amsmath} 

\usepackage{slashed}

\numberwithin{equation}{section}

\usepackage{etoolbox}
\usepackage{titlesec}
\usepackage{titletoc}

\titleformat{\section} 
	{
	\normalsize
	\large
	\bfseries
	} 
{\thesection.}       
{0.3em}                  
{}

\titlespacing{\section}
{0pt}                  
{20pt}                  
{13pt}                  

\titleformat{\subsection} 
[hang]                 
{
\bfseries
} 
{\thesubsection.}       
{0.3em}                  
{}                     

\titlespacing{\subsection}
{0pt}                  
{20pt}                  
{15pt}                  

\titleformat{\subsubsection} 
[hang]
{
\normalsize
\bfseries
} 
{\thesubsubsection.}    
{0.3em}                
{}                     

\titlespacing{\subsubsection}
{0pt}                  
{25pt}		        
{15pt}                  

\def\a{\alpha}
\def\b{\beta}
\def\s{\sigma}

\def\la{\lambda}
\def\La{\Lambda}

\def\vare{\varepsilon}
\def\e{\epsilon}

\def\rm{\mathrm}
\def\cal{\mathcal}
\def\scr{\mathscr}

\def\pa{\partial}

\def\be{\begin{equation}}
\def\ee{\end{equation}}
\def\br{\begin{eqnarray}}

\def\er{\end{eqnarray}}
\def\bsub{\begin{subequations}}
\def\esub{\end{subequations}}

\def\R{\mathrm{R}}

\def\lor{\, \colon \!}
\def\ror{\! \colon \,}

\def\Oint{O^{(\rm{int})}_{[2]}}

\def\Oincov{O^{(\rm{int})}}

\def\inter{\rm{int}}
\def\free{\rm{free}}

\def\lord{[}
\def\rord{]}

\def\bphi{\tilde\phi}

\def\bbS{{\mathbb S}}

\title{On the Dynamics of Protected Ramond Ground States in the D1-D5 CFT}

\date{} 

\usepackage{authblk}

\author[1]{\normalsize A.~A. Lima\thanks{andrealves.fis@gmail.com}}
\author[1]{\normalsize G.~M. Sotkov\thanks{gsotkov@gmail.com}}
\author[2]{\normalsize M. Stanishkov\thanks{marian@inrne.bas.bg}}

\affil[1]{\textit{\footnotesize Department of Physics, Federal University of Esp\'irito Santo, 29075-900, Vit\'oria, Brazil}}
\affil[2]{\textit{\footnotesize Institute for Nuclear Research and Nuclear Energy, Bulgarian Academy of Sciences, 1784 Sofia, Bulgaria}}

\begin{document}

\begin{titlepage}

\maketitle

\begin{abstract}

We examine the behavior  of the Ramond ground states in the D1-D5 CFT after a deformation of the free-orbifold sigma model on target space 
$({\mathbb T}^4)^N / S_N$ by a marginal interaction operator.
These states are compositions of Ramond ground states of the twisted and untwisted sectors.
They are characterized by a conjugacy class of $S_N$ and by the set of their ``spins'', including both R-charge and ``internal'' SU(2) charge.
We compute the four-point functions of an arbitrary Ramond ground state with its conjugate and two interaction operators,
for  genus-zero covering surfaces representing the leading orders in the large-$N$ expansion.
We examine short distance limits  of these four-point functions, shedding light on the dynamics of the interacting theory. We find the OPEs and a collection of structure constants of the ground states with the interaction operators and a set of resulting non-BPS twisted operators.
We also calculate the integrals of  the four-point functions over the positions of the interaction operators and show that they vanish. This provides an explicit demonstration that the Ramond ground states remain protected against deformations away of the free orbifold point, as expected from algebraic considerations using the spectral flow of the ${\cal N} = (4,4)$ superconformal algebra with central charge $c = 6N$.

{\footnotesize 
\bigskip
\noindent
\textbf{Keywords:}

\noindent
Symmetric product orbifold of $\mathcal {N}=4$ SCFT, marginal deformations, twisted Ramond ground states, correlation functions, non-renormalization.
}

\end{abstract}

\pagenumbering{gobble}

\end{titlepage}

\pagenumbering{arabic}

\tableofcontents

\section{Introduction} \label{sec:Introduction}%

The low energy limit of the bound states of the D1-D5 brane system is described by a two-dimensional supersymmetric conformal field theory (SCFT)  holographically dual to
${\rm{AdS}}_3 \times \bbS^3 \times {\mathbb T}^4$ \cite{Maldacena:1997re}.%
	\footnote{The torus ${\mathbb T}^4$ can be replaced by K3, but we will only consider the former.}
In the supergravity description, with the D1-branes wrapped around a large $\bbS^1 $ and the D5-branes wrapped around $\bbS^1 \times {\mathbb T}^4$, the system assumes the form of an asymptotically flat black ring at spatial infinity, with six large dimensions, whose geometry becomes
${\rm{AdS}}_3 \times \bbS^3 \times {\mathbb T}^4$
in the near-horizon scaling limit,%
	\footnote{See \cite{David:2002wn} for a review}
	and whose Bekenstein-Hawking entropy was derived microscopically by Strominger and Vafa \cite{Strominger:1996sh}.
In a certain point of its moduli space, the D1-D5 SCFT  becomes a free ${\cal N}=(4,4)$ supersymmetric sigma model on the orbifold $({\mathbb T}^4)^N/S_N$, where $S_N$ is the symmetric group of $N$ elements \cite{Larsen:1999uk,Seiberg:1999xz}. But this is not the same point in moduli space as the supergravity black hole description, so to make contact between the two one should turn on marginal deformations on the SCFT side. Supergravity properties, such as the entropy of the Strominger-Vafa black hole, can be obtained from protected objects in the SCFT which are unaffected by these deformations.
With the development of the fuzzball program
\cite{Lunin:2001jy,Mathur:2005zp,Kanitscheider:2007wq,Kanitscheider:2006zf,Skenderis:2008qn,Mathur:2018tib},
a dictionary between states of the free-orbifold SCFT and some `microstate geometries'  of supergravity was developed
\cite{Lunin:2001fv,Lunin:2004uu,
Giusto:2004id,Giusto:2004ip,Giusto:2004kj,Giusto:2011fy,Giusto:2012yz,Giusto:2013bda,Giusto:2013rxa,Mathur:2005zp,Skenderis:2008qn,Mathur:2012tj,Taylor:2007hs},
and  significant progress has been achieved in the construction of geometries supporting the horizon-scale scale structure necessary for addressing the information loss problem
\cite{Bena:2010gg,Bena:2013dka,Bena:2015bea,Bena:2016ypk,Bena:2018mpb,Warner:2019jll}.
This success in the supergravity side is one important motivation for an even deeper understanding of both the free orbifold SCFT and its deformation.

Even at the free-orbifold point, the D1-D5 SCFT is quite non-trivial.
The twisted boundary conditions of the symmetric orbifold lead to correlation functions with complicated monodromies, and one must resort to specific techniques to compute them
 \cite{Dixon:1986qv,Arutyunov:1997gt,Lunin:2000yv,Lunin:2001pw,Pakman:2009zz}. The  explicit construction  of these functions remains an active area of research
\cite{Burrington:2012yn,Burrington:2012yq,Burrington:2015mfa,Burrington:2018upk,deBeer:2019ioe,Galliani:2016cai,Galliani:2017jlg,Roumpedakis:2018tdb,Tormo:2018fnt,Giusto:2020mup,Dei:2019iym}.
In fact, the study of the $({\mathbb T}^4)^N/S_N$ orbifold SCFT has been associated with the development of an explicit realization of AdS$_3$/CFT$_2$ by mapping the D1-D5 system to the F1-NS5 system by S-duality, and working on the  
${\rm{AdS}}_3 \times \bbS^3 \times {\mathbb T}^4$
background with only (one unit of) NS fluxes \cite{Dabholkar:2007ey,Gaberdiel:2007vu,Dei:2019iym,Dei:2019osr,Eberhardt:2018ouy,Eberhardt:2019qcl,Eberhardt:2019ywk,Gaberdiel:2020ycd}.

Describing the deformation of this non-trivial theory is not an easy task, but progress can be made by working perturbatively with respect to the dimensionless coupling $\la$ of the deformation operator $\Oint$ \cite{Avery:2010er,Avery:2010hs,Pakman:2009mi,Carson:2014ena,Carson:2015ohj,Burrington:2017jhh,Carson:2016uwf,Guo:2019pzk,Hampton:2018ygz,Guo:2019ady,Guo:2020gxm}.
In doing perturbation theory, it is necessary to compute correlation functions at the free orbifold point, involving the deformation operator among other fields.
For the the analysis of some important fields (as in the case of the present paper), it is necessary to go to second order in perturbation theory, and hence to compute four-point functions, which are dynamical and not fixed by the symmetries. 
One complicating factor is that $\Oint$ has twist two, so it can join and split the ``twisted strings'' of the effective string description of the orbifold, and correlation functions have the associated non-trivial monodromies.
Meanwhile, based both on results from bulk supergravity in the D1-D5 system and from AdS$_3$/CFT$_2$ in the F1-NS5 system, non-renormalization theorems are believed to exist, with explicit proofs available in some cases, e.g.~for three-point functions \cite{deBoer:2008ss,Baggio:2012rr} and for the extremal fields in the NS chiral ring \cite{Pakman:2009mi}.
This latter proof is given at order $\la^2$ by computing explicitly the necessary integral of a four-point function giving the one-loop correction to the propagator. The result confirms the expectation that BPS operators are protected.

The states most relevant for fuzzball and black hole microstates are in the Ramond sector of the SCFT, in particular the Ramond ground states.
Since this sector is more complicated than the NS sector precisely because of the presence of `spin fields' and a non-trivial set of ground states, it is very convenient, for many purposes, to work on the NS sector and then perform a spectral flow \cite{Schwimmer:1986mf} of the ${\cal N}=4$ superconformal algebra.
Spectral flow maps the NS chiral ring to the set of degenerate Ramond ground states, a fact which has been extensively used to simplify computations, classify states, etc.


The purpose of this paper is to study the effects of the deformation on generic Ramond ground states of the full orbifold theory, that is the Ramond fields with
$(h,\tilde h) = (\frac14 N , \frac14 N)$,
by working directly on the Ramond sector, without recurring to spectral flow.
The generic composite Ramond states, which have the form
\be \label{CompleteRamondIntro}
\Big\lord
\prod_i (R^{\zeta_i}_{[n_i]})^{q_i}
\Big\rord
  , \qquad \sum_{i} n_i q_i = N ,
\ee
are in fact highly degenerate:
they are made of all the allowed compositions of $n_i$-twisted ground states whose cycles $(n_i)$ form a conjugacy class of $S_N$.
To obtain $S_N$-invariance, one must sum over the orbits of the cycles.
The labels $\zeta_i$ indicate the charges of the single-cycle components, which are doublets of the R-symmetry SU(2) groups and the SU(2) groups that form the automorphism of the ${\cal N} = (4,4)$ algebra.
Our goal is to compute the four-point functions of (\ref{CompleteRamondIntro}) with their conjugates and two deformation operators:
\be \label{GuIntro}
\Big\langle
\Big\lord
\prod_i (R^{\zeta_i}_{[n_i]})^{q_i}
\Big\rord^\dagger (z_1,\bar z_1)
\;
\Oint(z_2, \bar z_2)
\;
\Oint (z_3 , \bar z_3)
\;
\Big\lord
\prod_i (R^{\zeta_i}_{[n_i]})^{q_i}
\Big\rord (z_4,\bar z_4)
\Big\rangle .
\ee
From this four-point function one can derive several dynamical data following the same steps as in Refs.\cite{Lima:2020boh,Lima:2020kek,Lima:2020nnx,Lima:2020urq}.
Twisted correlators are associated with ramified coverings of the sphere, and their large-$N$ expansion is associated with an expansion in covering surfaces with higher genera \cite{Lunin:2000yv,Pakman:2009zz}. We compute (\ref{GuIntro}) at the leading order, i.e.~for genus-zero covering surfaces.

Our first result is to show that (\ref{GuIntro}) factorizes into a sum of \emph{connected} correlation functions
\be \label{GuIntro2}
\Big\langle
\Big\lord
R^{\zeta_1}_{[n_1]}
R^{\zeta_2}_{[n_2]}
\Big\rord^\dagger (z_1,\bar z_1)
\;
\Oint(z_2, \bar z_2)
\;
\Oint (z_3 , \bar z_3)
\;
\Big\lord
R^{\zeta_1}_{[n_1]}
R^{\zeta_2}_{[n_2]}
\Big\rord (z_4,\bar z_4)
\Big\rangle ,
\ee
with only \emph{double}-cycle Ramond fields.
This reduces the problem considerably --- in fact, the connected function above has already been considered recently in Ref.\cite{Lima:2020nnx} for Ramond fields with R-charged single-cycle components, for which (\ref{GuIntro2}) was computed with the stress-tensor method. Here we compute (\ref{GuIntro2}) for all the possible combinations of R-charged and R-neutral single-cycle constituents,
by using both the stress-tensor method \cite{Dixon:1985jw,Arutyunov:1997gt,Pakman:2009ab,Pakman:2009zz,Pakman:2009mi} and the Lunin-Mathur covering surface technique \cite{Lunin:2000yv,Lunin:2001pw}.
This is our second main result.

Knowledge of these four-point functions gives us important dynamical information about the D1-D5 CFT: one can take coincidence limits such as $z_3 \to z_4$, to find the fusion rules of the operator product expansions (OPEs) 
\be \label{OPEointRRIntro}
\Oint \times \big\lord R^{\zeta_1}_{[n_1]}  R^{\zeta_2}_{[n_2]} \big\rord 
=
\sum_{\frak a}
 Y^{\zeta_1\zeta_2}_{\frak a,  [n_1+n_2]} .
\ee
We are able to find the conformal dimensions of the twisted operators
$Y^{\zeta_1\zeta_2}_{\frak a, [n_1+n_2]}$
in the (two, ${\frak a}={\frak1},{\frak2}$) channels of these OPEs, and to compute a collection of structure constants.
Furthermore, using conformal perturbation theory, the function (\ref{GuIntro}) can be used to derive the $\la^2$-correction%
	\footnote{%
	The first-order correction vanishes automatically by $S_N$ selection rules obeyed by the relevant three-point functions.}
to the conformal dimension of the Ramond fields (\ref{CompleteRamondIntro}).
For that, one needs to integrate over the positions of the two deformation operators.
The factorization means that this integral must be computed for all the functions (\ref{GuIntro2}). These integrals are divergent, but can be regularized with the same framework developed in
Refs.\cite{Lima:2020boh,Lima:2020kek,Lima:2020nnx,Lima:2020urq}.
We show that the regularized integrals of every function (\ref{GuIntro2}) vanish --- a direct verification of the non-renormalization of the generic Ramond ground states.

\bigskip

The structure of the paper is the following.
In Sect.\ref{SectSCFT} we describe the relevant properties of the free orbifold SCFT and recall some aspects of conformal perturbation theory which we use throughout the paper.
In Sect.\ref{SectFactorization} we discuss the factorizations of (\ref{GuIntro}), and show that it factorizes into a sum of functions (\ref{GuIntro2}) at genus-zero order.
In Sect.\ref{SectDoublCyclFunc} we compute the function (\ref{GuIntro2}) for every combination of single-cycle Ramond field components.
In Sect.\ref{SectOPEs} we use these functions to extract conformal data from the fusion rules of the interaction operator with the Ramond fields.
In Sect.\ref{SectNonRenor} we compute the one-loop integrals which give the second-order correction to the conformal dimensions of the Ramond fields, and show that they vanish.
We conclude in Sect.\ref{SectConclusion} by putting our results in perspective.
Several auxiliary computations and examples, as well as lists of the structure constants, are presented in the appendices.

\section{The free orbifold theory and its deformation} \label{SectSCFT}


In this section, we describe the twisted Ramond fields which will be the main subject of our work, and the necessary tools for their description in the deformed theory.

\subsection{Twisted Ramond fields}

In the free orbifold point, the D1-D5 CFT is made by $N$ copies of the `seed' $\cal N = (4,4)$ superconformal field theory of four free bosons $X_I^{A\dot A} (z,\bar z)$, four free holomorphic fermions $\psi_I^{\a A} (z)$, and four anti-holomorphic fermions $\tilde \psi_I^{\dot \a \dot A}(\bar z)$, on which $S_N$ acts on the `copy index' $I = 1,\cdots,N$. 
The total central charge is $c_{orb} = (6N, 6N)$.
Each copy SCFT has has central charge $c = (6,6)$, R-symmetry group
$\rm{SU}(2)_L \times \rm{SU}(2)_R$ and an ``internal'' automorphism group 
$\rm{SO}(4)_I = \rm{SU}(2)_1 \times \rm{SU}(2)_2$.
Indices $\a = + , -$ and $\dot \a = \dot +, \dot -$ transform as a doublets of SU(2)$_L$ and SU(2)$_R$,  and $A=1,2$ and $\dot A=\dot1,\dot2$ transform as doublets of SU(2)$_1$ and SU(2)$_2$, respectively.
The fields satisfy the reality conditions
\be	\label{RealiCondXpsi}
( X_I^{\dot A A} )^\dagger =  - \e_{\dot A \dot B} \e_{A B} X_I^{\dot B B} \ , 
\quad
( \psi_I^{\a \dot A} )^\dagger = - \e_{\a \b} \e_{\dot A \dot B} \psi_I^{\b\dot B} .
\ee
The non-vanishing bosonic two-point functions are
\begin{align}
\Big\langle \pa X_I^{\dot 1 1}(z) ( \pa X_I^{\dot 1 1} )^\dagger (z') \Big\rangle 
	= 
	\frac{2 }{(z - z')^2} 
	=
- \Big\langle \pa X_I^{\dot 1 2}(z) ( \pa X_I^{\dot 1 2} )^\dagger (z') \Big\rangle ,
\label{twopntboconj}
\end{align}
and the fermionic two-point functions are
$\langle \psi_I^{\a \dot A} (z) \psi_I^{\b \dot B} (z') \rangle 
	= - \e^{\a\b} \e^{\dot A \dot B} / (z - z') $.	
Similar formulae hold in the anti-holomorphic sector.

It is convenient to work with bosonized fermions%
	\footnote{%
	For a detailed account of cocycles in the bosonization see Refs.\cite{Burrington:2012yq,Burrington:2015mfa}. Our notation follows closely what is used in these references; the most relevant changes are that we call $\phi_1, \phi_2$ the fields they call $\phi^5,\phi^6$,
	and we call $R^\a$, $R^{\dot A}$, etc.~the spin fields which they call ${\cal S}^\a$, ${\cal S}^{\dot A}$, etc.
	There is also a change of sign in the SU(2)$_2$ current, see Footnote \ref{FootfrakJ} below.
	}
\bsub	\label{FermionsBoson}
\begin{align}
\psi_I^{+ \dot 1}(z) &=  e^{-i \phi_{I,2}(z)} , 
&
\psi_I^{+ \dot 2}(z) &=  e^{i \phi_{I,}1(z)} ,	\label{Fermholoconline1}
\\
\psi_I^{- \dot 1}(z) &= e^{- i \phi_{I,1} (z)}  , 
&
\psi_I^{- \dot 2}(z) &= - e^{i\phi_{I,2} (z)}  , \label{Fermholoconline}
\end{align}\esub
The Ramond ground states are created from the NS vacuum by  `spin fields'.
and then the holomorphic spin fields are given explicitly by
\bsub\label{spinfields}\begin{align}
R_I^+(z) &= e^{+ \frac{i}{2} [ \phi_{I,1}(z) - \phi_{I,2} (z) ]} 	,
&
 R_I^-(z) &=  e^{- \frac{i}{2} [ \phi_{I,1}(z) - \phi_{I,2} (z) ]} , 	\label{spinfildpm}
\\
R_I^{\dot 1}(z) &=  e^{- \frac{i}{2} [ \phi_{I,1} (z) + \phi_{I,2} (z) ]} ,
&
R_I^{\dot 2}(z) &=  e^{+ \frac{i}{2} [ \phi_{I,1} (z) + \phi_{I,2} (z) ]} ,
		\label{Spinfield12}
\end{align}\esub
all with conformal weights $h = \frac14$. 
The R-charges and the internal SU(2)$_2$ charges of each of these fields are given in Table \ref{TabQuNaRam}.

The orbifold SCFT has `twisted sectors' created by `twist operators' $\s_g$, $g \in S_N$, 
with conformal weight
\cite{Dixon:1986qv,Lunin:2000yv}
\be\label{twistdim}
h_n^\s = \frac{1}{4} \Big( n - \frac{1}{n} \Big) = \tilde h^\s_n ,
\ee
which introduce the boundary conditions 
$
\scr O_I(e^{2\pi i} z ) \s_g(0) = \scr O_{g(I)}(z) \s_g(0) 
$
for an operator $\scr O_I$ in the copy $I$.
Let us denote by $(n) = (I_1, \cdots, I_{n})  \in {\mathbb Z}_{n}$ a generic cyclic permutation of length $n$.
Any $g \in S_N$ can be expressed as a product of disjoint cycles, 
\be	\label{gPartition}
{\textstyle g = \prod_{i =1} (n_i)^{q_i} ,	\qquad \sum_{i=1} n_i q_i = N ,}
\ee
with the partition of $N$ on the right defining the conjugacy class $[g]$.
As a consequence, single-cycle permutations can be seen as ``fundamental'' permutations, out of which the elements of $S_N$ are built.
The twist field $\s_{(n)}$ connects the $n$ copies of the seed CFT entering the cycle $(n)$. Then we have $n$ sets of free fields on a circle of radius $nR$, which can be interpreted as a string winding $n$ times around the $\bbS^1$ of radius $R$ parallel to the D1 branes. This is often called a `$n$-wound component string'.
On the the $n$-wound string, the Ramond ground states are created by  the $n$-twisted Ramond fields 
\bsub	\label{RamondFields}
\begin{align}
R^{\pm}_{(n)}(z) 
	&\equiv
	\exp \left( \pm \frac{i}{2n} 
	\sum_{I = 1}^n \big[ \phi_{1,I}(z) - \phi_{2,I}(z)  \big] 
	\right) \s_{(1 \cdots n)}(z)	
\label{Rampmnnoninv}
\\
R^{\dot1}_{(n)}(z) 
	&\equiv 
	\exp \left( - \frac{i}{2n} 
	\sum_{I = 1}^n \big[ \phi_{1,I}(z) + \phi_{2,I}(z)  \big] 
	\right) \s_{(1 \cdots n)}(z)	
\label{Rampmnnoninv}
\\
R^{\dot2}_{(n)}(z) 
	&\equiv 
	\exp \left( + \frac{i}{2n} 
	\sum_{I = 1}^n \big[ \phi_{1,I}(z) + \phi_{2,I}(z)  \big] 
	\right) \s_{(1 \cdots n)}(z)	
\label{Rampmnnoninv}
\end{align}\esub
where we have chosen the representative $n$-cycle to be $(n) = (1, \cdots ,n)$.
The spin fields (\ref{spinfields}) are the untwisted Ramond fields, i.e.~the Ramond fields with $n=1$.
To obtain $S_N$-invariant operators from the single-cycle fields, we sum over the orbits of $(n)$ to obtain its conjugacy class $[n]$, and divide by the appropriate combinatorial factor $\scr S_n(N)$ so that the field remains renormalized \cite{Lunin:2000yv,Pakman:2009zz}. Thus, for example,
\be	\label{RnpmInv}
 R^{\pm}_{[n]}(z) 
	\equiv
	\frac{1}{\scr S_n(N)}
	\sum_{h \in S_N}
	\exp \left( \pm \frac{i}{2n} 
	\sum_{I = 1}^n \big[ \phi_{1,h(I)}(z) - \phi_{2,h(I)}(z)  \big] 
	\right) \s_{h^{-1}(1 \cdots n) h}(z) 	
\ee

The Ramond fields $R^{\a}_{(n)}$ and $R^{\dot A}_{(n)}$ form two SU(2) doublets, according to the spinorial indices $\a = \pm$ and $\dot A = \dot1,\dot2$ in the fermions $\psi^{\a\dot A}$. The $R^\a_{(n)}$ are charged under the holomorphic R-symmetry group SU(2)$_L$, and the $R^{\dot A}_{(n)}$ charged under the internal symmetry SU(2)$_2$. The charges are respectively the eigenvalues 
$j^3$ and $\frak j^3$ of the currents
\begin{align}
J^3(z) &=  \frac{i}{2} \sum_{I = 1}^N \big[  \pa \phi_{1,I} (z)  -   \pa \phi_{2,I} (z)  \big] ,
\\
{\frak J}^3(z) &=  \frac{i}{2} \sum_{I=1}^N \big[  \pa \phi_{1,I} (z)  +   \pa \phi_{2,I}(z)  \big]  .
\end{align}
The right-moving fields $\tilde R^{\dot \a}_{(n)}(\bar z)$ and $\tilde R^{\dot A}_{(n)}(\bar z)$ are charged under the anti-holomorphic currents $\tilde J^3$ and $\tilde{\frak J}^3$ with charges $\tilde \jmath^3$ and $\tilde{\frak j}^3$.%
	\footnote{\label{FootfrakJ}%
	The total SU(2)$_2$ current is the \emph{sum}
	${\frak J}^3(z) + \tilde{\frak J}^3(\bar z)$, see \cite{Burrington:2015mfa}. 
	Note that the SU(2)$_2$ current ${\cal J}^3(z)$ defined in 
	\cite{Burrington:2015mfa} has its sign opposite from ours, 
	i.e.~${\cal J}^3(z) = - {\frak J}^3(z)$. 
	}
We will use the same notation for the left-moving fields and for the left-right moving operators $R^{\a}_{(n)}(z,\bar z) = R^\a_{(n)}(z) \tilde R^{\dot\a}_{(n)}(\bar z)$ and $R^{\dot A}_{(n)}(z,\bar z) = R^{\dot A}_{(n)}(z) \tilde R^{\dot A}_{(n)}(\bar z)$. 
The values of the charges for each Ramond field is given in Table \ref{TabQuNaRam}.
All $n$-twisted Ramond fields all have conformal weights
\be
h^\R_n = \frac{nc}{24} = \frac{n}{4} = \tilde h^\R_n
\ee
which is appropriate for the Ramond ground states of the $n$-wound string.
One can deduce this value by adding the weights of the exponentials in (\ref{RamondFields}) with the weight (\ref{twistdim}) of the twist fields. 
Of course, the $S_N$-invariant fields $R^\a_{[n]}$, $R^{\dot A}_{[n]}$ have the same charges and weights as their non-$S_N$-invariant components.

\begin{table}
\begin{center}
\begin{tabular}{r|| r r r r}
	& $R^+_{(n)}$ & $R^-_{(n)}$  & $R^{\dot 1}_{(n)}$ & $R^{\dot 2}_{(n)}$ 
\\
\hline
\hline
\text{\footnotesize (R-charge)} \ 
$j^3$ 
	& $+ \tfrac{1}{2} \ $ & $-\tfrac{1}{2} \ $ & $0 \ $ & $0 \ $
\\
\text{\footnotesize (R-charge)} \ 
$\tilde \jmath^3$ 
	&  $+ \tfrac{1}{2} \ $ & $- \tfrac{1}{2} \ $ & $0 \ $ & $0 \ $ 
\\
\text{\footnotesize (internal)} \ 
$\frak j^3$ 
	& $0 \ $ & $0 \ $  & $-\tfrac{1}{2} \ $ & $+\tfrac{1}{2} \ $  
\\
\text{\footnotesize (internal)} \ 
	$\tilde{\frak j}^3$ & $0 \ $ & $0 \ $ & $-\tfrac{1}{2} \ $ & $+\tfrac{1}{2} \ $
\end{tabular}
\caption{SU(2) charges of Ramond fields}
\label{TabQuNaRam}
\end{center}
\end{table}

The Ramond ground states of the orbifold with $c = 6N$ are compositions of single-cycle fields with \emph{disjoint} twists defining a conjugacy class of $S_N$,
\be	\label{CompleteRamond}
{\textstyle \prod_i (R^{\zeta_i}_{(n_i)})^{q_i}  , \qquad \sum_{i=1} n_i q_i = N } ,
\ee
where $\zeta_i = \pm, \dot1, \dot2$.
Since each component of the product is made from different copies (i.e.~the cycles $(n_i)$ are disjoint), when applying the stress-tensor or the SU(2) currents, we find that the dimension and the charges of the composite operator are the sums of the respective quantum numbers of the component strings, hence  
\be
h^\R = \frac{\sum_i q_i n_i}{4} = \frac{N}{4} = \tilde h^\R .
\ee
We define composite operators with sum-over-orbits as in (\ref{RnpmInv}) by \be
\prod_i (R^{\zeta_i}_{[n_i]})^{q_i}  
= \sum_{h \in S_N} 
\left\lord
 \prod_i \frac{1}{{\scr S}_{n_i}^{q_i}} (R^{\zeta_i}_{h^{-1} (n_i) h} )^{q_i}  
 \right\rord ,
\ee
i.e.~we permute all cycles with the same $h \in S_N$, ensuring that only disjoint cycles enter the products of twists.
This is similar to defining a ``normal ordering'', see \cite{Roumpedakis:2018tdb,Lima:2020nnx}, which we indicate by writing the composite operator inside square brackets.
As we will see later, a prominent r\^ole will be played by the double-cycle operator with only two twisted strings of lengths $n_1$ and $n_2$, each in its Ramond ground state; it can be expressed explicitly as
\begin{align}	\label{Rn1n2ppInv}
\begin{split}
&\big\lord R_{[n_1]}^{\dot1} R_{[n_2]}^{+} \big\rord (z) 
	= 
\\
&	\frac{1}{{\scr S}_{n_1} {\scr S}_{n_2}}
	\sum_{h\in S_N} 
	\exp \Bigg[ 
	-
	\frac{i}{2n_1}\sum_{I=1}^{n_1} \big( \phi_{1,h(I)} + \phi_{2,h(I)} \big) 
	- 
	\frac{i}{2n_2}
	\sum_{I=n_1+1}^{n_1+n_2} \big( \phi_{1,h(I)} -\phi_{2,h(I)} \big)
	\Bigg]
\\	&\qquad\qquad\qquad
	\times
	\sigma_{h^{-1}(1\cdots n_1)h} 
	\sigma_{h^{-1}(n_1+1\cdots n_1+n_2)h} .
\end{split}
\end{align} 
Cf.~Ref.\cite{Lima:2020nnx}. 
Note how the definition ensures that the cycles $(n_1)$ and $(n_2)$ appearing the product are always disjoint.
In the copies with trivial cycles there are, implicitly, identity operators.
As mentioned, the conformal weight and charges are the sums of the components, hence for (\ref{Rn1n2ppInv})
\be
h^\R_{n_1,n_2} = \frac{n_1 + n_2}{4} , 
\quad
j^3 = \tfrac12 ,
\quad
\frak j^3 = - \tfrac12 .
\ee
We can compose the 
 build a set of R-neutral operators not only from the R-neutral fields $R^{\dot A}_{[n]}$, but also by combining R-charged single-cycle fields as 
$\lord R_{[n_1]}^{\pm} R_{[n_2]}^{\mp} \rord$, 
which has $h^\R_{n_1,n_2} = \frac{n_1 + n_2}{4}$, $j^3 = 0$, and $\frak j^3 = 0$. 
 A complete list of the SU(2) charges of the double-cycle Ramond fields is given in Table \ref{TabQuNaRamComp}.

\begin{table}
\begin{center}
\begin{tabular}{r|| c c c c c c c}
& {\footnotesize
	$ R^\pm_{[n_1]} R^\pm_{[n_2]} $ }
& {\footnotesize
	 $ R^\pm_{[n_1]} R^\mp_{[n_2]} $ }
& {\footnotesize
	 $ R^{\dot2}_{[n_1]} R^{\dot2}_{[n_2]} $ }
& {\footnotesize
	 $ R^{\dot1}_{[n_1]} R^{\dot1}_{[n_2]} $ }
& {\footnotesize
	 $ R^{\dot2}_{[n_1]} R^{\dot1}_{[n_2]} $ }
& {\footnotesize
	 $ R^{\dot1}_{[n_1]} R^{\pm}_{[n_2]} $ }
& {\footnotesize
	 $ R^{\dot2}_{[n_1]} R^{\pm}_{[n_2]} $ }
\\
\hline
\hline
%
$j^3$ 
	& $\pm 1$ 
	& $0$ 
	& $0$ 
	& $0$
	& $0$
	& $\pm\tfrac12$
	& $\pm\tfrac12$
\\
$\frak j^3$ 
	& $0$ 
	& $0$ 
	& $+1$ 
	& $-1$
	& $0$
	& $-\tfrac12$
	& $+\tfrac12$
\end{tabular}
\caption{SU(2) charges of composite Ramond fields}
\label{TabQuNaRamComp}
\end{center}
\end{table}

\subsection{The interacting CFT and renormalization}

The interacting CFT is a deformation of the free orbifold by the scalar modulus
\begin{align}	\label{DeformwithMo}
\begin{split}
\la O^{(\inter)}_{[2]}(z_* , \bar z_* ) 
	&= \la \e_{A  B} G^{-  A}_{-\frac{1}{2}} \tilde G^{ \dot -  B}_{-\frac{1}{2}} 
	O^{(0,0)}_{[2]}(z_*, \bar z_*) 	
\\
	&= \la \e_{A  B} 
	\oint \frac{d z}{2\pi i}
	\oint \frac{d\bar z}{2\pi i}
	G^{-  A}(z) \tilde G^{ \dot -  B}(\bar z) 
	O^{(0,0)}_{[2]}(z_*, \bar z_*) 	
\end{split}
\end{align}
where $\la$ is the coupling parameter. 
This is one of the SUGRA moduli \cite{Avery:2010er}, an $S_N$-invariant singlet of the SU(2) symmetries, and it is marginal, with dimension $\Delta = h + \tilde h = 2$. The deformation is a descendant of the NS chiral operator 
$O^{(0,0)}_{[n]}$ with twist $n = 2$, which for generic $n$ have 
$j^3 = \frac{n-1}{2} = h$, after the action of the $-\frac12$-modes of the supercharges.

When the deformation is turned on, generic fields are renormalized while some selected others remain protected by algebraic conditions such as BPS bounds. 
Since the deformation is exactly marginal, a renormalized propagator (two-point function) is still fixed by conformal symmetry, but with a corrected conformal dimension. This correction can be obtained by computing the functional integral with the perturbed action
$
S_\inter = S_\free  + \la \int d^2z \, \Oint(z, \bar z) , 
$
and looking at the coefficient of logarithmic divergencies.
At first order in $\la$, the change in the conformal dimension
$\Delta_{\scr R} = h^{\scr R} + \tilde h^{\scr R}$ 
of an operator $\scr R(z,\bar z)$ is proportional to 
\be	\label{3ptStrcR}
\big\langle \scr R^\dagger(\infty) \Oint(1,\bar 1) \scr R(0) \big\rangle,
\ee
a structure constant in the free theory \cite{KADANOFF197939}.%
	\footnote{%
	Unless explicitly indicated otherwise, all correlators $\langle \, \cdots \rangle$ in this paper should be understood as evaluated on the free orbifold theory.}
When this structure constant vanishes, the second-order correction to the conformal dimension is given by the ``one-loop'' integral over the positions of the interaction operators in the four-point function
\be
\Big\langle 
	\scr R(z_1, \bar z_1) 
	\Oint(z_3, \bar z_3) 
	\Oint (z_4 , \bar z_4) 
	\scr R (z_2 , \bar z_2) 
\Big\rangle 
	=  
	\frac{G_{\scr R} (u, \bar u) 
	}{
	 |z_{13}|^2 |z_{32}|^2|z_{12}|^{2 (\Delta_{\scr R} - 1)}
	 } ,
	\label{4ptgenG}
\ee
where $z_{ij} \equiv z_i - z_j$ and $u$ is the anharmonic ratio
$u \equiv {z_{12}z_{34}} / { z_{13}z_{24} }$.
The integral over the position $z_3$ gives rise to the logarithmic divergence $\log \La$, with a cutoff $\La \ll 1$, and the integral over $z_4$, which can be exchanged for an integral over $u$, 
\be	\label{JscrR}
J_{\scr R} \equiv \int \! d^2u \, G_{\scr R}(u,\bar u) ,
\ee
gives the corrected dimension
\bsub	\label{Renormalization}
\be
\Delta_{\scr R}^{(ren)} = \Delta_{\scr R} - \tfrac{\pi}{2} \la^2 J_{\scr R} + \rm{O}(\la^3)
\ee
for the renormalized field
\be
{\scr R}^{(ren)} (z,\bar z) = \La^{\frac12 \pi \la^2 J_{\scr R}} {\scr R}(z,\bar z) .
\ee\esub
The integral (\ref{JscrR}) also gives the first-order expression for the structure constant (\ref{3ptStrcR}) in the interacting theory.
See \cite{Lima:2020kek} for details, and also \cite{Keller:2019yrr,Burrington:2012yq} for interesting expositions on conformal perturbation theory.

\section{Factorizations of the four-point function with Ramond ground states}
\label{SectFactorization}

Consider the Ramond ground state (\ref{CompleteRamond}). 
The four-point function
\be	\label{Gu}
G(u,\bar u) =
\Big\langle 
\Big\lord
\prod_i (R^{\zeta_i\dagger}_{[n_i]})^{q_i}
\Big\rord (\infty,\bar \infty) 
	\;
	\Oint(1, \bar 1) 
	\;
	\Oint (u , \bar u) 
	\;
\Big\lord
\prod_i (R^{\zeta_i}_{[n_i]})^{q_i} 
\Big\rord (0,\bar 0)
\Big\rangle 
\ee
is a sum of terms with specific twist cycles, which we denote by 
\be	\label{Fourpttofact}
\Big\langle 
\Big\lord
\prod_i (R^{\zeta_i\dagger}_{(n_i)_\infty})^{q_i}
\Big\rord
(\infty,\bar \infty) 
	\;
	\Oincov_{(2)_1}(1, \bar 1) 
	\;
	\Oincov_{(2)_u} (u , \bar u) 
	\;
\Big\lord
\prod_i (R^{\zeta_i}_{(n_i)_0})^{q_i} 
\Big\rord
(0,\bar 0)
\Big\rangle  .
\ee
That is, we label a cycle $(m)_z$ by the position $z$ where it is inserted. 
Let us recall some properties of twisted correlation functions \cite{Lunin:2000yv}.
In a generic $Q$-point function  the monodromy conditions impose that 
\begin{flalign}
&&
&\left\langle \scr O^1_{(n_1)}  (z_1 , \bar z_1) \scr O^2_{(n_2)}(z_2,\bar z_2) 
\cdots 
\scr O^Q_{(n_Q)}  (z_Q , \bar z_Q)  \right\rangle \neq 0
&&
\label{QptscrO}
\\
\text{only if}
&&
&(n_1) \cdots (n_Q) = \mathds 1. &&	\label{compto1}
\end{flalign}
The correlator is connected if no collection of cycles is disjoint from all the others; let us assume that this is the case in (\ref{QptscrO}).
The monodromies define a ramified surface $\Sigma$ with coordinates $(t,\bar t)$ covering the base Riemann sphere $S^2_{\rm{base}}$, with its coordinates $(z,\bar z)$. The covering surface $\Sigma$ has one ramification point of order $n_r-1$ for each cycle $(n_i)$ entering the correlator, irrespective of the position where the operators are inserted.
(That is, we can take $z_2 \to z_1$ in (\ref{QptscrO}), and the number of ramification points is the same.)
The genus $\bf g$ of $\Sigma$ is determined by $\bf s$, the number of \emph{distinct} copies entering the permutations, via the Riemann-Hurwitz formula
\be	\label{RHur}
{\bf g} = \frac{1}{2} \sum_{r=1}^Q (n_r - 1) - {\bf s} + 1
\ee
If we sum over the orbits of the cycles $(n_1), \cdots , (n_Q)$, the resulting $S_N$-invariant function depends on $N$ via normalization factors and combinatorial properties of the different solutions of Eq.(\ref{compto1}), and have the large-$N$ scaling  \cite{Lunin:2000yv,Pakman:2009zz}
\be	\label{Ndep}
\begin{split}
\left\langle \scr O^1_{[n_1]}  (z_1 , \bar z_1) \scr O^2_{[n_2]}(z_2,\bar z_2) 
\cdots 
\scr O^Q_{[n_Q]}  (z_Q , \bar z_Q)  \right\rangle
&\sim
N^{{\bf s} - \frac12 \sum_{r=1}^Q n_r} (1 + N^{-1} + \cdots)
\\
&=
N^{1 - {\bf g} - \frac12 Q} (1 + N^{-1} + \cdots)
\end{split}
\ee

It is easy to see that a generic term like (\ref{Fourpttofact}) factorizes because, by construction, the cycles in the Ramond fields $R^{(\zeta_i)}_{(n_i)}$ are all disjoint. Hence the overlapping of copy indices is governed by the twist-2 cycles of the interaction operators, which can ``sew together'' a very limited number of cycles.

\subsection{Factorization into double-cycle Ramond fields}	\label{SectFactTypes}

Let us examine in detail the function (\ref{Fourpttofact}). 
The cycles of each composite Ramond field are all disjoint; to make this more explicit, we can relabel the components by writing each cycle separately, i.e.~%
$\lord
\prod_i (R^{\zeta_i}_{(n_i)})^{q_i}
\rord
 = 
\lord
\prod_r R^{\zeta_r}_{(n_r)} 
\rord
$,
where now
$\sum_i q_i n_i \equiv \sum_{r} n_{r} = N 
$.
In this notation,  there are (in general) cycles of same length $n_{r} = n_{r'}$, but they are always disjoint: $(n_{r}) \neq (n_{r'})$.
Now, 
every one of the $N$ copies is present (only once) in the cycles of $\prod_{r} R^{\zeta_r}_{(n_{r})}$, hence \emph{each of the cycles $(2)_1$ and $(2)_u$ necessarily overlap with one or, at most, two of the cycles $(n_r)$.}
As a result, the possible factorizations of (\ref{Fourpttofact}) are the following.

\begin{itemize}[label={}, leftmargin=*]

\item
\underline{$\cdot$ Three-point functions:}

In principle, the $S_N$ rules allow for a factorization into three-point functions, 
\begin{flalign}	\label{3ptFact}
\begin{split}
&\Big\langle 
R^{\zeta_1\dagger}_{(M)_\infty} (\infty,\bar \infty)
	\;
	\Oincov_{(2)_1}(1, \bar 1) 
R^{\zeta_1}_{(n_1)_0} 
R^{\zeta_2}_{(n_2)_0} (0,\bar 0)
\Big\rangle 
\\
&\quad
\times 
\Big\langle 
R^{\zeta_1\dagger}_{(n_1)_0} 
R^{\zeta_2\dagger}_{(n_2)_0} (\infty,\bar \infty)
	\Oincov_{(2)_u}(u, \bar u) 
R^{\zeta_M}_{(M)_0} (0,\bar 0)
\Big\rangle  
\,
\prod_k
\Big\langle 
R^{\zeta_k\dagger}_{(n_k)_\infty} \Big| R^{\zeta_k}_{(n_k)_0} 
\Big\rangle  
\end{split}
\end{flalign}
where $M = n_1 + n_2$.
But charge conservation implies that these three-point functions vanish, since no double-cycle operator has the same charges as a single-cycle operator; see Tables \ref{TabQuNaRam} and \ref{TabQuNaRamComp}.

\item
\underline{$\cdot$ Single-Cycle:}
\be	\label{SingleCycle4pt}
\Big\langle 
R^{\zeta_1\dagger}_{(n)_\infty} (\infty,\bar \infty) 
	\;
	\Oincov_{(2)_1}(1, \bar 1) 
	\;
	\Oincov_{(2)_u} (u , \bar u) 
	\;
R^{\zeta_1}_{(n)_0} (0,\bar 0)
\Big\rangle  
\prod_s
\Big\langle 
R^{\zeta_s\dagger}_{(n_s)_\infty} \Big| R^{\zeta_s}_{(n_s)_0} 
\Big\rangle  
\ee
For this factorization to be possible, there must be ${\bf s} = n$ distinct copies entering the remaining, connected four-point function. The associated covering surface has $Q = 4$ ramification points, and the Riemann-Hurwitz formula gives ${\bf g} = 1$. 
The covering surface associated with these functions is therefore a torus.

Note that, in this case, we must necessarily have $n \geq 2$, because, by hypothesis, \emph{both} copy indices in $(2)_1$ have to overlap with indices of the cycle $(n)_\infty$. 
The fields with $n = 1$ all factorize into the double-cycle case described below.

\item
\underline{$\cdot$ Double-Cycle:}
\be	\label{DoubleCycle4pt}
\begin{split}
\Big\langle 
R^{\zeta_1\dagger}_{(n_1)_\infty} 
R^{\zeta_2\dagger}_{(n_2)_\infty} (\infty,\bar \infty) 
	\;
	\Oincov_{(2)_1}(1, \bar 1) 
	\;
	\Oincov_{(2)_u} (u , \bar u) 
	\;
R^{\zeta_1}_{(n_1)_0} 
R^{\zeta_2}_{(n_2)_0} (0,\bar 0)
\Big\rangle  \qquad
\\
\times
\prod_s
\Big\langle 
R^{\zeta_s\dagger}_{(n_s)_\infty} \Big| R^{\zeta_s}_{(n_s)_0} 
\Big\rangle  
\end{split}
\ee
We can normalize the twisted Ramond fields such that the factorized bra-kets are
$
\braket{R^{\zeta_s\dagger}_{(n_s)_\infty} | R^{\zeta_s}_{(n_s)_0} } = 1
$.
(In these factors we have always $(n_s)_\infty = (n_s)_0^{-1}$, in accordance with Eq.(\ref{compto1}).)
Because of Eq.(\ref{compto1}), the twists of the remaining four-point function in Eq.(\ref{DoubleCycle4pt}) must involve ${\bf s} = n_1 + n_2$ distinct copies. There are $Q = 6$ ramification points, corresponding to the six different cycles, and the Riemann-Hurwitz equation now gives ${\bf g} = 0$. So the covering surface associated with (\ref{DoubleCycle4pt}) is a sphere.

There is a subtlety here when one or both of the cycles has length one.
Say $n_2 = 1$ and $n_1 = n \leq 2$: the covering surface loses two ramification points, corresponding to $(n_2)_0$ and $(n_2)_\infty$, and we are left with four ramification points, as in (\ref{SingleCycle4pt}); but now the number of distinct copies is ${\bf s} = n + 1$, hence ${\bf g} = 0$.
If both $n_1 = n_2 = 1$, there are only two ramification points, corresponding to the cycles $(2)_1$ and $(2)_u = (2)_1^{-1}$; hence ${\bf s} = 2$  and Eq.(\ref{RHur}) gives again ${\bf g} =  0$.

\end{itemize}

In summary: 

\emph{All genus-zero contributions to the four-point function (\ref{Gu}) are given by the double-cycle factorization (\ref{DoubleCycle4pt}).}

\subsection{Sums over orbits} 	\label{SectSumoveOrb1}

We have analyzed the factorization of individual terms in the sum over orbits (\ref{Gu}). Now we must explain how these factorizations are organized when we perform the summation. 
It is very instructive to look at specific examples of configurations of $\lord \prod_i ( R^{\zeta_i}_{[n_i]})^{q_i} \rord$; we give detailed examples in Appendix \ref{AppSumoveOrb1}.
Here we develop the general argument.

For now we omit the positions of the fields for brevity; they will always follow the order $\infty, 1, u, 0$, and can be read in the labels of the cycles.
Start with fixing the cycles of the interaction operators, while summing over the orbits of the Ramond fields:
\be	\label{dobsumorbGen}
\Big\langle 
\Big\lord
\sum_{h \in S_N}
\prod_i (R^{\zeta_i\dagger}_{h (n_i)_\infty h^{-1}})^{q_i}
\Big\rord
	\;
	\Oincov_{(2)_1} 
	\;
	\Oincov_{(2)_u}
	\;
\Big\lord
\sum_{g \in S_N}
\prod_i (R^{\zeta_i}_{g (n_i)_0 g^{-1}})^{q_i} 
\Big\rord
\Big\rangle  .
\ee
In each term of the (double) sum (\ref{dobsumorbGen}), the copies entering the cycles 
$(2)_1$ and $(2)_u$ 
will select a pair of single-cycle Ramond fields from the  square brackets on the left, and another pair from the square brackets on the right. The remaining single-cycle Ramond fields, which do not share any copies with the interaction operators, will factorize out of the four-point function.
Thus (\ref{dobsumorbGen}) becomes a sum of terms like (\ref{DoubleCycle4pt}).
Since in the sum over orbits every Ramond field component will have its copies shuffled into every possible configuration, the fixed copies of 
$(2)_1$ and $(2)_u$ 
will select every possible pairings of fields at least once.
In fact, the same pairs will be selected many times: there are many rearrangements of the copies inside their cycles whose overlap with $(2)_1$ and $(2)_u$ satisfy Eq.(\ref{compto1}).
More precisely, the factorized terms will organize not as individual terms like (\ref{DoubleCycle4pt}), but rather as sums over orbits themselves, such as
\be	\label{DoubleCycle4ptSm}
\begin{split}
\Big\langle
\Big\lord
\sum_{h \in S_N}
R^{\zeta_1\dagger}_{h (n_1)_\infty h^{-1}} 
R^{\zeta_2\dagger}_{h (n_2)_\infty h^{-1}} 
\Big\rord
	\Oincov_{(2)_1}
	\Oincov_{(2)_u} 
\Big\lord
\sum_{g \in S_N}
R^{\zeta_1}_{g (n_1)_0 g^{-1}} 
R^{\zeta_2}_{g (n_2)_0 g^{-1}}
\Big\rord
\Big\rangle  
\\
\times
\prod_s
\sum_{h,g \in S_N}
\Big\langle 
R^{\zeta_s\dagger}_{h (n_s)_\infty h^{-1}} 
\Big| 
R^{\zeta_s}_{g (n_s)_0 g^{-1}} 
\Big\rangle  .
\end{split}
\ee
We now perform a sum over orbits for the cycles of the interactions, divide by the factors $\scr C_{n_1 n_2}$, $\scr S_2$, etc.%
		\footnote{\label{FootSr}%
		See Eqs.(\ref{RnpmInv}) and (\ref{Rn1n2ppInv});
		note that the factors are ``extensive'': 
		${\scr C}_{n_1n_2} = {\scr S}_{n_1} {\scr S}_{n_2}$, etc. so the normalization of the full Ramond field is obtained with $\prod {\scr S}_{n_i}$.
		In (\ref{DoubleCycle4ptSmCl}), most ${\scr S}_{n_i}$ have gone into the normalized two-point functions in the second line of (\ref{DoubleCycle4ptSm}), which are \emph{not} normalized.
				} 
Eq.(\ref{dobsumorbGen}) becomes (\ref{Gu}), 
\be	\label{GuShor}
G =
\Big\langle 
\big\lord
\prod_i (R^{\zeta_i}_{[n_i]_\infty})^{q_i}
\big\rord^\dagger
	\Oincov_{[2]_1}
	\Oincov_{[2]_u} 
\big\lord
\prod_i (R^{\zeta_i}_{[n_i]_0})^{q_i} 
\big\rord
\Big\rangle 
\ee			
while (\ref{DoubleCycle4ptSm}) becomes
\be	\label{DoubleCycle4ptSmCl}
\Big\langle
\big\lord
R^{\zeta_1}_{[n_1]_\infty}
R^{\zeta_2}_{[n_2]_\infty}
\big\rord^\dagger
	\Oincov_{[2]_1}
	\Oincov_{[2]_u} 
\big\lord
R^{\zeta_1}_{[n_1]_0}
R^{\zeta_2}_{[n_2]_0}
\big\rord
\Big\rangle  
\equiv 
G^{n_1,n_2}_{\zeta_1,\zeta_2}.
\ee
where we have used the fact that
$\langle  R^{\zeta_s\dagger}_{[n_s]_\infty} | R^{\zeta_s}_{[n_s]_0} \rangle  
= 1$.%
	\footnote{See Footnote \ref{FootSr}.}
In short, we have found that 
\be	\label{GuFact}
G = \sum_{\text{pairings}} G^{n_j,n_k}_{\zeta_j,\zeta_k}.
\ee
where the sum is over all possible pairings of 
single-cycle components of the multi-cycle Ramond field
$\lord\prod_i (R^{\zeta_i}_{[n_i]})^{q_i}\rord$.

For a generic configuration of the Ramond ground state, 
there will be many terms in the r.h.s.~of Eq.(\ref{GuFact}) which are the same function. 
For example, if the Ramond ground state is
$\lord (R^{+}_{[n]})^{N/n}\rord$,
every pairing will give the same function 
$G^{n,n}_{+,+}$, and the r.h.s.~of Eq.(\ref{GuFact}) will result in 
${\scr P}^2(N/n) G^{n,n}_{+,+}$, where ${\scr P}(q)$ is the number of ways of pairing $q$ objects. 
This number can be written as
\be	\label{scrPq}
{\scr P}(q) \equiv 
\frac{(2 \lfloor q \rfloor)!
		}{ 
	( \lfloor \frac12 q \rfloor )! 
	\ 2^{\lfloor \frac12 q \rfloor}
		}
+
2 (\tfrac12 q - \lfloor \tfrac12 q \rfloor) \lfloor \tfrac12 q \rfloor ,
\ee
see Appendix \ref{AppSumoveOrb1}.
Note that two functions 
$G^{n_1,n_2}_{\zeta_1,\zeta_2}$
and
$G^{n_1',n_2'}_{\zeta_1',\zeta_2'}$
are equal iff
$(n_1,\zeta_1) = (n_1',\zeta_1')$
and
$(n_2,\zeta_2) = (n_2',\zeta_2')$.
Let us write the most general Ramond field as
\bsub	\label{GenRam}
\begin{flalign}
&&
&
\Big\lord
\prod_i 
( R^+_{[n_i]})^{q^+_i} 
( R^-_{[n_i]})^{q^-_i}
( R^{\dot1}_{[n_i]})^{q^{\dot1}_i}
( R^{\dot2}_{[n_i]})^{q^{\dot2}_i}
\Big\rord
=
\Big\lord
\prod_i \prod_{\zeta = +,-,\dot1,\dot2}
( R^\zeta_{[n_i]})^{q^\zeta_i} 
\Big\rord ,
&&
\\
\text{where}
&&
&
\sum_i (q^+_i + q^-_i + q^{\dot1}_i + q^{\dot2}_i) n_i 
=
\sum_i \sum_\zeta q^\zeta_i n_i
= N .
&&
\end{flalign}
\esub
We have thus shown that its four-point function with the interaction operator factorizes as
\begin{align}	\label{Gufact}
\begin{split}
& G(u,\bar u) 
	= 
\Big\langle
\Big\lord
\prod_{i,\zeta}
( R^\zeta_{[n_i]})^{q^\zeta_i} 
\Big\rord^\dagger \!\!
	(\infty,\bar \infty) 
\,	\Oint(1,\bar 1) 
\,	\Oint (u,\bar u)
\Big\lord
\prod_{i,\zeta}
( R^\zeta_{[n_i]})^{q^\zeta_i} 
\Big\rord
	(0,\bar 0)
\Big\rangle 
\\
&
	= 
\sum_{i>j} \sum_{\zeta, \zeta' \neq \zeta}
(q_i^\zeta q_j^{\zeta'})^2
\Big\langle
\big\lord
	R^{\zeta}_{[n_i]} 
	R^{\zeta'}_{[n_j]}
\big\rord^\dagger 
			(\infty,\bar \infty) 
\,	\Oint(1,\bar 1) 
\,	\Oint (u,\bar u)
\big\lord
	R^{\zeta}_{[n_i]} 
	R^{\zeta'}_{[n_j]} 
\big\rord
			(0,\bar 0)
\Big\rangle 
\\
&
+
\sum_{i} \sum_{\zeta, \zeta' \neq \zeta}
(q_i^\zeta q_i^{\zeta'})^2
\Big\langle
\big\lord
	R^{\zeta}_{[n_i]} 
	R^{\zeta'}_{[n_i]}
\big\rord^\dagger 
			(\infty,\bar \infty) 
\,	\Oint(1,\bar 1) 
\,	\Oint (u,\bar u)
\big\lord
	R^{\zeta}_{[n_i]} 
	R^{\zeta'}_{[n_i]} 
\big\rord
			(0,\bar 0)
\Big\rangle 
\\
&
+
\sum_{i,\zeta} 
{\scr P}^2(q_i^\zeta)
\Big\langle
\big\lord
	R^{\zeta}_{[n_i]} 
	R^{\zeta}_{[n_i]}
\big\rord^\dagger
			(\infty,\bar \infty) 
\,	\Oint(1,\bar 1) 
\,	\Oint (u,\bar u)
\big\lord
	R^{\zeta}_{[n_i]} 
	R^{\zeta}_{[n_i]} 
\big\rord
			(0,\bar 0)
\Big\rangle 
\end{split}
\end{align}
where ${\scr P}(q)$ is given by Eq.(\ref{scrPq}), and the factors $q^\zeta q^{\zeta'}$ are the number of ways of pairing a set of $q^\zeta$ objects with the a set of $q^{\zeta'}$ different objects.
The coefficients are squared because is if have, e.g.~${\scr P}(q)$ pairings on the left, we also have ${\scr P}(q)$ different pairings on the right, hence ${\scr P}(q) \times {\scr P}(q)$ equivalent factorizations.
Again, we refer to Appendix \ref{AppSumoveOrb1} for more details.

\subsection{$N$-scaling}	\label{SectNscaling}

When restricted to correlators with ${\bf g} = 0$, every function appearing in the r.h.s.~of Eq.(\ref{Gufact}) is connected. Their scaling in the large-$N$ limit is given by Eq.(\ref{Ndep}), along with the analysis for double-cycle factorizations below Eq.(\ref{DoubleCycle4pt}).
When $n_i > 1$ for all $q_i \neq 0$, i.e.~when every component of the Ramond field (\ref{GenRam}) is twisted, then every correlator in the r.h.s.~of Eq.(\ref{Gufact}) is a function with $Q = 6$ ramification points, and from Eq.(\ref{Ndep}) we have that
\be
G(u,\bar u) |_{{\bf g} = 0} \sim N^{-2} , 
\qquad
\text{if}
\quad
n_i > 1 \quad \forall \quad q_i^\zeta \neq 0 .
\ee

When exactly one of the cycles has length one, i.e.~when $n_k = 1$ with $q_k^{\zeta'} = 1$ and $n_i > 1$ for all $q_i^\zeta \neq 0$ and $i \neq k$, $\zeta\neq\zeta'$ then there will be terms like
\begin{align*}
\Big\langle
\big\lord
	R^{\zeta'}_{[1]_\infty} 
	R^{\zeta}_{[n_i]_\infty}
\big\rord^\dagger
	\Oincov_{[2]_1} 
	\Oincov_{[2]_u}
\big\lord
	R^{\zeta'}_{[1]_0} 
	R^{\zeta}_{[n_i]_0}
\big\rord
\Big\rangle_{{\bf g} = 0} 
	&\sim N^{1 - {\bf g} - 2} 
	= N^{-1} 
\end{align*}
because this correlators have only $Q = 4$ ramification points.
Since the other functions scale as $N^{-2}$, the components with the untwisted field dominate at large $N$.

When there is \emph{at least two} untwisted fields in the operator (\ref{GenRam}), there will be correlators pairing these two untwisted components; the covering surface associated with these correlators has only $Q = 2$ ramification points, the ones given by the cycles $(2)_1$ and $(2)_u$ of the interaction. Then Eq.(\ref{Ndep}) gives
\begin{align*}
\Big\langle
\big\lord
	R^{\zeta'}_{[1]_\infty} 
	R^{\zeta}_{[1]_\infty}
\big\rord^\dagger
	\Oincov_{[2]_1} 
	\Oincov_{[2]_u}
\big\lord
	R^{\zeta'}_{[1]_0} 
	R^{\zeta}_{[1]_0}
\big\rord
\Big\rangle_{{\bf g} = 0} 
	&\sim N^{1 - {\bf g} - 1} 
	= \rm{const.}
\end{align*}
So these functions do not scale with $N$ in the large-$N$ limit.
(This is actually expected: these are simply two-point functions of $\Oint$ in an untwisted Ramond state, and the normalization factor $\scr S_n(N)$ is such that two-point functions are $N$-independent.)

In summary, we can sketch the $N$-scaling of a general four-point function as
\be
G(u,\bar u) |_{{\bf g}=0} \sim
\underbrace{ N^0 }_{\substack{ \text{Terms with} \\ \text{two untwisted} \\ \text{components}}}
+
\underbrace{ N^{-1} }_{\substack{ \text{Terms with} \\ \text{one twisted and} \\ \text{one untwisted} \\ \text{components}}}
+
\underbrace{ N^{-2} }_{\substack{ \text{Terms with} \\ \text{two twisted} \\ \text{components}}}
\ee
Hence we can see that, because of the factorizations, for the four-point function (\ref{Gufact}) the genus expansion is not exactly the same thing as the large-$N$ expansion. 
In what follows, we will work at the genus-zero order, and evaluate all possible contributions to $G(u,\bar u)$. 
That is, we will compute every four-point functions containing two interaction operators and two \emph{double-cycle} Ramond ground states.

\section{Double-cycle connected four-point functions}	\label{SectDoublCyclFunc}

We have shown that the generic multi-cycle Ramond ground state four-point function (\ref{Gu}) factorizes into functions  with only double-cycle composite Ramond fields as in Eq.(\ref{GuFact}). 
We are now going to compute the $S_N$-invariant functions
\be	\label{Gcfull}
G_{\zeta_1, \zeta_2}(u,\bar u)
=
\Big\langle 
\big\lord
R^{\zeta_1}_{[n_1]}
R^{\zeta_2}_{[n_2]} 
\big\rord^\dagger
(\infty,\bar \infty) 
	\;
	\Oincov_{[2]}(1, \bar 1) 
	\;
	\Oincov_{[2]} (u , \bar u) 
	\;
\big\lord
R^{\zeta_1}_{[n_1]} 
R^{\zeta_2}_{[n_2]}
\big\rord
 (0,\bar 0)
\Big\rangle  
\ee
for all possible combination of $\zeta_i = \pm, \dot1, \dot2$, with ${\bf g} = 0$.
(From now on, for the sake of simplicity, we drop the labels $n_1,n_2$ used in Eq.(\ref{DoubleCycle4ptSmCl}).)

We are going to derive $G_{\zeta_1 \zeta_2}(u,\bar u)$ using two different methods: the so-called `stress tensor method'  \cite{Dixon:1986qv} and the path-integral Lunin-Mathur (LM) construction \cite{Lunin:2000yv,Lunin:2001pw}.
Both have been widely used in the context of the $\cal N = (4,4)$ $S_N$ orbifold, see e.g.~\cite{Arutyunov:1997gi,Arutyunov:1997gt,Pakman:2009zz,Pakman:2009ab,Pakman:2009mi,Lima:2020boh,Lima:2020kek,Lima:2020nnx,Lima:2020urq} for the former and \cite{Burrington:2012yq,Burrington:2012yn,Burrington:2015mfa,Burrington:2017jhh,Tormo:2018fnt,Burrington:2018upk,deBeer:2019ioe}
for the latter.
They provide complementary points of view of the same problem, and each has its advantages.
The LM technique computes the four-point function directly, 
but it requires a complicated regularization of the path integral around the ramification points. 
The stress-tensor method 
makes use of conformal symmetry only, 
but it requires that one solves a first-order differential equation to obtain the four-point function.

\subsection{Covering map for connected  functions}	\label{SectCovMaps}

The non-trivial monodromies of a twisted correlation function defined on the `base' Riemann sphere $\bbS^2_{\rm{base}}$ can be implemented \cite{Lunin:2000yv} by lifting $\bbS^2_{\rm{base}}$ to a ramified covering surface $\Sigma_c$, where there is only one copy of the $\cal N = (4,4)$ SCFT --- a single field on multiple sheets of $\Sigma_c$ being mapped to different-copy fields on $\bbS^2_{\rm{base}}$.
We are interested in genus-zero covering surfaces, hence $\Sigma_c = \bbS^2_{\rm{cover}}$.
This surface is defined by a covering map 
$\{z \in \bbS^2_{\rm{base}} \}\mapsfrom \{t \in \bbS^2_{\rm{cover}}\}$, 
with one ramification point for each twist on the base; for the connected four-point function, we have six points:
 $t_* = \{0, t_0, t_1, x\}$, the point at infinity, and the finite images $t_\infty$ of $z=\infty$, where
\bsub	\label{localzt}
\begin{align}
z(t) &= z_* + b_{t_*} (t-t_*)^{n_*} + \cdots ,	
&& t \to t_*
\\
z(t) &= b_\infty t^{n_\infty} + \cdots, 
&& t \to \infty
\\
z(t) &= b_{t_\infty} (t - t_\infty)^{- n_{t_\infty}} + \cdots ,
&& t \to t_\infty
\end{align}
\esub
The coefficients $b_*$ enter the lifting of fermions, see Eqs.(\ref{rim}) below.
For example,  the lifting of $\Oint$ is%
	\footnote{%
	Note that  the contour integrals in (\ref{DeformwithMo}) disappear on the covering, because they simply pick up a residue. The power of $b_{t_*}$ also gets a contribution from the Jacobian of the transformation from $z$ to $t$. Finally, one must take into account the cocycles (on the covering surface) to obtain correct signs; see e.g.~\cite{Burrington:2012yq} for a detailed derivation.
	}
\be
\begin{split}
 \Oincov(t_*,\bar t_*)  =
  		&
		2 |b_{t_*}|^{-\frac54}
  		 \Big[
		 \lor \pa X^{\dot 1 1} \, e^{+\frac{i}{2} (\phi_1 + \phi_2)} 
			\left( \bar \pa X^{\dot 1 2} e^{+\frac{i}{2} (\tilde \phi_1 + \tilde \phi_2)} 
				- (\bar \pa X^{\dot 1 1})^\dagger e^{- \frac{i}{2} (\tilde \phi_1 + \tilde \phi_2)} 
			\right) \ror	
\\
		- &	\lor \pa X^{\dot 1 2} \, e^{+\frac{i}{2} (\phi_1 + \phi_2)} 
			\left( (\bar \pa X^{\dot 1 2})^\dagger e^{- \frac{i}{2} (\tilde \phi_1 + \tilde \phi_2)} 
				+ \bar \pa X^{\dot 1 1} e^{+ \frac{i}{2} (\tilde \phi_1 + \tilde \phi_2)} 
			\right) \ror	
\\
		+ &	\lor (\pa X^{\dot 1 1})^\dagger e^{- \frac{i}{2} (\phi_1 + \phi_2)} 
			\left( (\bar \pa X^{\dot 1 2})^\dagger e^{-\frac{i}{2} (\tilde \phi_1 + \tilde \phi_2)} 
				+ \bar \pa X^{\dot 1 1} e^{+\frac{i}{2} (\tilde \phi_1 + \tilde \phi_2)} 
			\right) \ror	
\\
		+ &	\lor (\pa X^{\dot 1 2})^\dagger \, e^{-\frac{i}{2} (\phi_1 + \phi_2)} 
			\left( \bar \pa X^{\dot 1 2} e^{+ \frac{i}{2} (\tilde \phi_1 + \tilde \phi_2)} 
				- (\bar \pa X^{\dot 1 1})^\dagger e^{- \frac{i}{2} (\tilde \phi_1 + \tilde \phi_2)} 
			\right) \ror	
			\Big] 
\end{split}	\label{InteraOpera}
\ee  
where $(t_*,\bar t_*)$ are coordinates on the covering with $n_* =2$. We drop the twist index in the lifted operator $\Oincov$.

For the fully-connected four-point function (\ref{Gcfull}), the covering surface $\bbS^2_{\rm{cover}}$ is given by the covering map \cite{Lima:2020nnx}
\be		\label{coverm}
z(t) = \left( \frac{t}{t_1}\right)^{n_1} 
	\left( \frac{t-t_0}{t - t_\infty} \right)^{n_2} 
	\left( \frac{t_1-t_\infty }{t_1-t_0} \right)^{n_2}.
\ee
One can see that the two independent twist cycles $(n_1)_0$ and $(n_2)_0$ inserted at the same point $z = 0$ on the base are reproduced on the covering by two separate ramification points $t = 0$ and $t = t_0$. The same goes for the twists at $z = \infty$, lifted to $t = \infty$ and $t = t_\infty$.
The twists $(2)_1$ and $(2)_u$ impose the condition of a vanishing derivative, 
$z'(t) \sim (t - t_1)(t-x)$,
where we have defined the point $x \in \bbS^2_{\rm{cover}}$ by $u \equiv z(x)$.
Thus $t_1$ and $x$ must be the solutions of the quadratic equation $z'(t) = 0$,
which imposes two relations between the four parameters $t_0$, $t_\infty$, $t_1$ and $x$.
We are free to choose one final relation between them, and then three of the parameters become functions of the remaining one. 
We choose the following \cite{Lima:2020nnx} parameterization in terms of the ``free-roving'' coordinate $x$,
\begin{align}		\label{tim}
\begin{split}
t_0 &= x-1 ,
\\
t_1 &= \frac{(x-1) (n_1+ n_2 x- n_2)}{n_1 + n_2 x} ,
\\
t_\infty &= x- \frac{n_2 x }{n_1+n_2 x}
\end{split}
\end{align}
As in the case of the map chosen in \cite{Arutyunov:1997gt}, this parameterization is such that the the function $u(x) \equiv z(x)$ is rational,
\be
u(x) = \Bigg( \frac{x+ \frac{n_1}{n_2}}{x-1} \Bigg)^{n_1+n_2} 
	\Bigg( \frac{x}{x - 1 + \frac{n_1}{n_2} } \Bigg)^{n_1-n_2}  .
	\label{uxm}
\ee
The obvious asymmetry between $n_1$ and $n_2$ in our parameterization of the covering surface was introduced in Eq.(\ref{coverm}), when we chose to lift $\s_{(n_1)}$ to the origin of $\bbS^2_{\rm{cover}}$. We could just as well have chosen the opposite, lifting $\s_{(n_2)}$ to the origin, and obtaining a different map, with $n_1$ and $n_2$ exchanged. These two maps are isomorphic \cite{Lima:2020nnx}, and, in any case, when we map the four-point functions back to the base sphere the results are symmetric in $n_1$ and $n_2$.

The multiple inverses of (\ref{uxm}) cannot be found in general, but we can find the asymptotic functions $x^{u_*}_{\frak a}(u)$ near the points $u = u_*$ where there are coincidence limits in the four-point function (\ref{Gcfull}).
As discussed above, we can assume that 
\be
n_1 > n_2 
\ee
without loss of generality.
Then for $u \to u_* = 0$,  we get the two functions
\bsub	\label{xfraka}
\be
(u \to 0) \quad
	\begin{sqcases}
x \to 0,  
	& x^0_{\frak1}(u) 
	\approx  \left( 1 - \tfrac{n_1}{n_2} \right) 
			\left( \tfrac{n_1}{n_2} \right)^{- \frac{n_1+n_2}{n_1-n_2}} 
			u^{\frac{1}{n_1-n_2}} + \cdots
\\
x \to - \tfrac{n_1}{n_2} , 
	&x^0_{\frak2}(u) 
	\approx - \tfrac{n_1}{n_2}
	+ \left(1 + \tfrac{n_1}{n_2} \right)
		\left(\tfrac{n_1}{n_2} \right)^{- \frac{n_1-n_2}{n_1+n_2}}
		 u^{\frac{1}{n_1+n_2}}
\end{sqcases}	\label{xto0minnuxComp}
\ee
When $u \to u_* =1$, we have $x = \infty$ and
$x =  -\tfrac{n_1-n_2 }{2n_2}$;	
the behavior of $u(x)$ near $x = \infty$ can be found with the conformal transformation $x = 1 / \vare$,
\begin{align}
(u \to 1) \quad
	\begin{sqcases}
	x \to \infty , 
	& x^1_{\frak1}(u) \approx - 4n_1 (1-u)^{-1} 
\\
\\
	x \to - \tfrac{n_1-n_2 }{2n_2} , 
 	& x^1_{\frak2}(u) \approx - \tfrac{n_1-n_2 }{2n_2}
\\
&\quad\quad\quad	
				+ 3^{\frac13} 2^{-2}
				 (n_1-n_2)^{\frac23} 
				 (n_1 + n_2)^{\frac23}
				 n_1^{-\frac13}
				 n_2^{-\frac43}
				 (1 - u)^{\frac{1}{3}} 
	\end{sqcases} \label{xaxp4nsComp}
\end{align}
\esub

\subsection{Four-point functions} 	\label{SectStressTensor}

In the LM approach \cite{Lunin:2000yv,Lunin:2001pw}, we compute $G^c_{\scr R}$ as a path integral. 
Choosing fiducial metrics $ds^2_{\rm{base}} = dz d\bar z$ and $ds^2_{\rm{cover}} = dt d \bar t$ on the two spheres, the covering map relates the base and the covering surfaces by a Weyl transformation
\be
ds^2_{\rm{base}} = e^\Phi ds^2_{\rm{cover}} ,
\ee
so the path integrals on each surface are related,
\begin{flalign}
&&
&G_{\zeta_1\zeta_2} \big (x,\bar x) |_{(\rm{base})} = e^{S_L} \; G_{\zeta_1\zeta_2}^{(\rm{cover})} (x, \bar x),		\label{CoverBaseG}
&&
\\
\text{where}
&&
&
S_L = \frac{6}{96 \pi} \int \! d^2t \sqrt{g_{\rm{cover}}} \Big[ g_{\rm{cover}}^{ab} \pa_a \Phi \pa_b \Phi + 2 \Phi R(g_{\rm{cover}}) \Big]
&&
\end{flalign}
is the Liouville action \cite{Friedan:1982is},
and $G_{\zeta_1,\zeta_2}^{(\rm{cover})}(x,\bar x)$ is the untwisted correlator of the fields lifted to the covering surface 
(where we drop twist indices of operators),
\be	\label{GcoverxMain}
\begin{split}
&G_{\zeta_1,\zeta_2}^{\rm{(cover)}}(x,\bar x)
\\
&	= 
	\Big\langle
	R^{\zeta_1 \dagger}(\infty,\bar \infty) 
	R^{\zeta_2 \dagger} (t_\infty, \bar t_\infty) 
	\Oincov (t_1,\bar t_1) 
	\Oincov (x, \bar x)
	R^{\zeta_1} (t_0,\bar t_0) 
	R^{\zeta_2}(0,\bar 0) 
	\Big\rangle	,
\end{split}
\ee
Note that the connected four-point functions on the base lift to six-point functions on the covering, since the composite operator
$\lord R^{\zeta_1}_{[n_1]}  R^{\zeta_2}_{[n_2]}\rord$
 is lifted to two operators at different points by the covering map.

The functions at the r.h.s.~of Eq.(\ref{CoverBaseG}) are naturally parameterized by $x$, hence the base-sphere function at the l.h.s.~appears parameterized by $x$ instead of $u$.
The inverse maps $x_{\frak a}(u)$ encode the restoration of the  twisted boundary conditions of the base-sphere function,
\begin{align}
G_{\zeta_1\zeta_2}(u,\bar u) |_{\bf g = 0} &= 	
	\frac{\varpi(n_1 n_2)}{N^2}
	\sum_{\frak a = \frak1}^{2 \max(n_1,n_2)} G_{\zeta_1\zeta_2} ( x_{\frak a}(u),\bar x_{\frak a}(\bar u))  .
\label{Guufromx}
\end{align}
Here $2 \max(n_1, n_2)$ is the number of inverses of $u(x)$, i.e.~the number of solutions $x_{\frak a}(u)$ of $u(x) = u_*$ for a general $u_*$;
 $\varpi$ is a combinatoric factor; see Appendix B of Ref.\cite{Lima:2020nnx} for a detailed discussion.

Let us now compute the r.h.s.~of Eq.(\ref{CoverBaseG}).
The Liouville factor depends only on the twists.
The Liouville field $\Phi = \log | dz/dt |^2$ must be carefully regularized around the ramification points (\ref{localzt}), and the Liouville action is fixed by the local behavior%
	\footnote{%
	See  Eq.(D.63) of Ref.\cite{Avery:2010qw} for a formula like (\ref{SL}), with twists  inserted at infinity.
	}
\cite{Lunin:2000yv}
\begin{align}	\label{SL}
\begin{split}
S_L = 
	- \frac{1}{2} 
	&\Bigg[ 
	\sum_* \frac{n_*-1}{n_*} \log | b_*| 
	+ 
	\frac{n_{t_\infty} +1}{n_{t_\infty}} \log |b_{t_\infty}|
	- 
	\frac{n_\infty - 1}{n_\infty} \log | b_{\infty}|
\\
&\quad
	+
	\sum_* (n_* -1) \log n_* 
	-
	(n_{t_\infty} +1) \log n_{t_\infty}
	-
	(n_\infty + 3) \log n_\infty
\\
&\quad
	+ \text{Regularization factors}	
	\Bigg] .
\end{split}
\end{align}
The coefficients in (\ref{localzt}) have to be expressed as functions of $x$ (which requires knowledge of the complete function $z(t)$). This is easily found by expanding (\ref{coverm}),
\bsub	\label{bofx}
\begin{align}
b_0 &= 
		x^{-n_2}
		(x-1)^{-n_1}
		(x + \tfrac{n_1}{n_2})^{n_1 + n_2}
		(x + \tfrac{n_1}{n_2} - 1)^{-n_1}
\\
b_{t_0} &=
		(- \tfrac{n_1}{n_2} )^{-n_2}
		(x-1)^{-n_2}
		(x + \tfrac{n_1}{n_2})^{n_1 + n_2}
		(x + \tfrac{n_1}{n_2} - 1)^{n_2-n_1}
\\
b_{t_1}
	&= 
		- n_1
		(x-1)^{-2}
		(x + \tfrac{n_1}{n_2})^{2}
		(x + \tfrac{n_1}{n_2} - 1)^{-2}
		(x + \tfrac{n_1 - n_2}{2n_2})
\\
\begin{split}
b_{x}
	&= 
		n_1
		x^{n_1-n_2-2}
		(x-1)^{-(n_1+n_2)}
		(x + \tfrac{n_1}{n_2})^{n_1+n_2}
\\
&\qquad\qquad\qquad\qquad\quad
	\times
		(x + \tfrac{n_1}{n_2} - 1)^{n_2-n_1}
		(x + \tfrac{n_1 - n_2}{2n_2})
\end{split}
\\
b_{t_\infty} &=
		(\tfrac{n_1}{n_2} )^{n_2}
		x^{n_1}
		(x-1)^{-(n_1+n_2)}
		(x + \tfrac{n_1}{n_2})^{-n_2}
		(x + \tfrac{n_1}{n_2} - 1)^{n_2}
\\
b_{\infty} &=
		(-1)^{n_2}
		(x-1)^{-(n_1+n_2)}
		(x + \tfrac{n_1}{n_2})^{n_1}
		(x + \tfrac{n_1}{n_2} - 1)^{n_2-n_1}
\end{align}
\esub
We can now find $e^{S_L}$ inserting (\ref{bofx}) into the Liouville action (\ref{SL}),
\begin{align}	\label{SLofx}
\begin{split}
S_L(x)    &=
	- 
	\frac{
	2 n_2^2 + (2+3 n_2) (n_1 - n_2) n_1
	}{
	4 n_1 n_2}
	\log |x|
\\
&\quad
	+
	\frac{
	2 n_2^2 + (2+3 n_2) (n_1 + n_2) n_1
	}{
	4 n_1 n_2}
	\log | x-1 |
\\
&\quad
	+
	\frac{
	2 n_2^2 + (2-3 n_2) (n_1 + n_2) n_1
	}{
	4 n_1 n_2}
	\log | x + \tfrac{n_1}{n_2} |
\\
&\quad
	-
	\frac{
	2 n_2^2 + (2-3 n_2) (n_1 - n_2) n_1
	}{
	4 n_1 n_2}
	\log | x + \tfrac{n_1}{n_2} -1 |
\\
&\quad
	-
	\frac{1}{2}
	\log | x + \tfrac{n_1-n_2}{2n_2}  |
	+
	\frac{3-n_2}{4} \log n_1	
	+
	2 \log n_2
\end{split}
\end{align}
The regularization factors of (\ref{SL}), omitted in (\ref{SLofx}) are universal and can be absorbed into a definition of the twist fields via the path integral.
The two last $x$-independent terms can also be changed by different normalizations of the twist fields, so we will replace them for a ($x$-independent) constant in what follows.
After these scalings, the non-trivial terms in the Liouville factor give the correlator of bare twists:
\be	\label{gsig}
\begin{split}
	\Big\langle 
	\lord \s_{[n_1]} \s_{[n_2]} \rord (\infty, \bar \infty) \,
	\s_{[2]}(1, \bar 1) \,
	 \s_{[2]}(u,\bar u) \,
	 \lord \s_{[n_1]} \s_{[n_2]} \rord (0,\bar 0) 
	 \Big\rangle 
\qquad\qquad
\\
=
	\frac{\varpi(n_1 n_2)}{N^2}
	\sum_{\frak a = \frak1}^{2 \max(n_1,n_2)} \exp S_L ( x_{\frak a}(u),\bar x_{\frak a}(\bar u)) .
\end{split}
\ee

The function (\ref{GcoverxMain}) does, of course, depend on the specific Ramond fields.
For the purpose of clarity, we now proceed by working with the composite operator with one R-neutral and one R-charged ground states, namely 
$\lord R^{\dot1}_{[n_1]}  R^{+}_{[n_2]} \rord$, and then we state the final result for the other double-cycle fields.
A general computation is given in Appendix \ref{General4pt}.
Fermionic exponentials inserted at the ramification points, when lifted to the covering surface, carry a factor of $b$,  see \cite{Lunin:2001pw}; thus
\begin{align}
R^{\dot1\dagger}(\infty) 
R^{+\dagger} (t_\infty) 
&=
	b_\infty^{\frac{1}{4n_1}}
	b_{t_\infty}^{\frac{1}{4n_2}}
	e^{
	 \frac{i}{2} [ \phi_{1} (\infty) +  \phi_{2} (\infty) ] 
	 }
	\
	e^{
	- \frac{i}{2}
	[  \phi_{1}(t_\infty) - \phi_{2}(t_\infty) ]
	} ,
\\
R^{\dot1}(0) 
R^{+} (t_0) 
&=
	b_0^{-\frac{1}{4n_1}}
	b_{t_0}^{-\frac{1}{4n_2}}
	e^{
	- \frac{i}{2} [  \phi_{1} (0) +  \phi_{2} (0) ] 
	}
	\
	e^{
	\frac{i}{2}	[  \phi_{1}(t_0) -  \phi_{2}(t_0) ]
	} .
\end{align}
The lifted interaction operators (\ref{InteraOpera}) also carry factors $b_{t_1}$ and $b_x$.
Bosonic currents $\pa X^{A\dot A}_I(z)$, $\bar \pa X^{A\dot A}_I(\bar z)$ simply lift to $\pa X^{A\dot A}(t)$, $\bar \pa X^{A\dot A}(\bar t)$.
The product of two interaction operators has several terms, but many of them do not contribute to the four-point function:
the bosonic currents factorize as two-point functions, all of which vanish except for the ones in (\ref{twopntboconj}),
hence
\be	\label{Gbosn}
\begin{split}
G_{\dot1 +}^{\rm{(cover)}} 
	&= 
	4 
	\left|
	b_\infty^{\frac{1}{4n_1}}	
	b_{t_\infty}^{\frac{1}{4n_2}}
	b_{t_1}^{- \frac{5}{8}}
	b_{x}^{- \frac{5}{8}}
	b_0^{-\frac{1}{4n_1}}	
	b_{t_0}^{-\frac{1}{4n_2} }
	\right|^2
	 \left| \frac{2}{ (t_1 - x)^{2} } \right|^2 
	\times 
		  G^F_{\dot1+} .
\end{split}	
\ee
The remaining factor, $G_F$, involves only the fermionic exponentials in the terms which did not vanish due to the bosonic factors.
They group into two different contributions:
\be	\label{GFIIImain}
\begin{split}
G^F_{\dot1+} = 
\Big\langle
&	e^{  \frac{i}{2} [  \phi_{1}  +  \phi_{2}  ] } (\infty)
	e^{  \frac{i}{2} [  \bphi_{1}  +  \bphi_{2}  ] } (\bar\infty) 
	e^{- \frac{i}{2} [  \phi_{1} - \phi_{2} ]} (t_\infty)
	e^{- \frac{i}{2} [  \bphi_{1} - \bphi_{2} ]} (\bar t_\infty)
\\
&\quad
\times
\Big(I + II \Big)
	e^{- \frac{i}{2} [  \phi_{1} +  \phi_{2}  ] } (0)
	e^{- \frac{i}{2} [  \bphi_{1} +  \bphi_{2}  ] } (\bar0)
	e^{\frac{i}{2}[  \phi_{1}  - \phi_{2}]} (t_0)
	e^{\frac{i}{2}[  \bphi_{1} -  \bphi_{2}]} (\bar t_0)
\Big\rangle
\end{split}
\ee
where
\begin{align}
\begin{split}		 \label{ints}
I &= 
	2
	e^{- \frac{i}{2}(\phi_1 + \phi_2)}(t_1)
	\,
	e^{ \frac{i}{2}(\phi_1 + \phi_2 )} (x)
\\
&\quad
\times
\bigg[ 
	e^{ \frac{i}{2}(\bphi_1 + \bphi_2)} (\bar t_1)
	\,
	e^{- \frac{i}{2}(\bphi_1 + \bphi_2)} (\bar x)
	+
	e^{- \frac{i}{2}(\bphi_1 + \bphi_2)} (\bar t_1)
	\,
	e^{ \frac{i}{2}(\bphi_1 + \bphi_2)} (\bar x)
\bigg]
\\
II &=
	2
	e^{ \frac{i}{2}(\phi_1 + \phi_2)} ( t_1)
	\,
	e^{- \frac{i}{2}(\phi_1 + \phi_2)} (x)
\\
& \quad
\times 
	\bigg[ 
	e^{- \frac{i}{2}(\bphi_1 + \bphi_2)} (\bar t_1)
	\,
	e^{ \frac{i}{2}(\bphi_1 + \bphi_2)} (\bar x)
	+
	e^{ \frac{i}{2}(\bphi_1 + \bphi_2)} (\bar t_1)
	e^{- \frac{i}{2}(\bphi_1 + \bphi_2)} (\bar x)
	\bigg]
\end{split}
\end{align}
Explicit computation of the correlators gives
\begin{align}	\label{GFfinalMain}
\begin{split}
&G^F_{\dot1+}
=
	2
	\Bigg|
	(  t_\infty -  t_0 )^{ - \frac12 } ( t_1 -  x)^{- \frac{1}{2} }
\Big[
	\Big( \frac{x}{t_1} \Big)^{ \frac{1}{2} }
+
	\Big( \frac{x}{t_1} \Big)^{ - \frac{1}{2} }
\Big]	
\Bigg|^2
\end{split}
\end{align}	
Combining Eqs.(\ref{GFfinalMain}), (\ref{Gbosn}), and  (\ref{SLofx}), and using Eqs.(\ref{tim}), Eqs.(\ref{bofx}), we get
\begin{align}
\begin{split}
G_{\dot1+}(x,\bar x) |_{\rm{base}}
= \Bigg| 
	C \
	 \frac{
	x^{1-n_1+n_2}
	(x-1)^{1+n_1+n_2}
	( x+ \frac{n_1}{n_2} )^{ 1 - n_1 - n_2 }
	(x -1 + \frac{n_1}{n_2} )^{ 1 + n_1 - n_2 }
	}{
	( x + \frac{n_1-n_2}{2n_2} )^4
	}
	&
\\
\times	
	\Big[
	(x-1)(x - 1 + \tfrac{n_1}{n_2}) + x ( x + \tfrac{n_1}{n_2} )	
	\Big]  
	\Bigg|^2 .
	&
\end{split}	
\end{align}

The computations for the other operators are very similar. 
From now on, we drop the indication $|_{\rm{base}}$ of the correlation functions, which for all double-cycle fields can be expressed as
\be
G_{\zeta_1\zeta_2}(x,\bar x) = \big| G_{\zeta_1\zeta_2}(x) \big|^2 ,
\ee
where
\begin{align} \label{Gm21p}
\begin{split}
&G_{\dot1 +}(x) = 
	C \
	 \frac{
	x^{1-n_1+n_2}
	(x-1)^{1+n_1+n_2}
	( x+ \frac{n_1}{n_2} )^{ 1 - n_1 - n_2 }
	(x -1 + \frac{n_1}{n_2} )^{ 1 + n_1 - n_2 }
	}{
	( x + \frac{n_1-n_2}{2n_2} )^4
	}
\\
&\qquad\qquad\qquad\qquad\qquad\qquad\qquad \ \;
\times	
	\Big[
	(x-1)(x - 1 + \tfrac{n_1}{n_2}) + x ( x + \tfrac{n_1}{n_2} )	
	\Big]
\\
&G_{\dot1-} (x) = G_{\dot1+}(x)
\end{split}	
\end{align}
\begin{align}	\label{G2112}
\begin{split}
G_{\dot1\dot2}(x) = 
	C \
	\frac{
	x^{1-n_1+n_2}
	(x-1)^{1+n_1+n_2}
	( x+ \frac{n_1}{n_2} )^{ 1 - n_1 - n_2 }
	(x -1 + \frac{n_1}{n_2} )^{ 1 + n_1 - n_2 }
	}{
	( x + \frac{n_1-n_2}{2n_2} )^4
	}
	&
\\
\times	
	\Big[
	x^2  + \left( x -1 + \tfrac{n_1}{n_2} \right)^2 	
	\Big]
	&
\end{split}	
\end{align}
\begin{align}	\label{G2211}
\begin{split}
G_{\dot1\dot1}(x) = 
	C 
	\frac{
	x^{1-n_1+n_2}
	(x-1)^{1+n_1+n_2}
	( x+ \frac{n_1}{n_2} )^{ 1 - n_1 - n_2 }
	(x -1 + \frac{n_1}{n_2} )^{ 1 + n_1 - n_2 }
	}{
	( x + \frac{n_1-n_2}{2n_2} )^4
	}
	&
\\
\times	
	\Big[
	(x - 1)^2  + \left( x +  \tfrac{n_1}{n_2} \right)^2
	\Big]
	&
\end{split}	
\end{align}
\begin{align}	\label{Gmpmp}
\begin{split}
G_{+-}(x) = 
	2
	C
	 \frac{
	x^{2-n_1+n_2}
	(x-1)^{1+n_1+n_2}
	( x+ \frac{n_1}{n_2} )^{ 1 - n_1 - n_2 }
	(x -1 + \frac{n_1}{n_2} )^{ 2 + n_1 - n_2 }
	}{
	( x + \frac{n_1-n_2}{2n_2} )^4
	}
\end{split}	
\end{align}
\begin{align}	\label{Gmmpp}
\begin{split}
G_{++}(x) = C
	 \frac{
	x^{1-n_1+n_2}
	(x-1)^{2+n_1+n_2}
	( x+ \frac{n_1}{n_2} )^{ 2 - n_1 - n_2 }
	(x -1 + \frac{n_1}{n_2} )^{ 1 + n_1 - n_2 }
	}{
	( x + \frac{n_1-n_2}{2n_2} )^4
	}
\end{split}	
\end{align}
We can find all these expressions from formula (\ref{Gcofx}), by using the parameters in Table \ref{ZNSR}.
These functions exhaust all possible combinations of two twisted Ramond fields; recall that
$G_{\dot2+}(x,\bar x) = [G_{\dot1-}(x,\bar x)]^* =  G_{\dot1-}(x,\bar x)$,
$G_{\dot2\dot2}(x,\bar x)  =  G_{\dot1\dot1}(x,\bar x)$,
etc.

Formulae (\ref{Gmpmp}) and (\ref{Gmmpp}) were derived in \cite{Lima:2020nnx} using the stress-tensor method \cite{Dixon:1986qv,Arutyunov:1997gi,Arutyunov:1997gt,Pakman:2009ab,Pakman:2009zz} instead of the Lunin-Mathur technique. These two functions are simpler than the others (the sum of terms in square brackets is absent) because of a convenient cancelation of terms, so the their derivation cannot be immediately generalized. It is instructive to have a general computation using the stress-tensor, and this is given in the Appendix, in \S\ref{AppStressTensor}.
Here let us briefly outline the procedure.

In the stress-tensor method \cite{Dixon:1986qv}, there is no reference to the path integral: we use only the conformal Ward identity to derive a first-order differential equation,
\be	\label{methodbaseMain}
\begin{split}
&
\pa_u \log G_{\zeta_1\zeta_2}(u) 
\\
&
= 
\underset{z = u}{ \rm{Res} } \,
	\frac{
\big\langle
T(z)
\lord 
R^{(\zeta_1)\dagger}_{[n_1]} 
R^{(\zeta_2)\dagger}_{[n_2]} 
\rord (\infty,\bar \infty)
\Oint (1,\bar 1) 
\Oint (u,\bar u)
\lord 
R^{(\zeta_1)}_{[n_1]} 
R^{(\zeta_2)}_{[n_2]} 
\rord (0,\bar 0)
\big\rangle
	}{
\big\langle
\lord 
R^{(\zeta_1)\dagger}_{[n_1]} 
R^{(\zeta_2)\dagger}_{[n_2]} 
\rord (\infty,\bar \infty)
\Oint (1,\bar 1) 
\Oint (u,\bar u)
\lord 
R^{(\zeta_1)}_{[n_1]} 
R^{(\zeta_2)}_{[n_2]} 
\rord (0,\bar 0)
\big\rangle
	}		.
\end{split}
\ee
Again, the covering map can be used as a facilitator for dealing with the monodromies \cite{Arutyunov:1997gi,Arutyunov:1997gt,Pakman:2009zz,Pakman:2009ab,Pakman:2009mi}.
To find the r.h.s.~of Eq.(\ref{methodbaseMain}) we compute the equivalent function on the covering, namely
\be\label{methodcovChCompMain}
\begin{split}
&
F^{\zeta_1\zeta_2}_{\rm{cover}} (t)
=
	 \frac{
\big\langle
T(t)
R^{(\zeta_1)\dagger}(\infty) 
R^{(\zeta_2)\dagger} (t_\infty) 
\Oincov (t_1,\bar t_1) 
\Oincov (x,\bar x)
R^{(\zeta_1)}(0) 
R^{(\zeta_2)} (t_0) 
\big\rangle
	}{
\big\langle
R^{(\zeta_1)\dagger}(\infty) 
R^{(\zeta_2)\dagger} (t_\infty) 
\Oincov (t_1,\bar t_1) 
\Oincov (x,\bar x)
R^{(\zeta_1)}(0) 
R^{(\zeta_2)} (t_0) 
\big\rangle
		} ,
\end{split}
\ee
The function in the denominator is simply $G_{\zeta_1\zeta_2}^{\rm{(cover)}}(x,\bar x)$ in Eq.(\ref{GcoverxMain}).
A crucial difference from the LM method is that here the factors $b_*$ are irrelevant, as they cancel in the fraction.
Also, $G_{\zeta_1\zeta_2}^{\rm{(cover)}}(x,\bar x)$ often cancels in the fraction as well. 
One must compute the contraction of the operators in the numerator with $T(z)$, and manipulate the result conveniently, see Refs.\cite{Lima:2020boh,Lima:2020kek,Lima:2020nnx,Lima:2020urq}.
For example, for  the field
$\lord R^{\dot1}_{[n_1]}  R^{+}_{[n_2]} \rord$ the final result is
\begin{align}
\begin{split}
F^{\dot1+}_{\rm{cover}}  (t)
&=
\frac{(t_1 - x)^2}{(t-t_1)^2 (t - x)^2} 
+
	\frac{1}{4} \Bigg[ 
		\left( \frac{1}{t} \right)^2
		+
		\left( \frac{1}{t - t_0} - \frac{1}{t-t_\infty} \right)^2
\\
&\qquad\qquad\qquad
		+
		\left( \frac{1}{t - t_1} - \frac{1}{t-x} \right)^2
		+
		\frac{2(t_1 - x)^2}{t ( t - t_1) (t - x) (t_1 + x)}
		\Bigg] .
\end{split}
\end{align}	
Once we have $F^{\zeta_1\zeta_2}_{\rm{cover}} (t)$, we go back to the base surface by mapping $t \mapsto z$. The stress-tensor transforms with the Schwarzian derivative $\{t, z\}$, and one must sum over the different copies/sheets of the covering. Near the point $z = u$, where there is a twist-two operator, there are two copies, so 
\be
\begin{split}
\pa_u \log G_{\zeta_1\zeta_2}(u) 
&
= 
2 \underset{z = u}{ \rm{Res} } 
	\Bigg[
	\{t,z\} + \left( \frac{dt}{dz} \right)^2 F^{\zeta_1\zeta_2}_{\rm{cover}}(t(z))
	\Bigg] .
\end{split}
\ee
The asymptotic form of the map $t(z)$ can be found in Ref.\cite{Lima:2020nnx}. 
The r.h.s.~is expressed as a function of $x$, which appears as a parameter in $t_1,t_0,t_\infty$, etc. Hence instead of $G_{\zeta_1\zeta_2}(u,\bar u)$, we are only able to solve the differential equation after a change of coordinates,
\be
\begin{split}
\pa_x \log G_{\zeta_1\zeta_2}(x) 
&
= 
2 \left( \frac{du}{dx} \right)
\underset{z = u}{ \rm{Res} } 
	\Bigg[
	\{t,z\} + \left( \frac{dt}{dz} \right)^2 F^{\zeta_1\zeta_2}_{\rm{cover}}(t(z))
	\Bigg]
\end{split}
\ee
where $u(x)$ is given in Eq.(\ref{uxm}). Integration gives immediately the functions (\ref{Gm21p})-(\ref{Gmmpp}), with $C$ as an integration constant. Repeating the procedure with $\tilde T(\bar z)$, we obtain the anti-holomorphic part $\bar G(\bar x)$, whence $G(x,\bar x) = G(x) \bar G(\bar x)$.


\section{Dynamics and OPE limits}	\label{SectOPEs}

We can obtain the fusion rules and the structure constants $C_{12k}$ from the OPEs  of fields $\scr O_k(u)$ with dimensions $\Delta_k$,
\be
 \scr O_1 (u,\bar u) \scr O_2(0,\bar 0) = \sum_k  C_{12k} \, |u|^{\Delta_k - \Delta_1 - \Delta_2} \scr O_k(0,\bar 0) + \text{descendants},
 \label{OPE}
 \ee
Charges are conserved: $j^3_k = j^3_1+ j^3_2$.
If we normalize the operators such that
\be
\big\langle \scr O_\ell^\dagger (u,\bar u) \scr O_k(0,\bar 0) \big\rangle
 = \delta_{k\ell} |u|^{-2 \Delta_k} ,
\quad
\text{i.e.}
\quad
\big\langle \scr O_\ell^\dagger (1,\bar 1) \scr O_k(0,\bar 0) \big\rangle
 = \delta_{k\ell},
\ee
 then taking the bracket of the OPE (\ref{OPE}) with $\scr O_k^\dagger$ we find that the coefficient of the leading term in the expansion, i.e.~the structure constant, can be written as 
\be
 C_{12k} = \big\langle \scr O_k^\dagger (\infty,\bar \infty) \scr O_1(1,\bar 1) \scr O_2(0,\bar 0)\big\rangle
\equiv
\big\langle \scr O_k^\dagger \scr O_1\scr O_2  \big\rangle .
\ee
We can obtain this conformal data from the connected four-point functions computed in Sect.\ref{SectDoublCyclFunc}.

\subsection{Different twists $n_1 \neq n_2$}

\begin{table}[t]
\begin{center}
\begin{tabular}{r|| c c c c }
{$G_{\zeta_1\zeta_2}(x)$}	
&{ $G_{\dot1\pm}$} 
&{ $G_{\dot1\dot2}$} 
&{ $G_{\dot1\dot1}$} 
&{ $G_{+-}$} 
\\
\hline
\hline
{ $A_{\zeta_1\zeta_2}$ }
& $\frac{n_2 - n_1}{2n_2}$ 
& $ \frac{(n_1 - n_2)^2}{2n_2^2}$ 
& $ \frac{n_1^2 + n_2^2}{2n_2^2}$ 
& 0
\\
Eq.
&
(\ref{Gm21p})
&
(\ref{G2112})
&
(\ref{G2211})
&
(\ref{Gmpmp})
\end{tabular}
\caption{Values of $A_{\zeta_1\zeta_2}$ in the different four-point functions with neutral fields.} \label{AGEqs}
\end{center}
\end{table}

We are interested in four-point functions 
\be	\label{Gcfull2}
G_{\zeta_1, \zeta_2}(u,\bar u)
=
\Big\langle 
\big\lord
R^{\zeta_1}_{[n_1]}
R^{\zeta_2}_{[n_2]}
\big\rord^\dagger
 (\infty,\bar \infty) 
	\;
	\Oincov_{[2]}(1, \bar 1) 
	\;
	\Oincov_{[2]} (u , \bar u) 
	\;
\big\lord
R^{\zeta_1}_{[n_1]} 
R^{\zeta_2}_{[n_2]}
\big\rord
 (0,\bar 0)
\Big\rangle  
\ee
from which the OPEs between  Ramond fields and interaction operators can be found by taking coincidence limits $u \to 0,1,\infty$, where the inverses of the map $u(x)$ are given by Eqs.(\ref{xfraka}).
The functions (\ref{Gm21p})-(\ref{Gmpmp}) can all be written in the following way:
\bsub	\label{GcompNeuGen}
\begin{align}
& G_{\zeta_1\zeta_2}(x) =
	\Big[	A_{\zeta_1\zeta_2} + x(x - 1 + \tfrac{n_1}{n_2}) \Big]	\cal F(x)
\\	
& \cal F(x)
	= C
	 \frac{
	x^{1-n_1+n_2}
	(x-1)^{1+n_1+n_2}
	( x+ \frac{n_1}{n_2} )^{ 1 - n_1 - n_2 }
	(x -1 + \frac{n_1}{n_2} )^{ 1 + n_1 - n_2 }
	}{
	( x + \frac{n_1-n_2}{2n_2} )^4
	}
\end{align}
\esub
being distinguished only by the constants $A_{\zeta_1\zeta_2}$, listed in Table \ref{AGEqs}.
The dynamics of the double-cycle fields involving at least one R-neutral field  have therefore the same qualitative features.
For $u \to 0$, we obtain the conformal data of the fusion rules
\be	\label{fusionOr}
\Oint \times  \lord R^{\dot1}_{[n_1]}  R^{+}_{[n_2]} \rord  ,
\quad
\Oint \times  \lord R^{\dot1}_{[n_1]}  R^{\dot2}_{[n_2]} \rord ,
\quad
\Oint \times  \lord R^{\dot1}_{[n_1]}  R^{\dot1}_{[n_2]} \rord  ,
\ee
so the universality of (\ref{GcompNeuGen}) already shows that these OPEs  all give rise to operators with the \emph{same} conformal dimensions, although they of course differ by the SU(2) charges, which are conserved. 
The function (\ref{Gmpmp}) for the neutral field 
$\lord R^{+}_{[n_1]}  R^{-}_{[n_2]} \rord$,
must be treated separately, because when $A_{+-} = 0$ the expansion around $x = 0$ (in the limit $u \to 0$) changes.
The function (\ref{Gmmpp}) with the composite field 
$\lord R^{+}_{[n_1]}  R^{+}_{[n_2]} \rord$
does not follow the structure (\ref{GcompNeuGen}), although it can also be related to $\cal F$. 
These two latter cases were discussed in Ref.\cite{Lima:2020nnx}.
From now on, unless otherwise specified, $G_{\zeta_1\zeta_2}$ should be understood to hold the structure (\ref{GcompNeuGen}).

In the limit $u \to 1$ we obtain the fusion rule $[\Oint] \times [\Oint]$, which must agree with the results found elsewhere \cite{Lima:2020boh,Lima:2020kek,Lima:2020nnx,Lima:2020urq} in other correlation functions involving two deformation operators.
Indeed, for $n_1 > n_2 > 1$, we can immediately see that $G_{\zeta_1\zeta_2}(x)$ has singular points at $x = \{0,- \frac{n_1}{n_2}\}$ and $x = \{ - \frac{n_1-n_2}{2n_2}, \infty\}$, the coincidence limits where $u = 0$ and $u = 1$, respectively, as shown in (\ref{xto0minnuxComp})-(\ref{xaxp4nsComp}), and, asymptotically,
\bsub	\label{Ggenchannu0}
\begin{align}
G(x^1_{\frak1}(u) ) 
	&= \frac{16 n_1^2 C}{(1-u)^2} + \text{non-sing.}	\label{Gx11u}
\\
\begin{split}	 \label{GgenAs3chan}
G(x^1_{\frak2}(u)) &= 
	- 2^{-2} 3^{-\frac23} 
	(n_1^2-n_2^2)^{-\frac23} 
	\big[ (n_1-n_2)^2 - 4 A n_2^2 \big]
	n_1^{-\frac23}  
	n_2^{-\frac23} \ 
	\frac{1}{(1-u)^{4/3}}
\\
&\quad	
	+
	\frac{\rm{const.}}{(1-u)^{2/3}}	
	+
	\frac{\rm{const.}}{(1-u)^{1/3}}		
	+
	\text{non-sing.}
\end{split}
\\
\begin{split}
G(x^0_{\frak1}(u)) 
&= 
	A
	(n_1 - n_2)^{-2}
	n_1^{- \frac{2n_1}{n_1-n_2}}
	n_2^{ \frac{2n_1}{n_1-n_2}}
	\
	u^{-1 + \frac{1}{n_1-n_2} } 
	 \left[ 1 +  \rm{O}(  u^{\frac{1}{n_1-n_2}} ) \right]
\label{Ggenchannu01}
\end{split}
\\
\begin{split}
G(x^0_{\frak2}(u)) 
&= 
	(A + \tfrac{n_1}{n_2})
	(1 + \tfrac{n_1}{n_2})^{-2}
	n_1^{- \frac{2n_1}{n_1+n_2}}
	n_2^{-\frac{2n_1}{n_1+n_2}}
	\
	 u^{-1 + \frac{1}{n_1+n_2} } 
	 \left[ 1 +  \rm{O}( u^{\frac{1}{n_1+n_2}} ) \right]
\end{split}
\label{Ggenchannu02}
\end{align}
\esub
In Eqs.(\ref{GgenAs3chan})-(\ref{Ggenchannu02}) we have taken 
\be
C = \frac{1}{16n_1^2},
\ee
so that the coefficient of the singularity in Eq.(\ref{Gx11u}) is unity.
Note that  $\cal F (x) \to C$ when $x \to \infty$, so only the universal term in Eq.(\ref{GcompNeuGen}) survives.
This ensures the normalization of the interaction operators, as extensively discussed in \cite{Lima:2020boh,Lima:2020kek,Lima:2020nnx,Lima:2020urq}.
In fact, the two channels (\ref{Gx11u}) and (\ref{GgenAs3chan}) correspond to the OPE
\be	\label{OPEoints}
\Oint \times \Oint = \mathds1 + \s_3 .
\ee
which can be deduced from the powers of the leading singularities, giving an operator of dimension zero in (\ref{Gx11u}) and an operator of dimension \be
\Delta^\s_3 = h^\s_3 + \tilde h^\s_3 = \tfrac23 + \tfrac23.
\ee
There is no operator of dimension zero in any of the other channels. 

The leading coefficients in the expansions (\ref{Ggenchannu0}) give products of structure constants of the operators involved in the OPEs. For example, in the $\s_{(3)}$ channel the OPE is
\be
\Oincov_{(2)}(u,\bar u) \Oincov_{(2)}(1,\bar 1) = 
	\frac{\big\langle \Oincov_{(2)}  \s_{(3)} \Oincov_{(2)} \big\rangle 
	}{
	|1-u|^{-2 ( \frac23 - 2)}
	}
	\s_{(3)}(1,\bar1)
	+
	\cdots
\ee	
Inserting this back into the four-point function (\ref{Gcfull2}) and comparing with (\ref{GgenAs3chan}), we find that%
	\footnote{%
	Recall that $G_{\zeta_1\zeta_2}(u,\bar u) = G_{\zeta_1\zeta_2}(x(u)) \bar G_{\zeta_1\zeta_2}(\bar x(\bar u))$.
The expression in the r.h.s.~is the absolute value of the leading coefficient in Eq.(\ref{GgenAs3chan}).}
\begin{align}	\label{OsORsR}
\begin{split}
&	
\big\langle \Oincov_{(2)}  \s_{(3)} \Oincov_{(2)} \big\rangle 
\big\langle 
\lord
R^{\zeta_1}_{(n_1)}
R^{\zeta_2}_{(n_2)}  
\rord^\dagger
\s_3 
\lord
R^{\zeta_1}_{(n_1)}
R^{\zeta_2}_{(n_2)}
\rord
\big\rangle 
\\
&\qquad\qquad\qquad
	=
\Big|	 2^{-2} 3^{-\frac23} 
	(n_1^2-n_2^2)^{-\frac23} 
	\big[ (n_1-n_2)^2 - 4 A_{\zeta_1\zeta_2} n_2^2 \big]
	n_1^{-\frac23} 
	n_2^{-\frac23} 
\Big|^2
\end{split}	
\end{align}
Note that the operators above are not sums over orbits --- only some of the products $\Oincov_{(2)} \times \Oincov_{(2)}$ result in a $\s_{(3)}$; other cycles result in $\mathds 1$. The function $G_{\zeta_1\zeta_2}(x^1_{\frak2}(u))$ gives us the behavior of one representative of the class of permutations $(2)(2) = (3)$, and all other representatives behave the same way. %
The analysis can be repeated for $G_{+-}(x)$ and $G_{++}(x)$.
The structure constant $\langle \Oincov_{(2)}  \s_{(3)} \Oincov_{(2)} \rangle$ has been computed in \cite{Lima:2020kek}\footnote{See Eq.(C.5) ibid.}
\be	\label{stcconsOsO}
\langle \Oincov_{(2)}  \s_{(3)} \Oincov_{(2)} \big\rangle = 2^{\frac{13}{3}} 3^{4} ,
\ee
yielding the structure constants 
$
\langle 
\lord
R^{\zeta_1}_{(n_1)}
R^{\zeta_2}_{(n_2)}  
\rord^\dagger
\s_{(3)} 
\lord
R^{\zeta_1}_{(n_1)}
R^{\zeta_2}_{(n_2)}
\rord
\rangle 
$, listed in App.\ref{AppLists}, Eq.(\ref{strucconss3}).

\bigskip

In the limit $u \to 0$, we find the fusion rules (\ref{fusionOr}). 
The two channels $x^0_{\frak1}$ and $x^0_{\frak2}$ give us two operators,
$ Y^{\zeta_1\zeta_2}_{\frak1, \ [n_1+n_2]}$ and $Y^{\zeta_1\zeta_2}_{\frak2, \ [n_1+n_2]}$ respectively, both with twist $n_1+n_2$,
\be	\label{OPEointRR}
\Oint \times \lord R^{\zeta_1}_{[n_1]}  R^{\zeta_2}_{[n_2]} \rord  
= 
 Y^{\zeta_1\zeta_2}_{\frak1, \ [n_1+n_2]}  
+ 
 Y^{\zeta_1\zeta_2}_{\frak2, \ [n_1+n_2]} 
\ee
Now the OPEs read as
\be	\label{OPEORO}
\Oincov_{(2)}(u,\bar u) \lord R^{\zeta_1}_{(n_1)}  R^{\zeta_2}_{(n_2)} \rord(0,\bar 0) 
	=
	\frac{
	\langle 
	Y^{\zeta_1\zeta_2 \dagger}_{\frak a, (n_1+n_2)} 
	\Oincov_{(2)} 
	\lord R^{\zeta_1}_{(n_1)}  R^{\zeta_2}_{(n_2)} \rord
	\rangle
	}{
	 |u|^{- \Delta^{\zeta_1\zeta_2}_{\frak a} + 2 + \frac{n_1 + n_2}{2}  }
	 }
	 Y^{\zeta_1\zeta_2}_{\frak a,  (n_1+n_2)} (0,\bar 0)
	 +\cdots
\ee	 
with $\frak a =1,2$.
The dimensions $\Delta^{\zeta_1\zeta_2}_{\frak 1}$ of $Y^{\zeta_1\zeta_2}_{\frak 1,  [n_1+n_2]}$ follow from the powers of $u$ in the channel (\ref{Ggenchannu01}),
\bsub \label{DeltO1}
\begin{align}
\Delta^{\dot1+}_{\frak1}
=
\Delta^{\dot1\dot2}_{\frak1}
=
\Delta^{\dot1\dot1}_{\frak1}
& =  \frac{2}{n_1-n_2}  + \frac{n_1 + n_2}{2}, 	\label{DeltO1a}
\\
\Delta^{+-}_{\frak1}
&=
\frac{4}{n_1+n_2}  + \frac{n_1 + n_2}{2}
\\
\Delta^{++}_{\frak1}
&=
\frac{2}{n_1-n_2}  + \frac{n_1 + n_2}{2}
\end{align}\esub
while the dimensions of $Y^{\zeta_1\zeta_2}_{\frak 2,  [n_1+n_2]}$ follow from the channel (\ref{Ggenchannu02}),
\bsub \label{DeltO2}
\begin{align}
\Delta^{\dot1+}_{\frak2}	\label{DeltO2a}
=
\Delta^{\dot1\dot2}_{\frak2}
=
\Delta^{\dot1\dot1}_{\frak2}
&=
\frac{2}{n_1+n_2}  + \frac{n_1 + n_2}{2}
\\
\Delta^{+-}_{\frak2}
&=
\frac{2}{n_1-n_2}  + \frac{n_1 + n_2}{2}	\label{DeltO2b}
\\
\Delta^{++}_{\frak2}
&=
\frac{4}{n_1+n_2}  + \frac{n_1 + n_2}{2}	\label{DeltO2c}
\end{align}\esub
(We recall that $\Delta^{+\pm}_{\frak a}$ must be computed separately.)
As already mentioned, all operators with at least one R-neutral single-cycle Ramond ground state have the same dimension.

Inserting back into (\ref{Gcfull2}) and comparing with (\ref{Ggenchannu01})-(\ref{Ggenchannu02}) we can determine the structure constants.
Explicitly, inserting (\ref{OPEORO}) back into the four-point function we get
\be	
\begin{split}
&
\lim_{u\to0}
\Big\langle
\lord R^{\zeta_1}_{(n_1)}  R^{\zeta_2}_{(n_2)} \rord^\dagger (\infty, \bar \infty)
\Oincov_{(2)}(1,\bar 1) 
\Oincov_{(2)}(u,\bar u) \lord R^{\zeta_1}_{(n_1)}  R^{\zeta_2}_{(n_2)} \rord (0,\bar 0) 
\Big\rangle
\\
&\qquad	
=
	\frac{
	\big\langle 
	Y^{\zeta_1\zeta_2 \dagger}_{\frak a,  (n_1+n_2)}
	\Oincov_{(2)} 
	\lord R^{\zeta_1}_{(n_1)}  R^{\zeta_2}_{(n_2)} \rord
	\big\rangle
	}{
	 |u|^{- \Delta^{\zeta_1\zeta_2}_{\frak a} + 2 - \frac{n_1 + n_2}{2}  }
	 }
\big\langle
\lord R^{\zeta_1}_{(n_1)}  R^{\zeta_2}_{(n_2)} \rord^\dagger 
\Oincov_{(2)}
Y^{\zeta_1\zeta_2}_{\frak a,  (n_1+n_2)}
\big\rangle
	 +\cdots
\end{split}
\ee	 
Hence the leading coefficients in the r.h.s.~of Eqs.(\ref{Ggenchannu01})-(\ref{Ggenchannu02}) give
\bsub	\label{ProdRO}
\begin{align}
\begin{split}	\label{ProdRO1}
&	
\Big|
\big\langle 
\Oincov_{(2)} 
\lord R^{\zeta_1}_{(n_1)}  R^{\zeta_2}_{(n_2)} \rord^\dagger
Y^{\zeta_1\zeta_2}_{\frak 1,  (n_1+n_2)} 
\big\rangle
\Big|^2
	=
\Big|	
	A_{\zeta_1\zeta_2}
	(n_1 - n_2)^{-2}
	n_1^{- \frac{2n_1}{n_1-n_2}}
	n_2^{ \frac{2n_1}{n_1-n_2}}
\Big|^2
\end{split}	
\\
\begin{split}	\label{ProdRO2}
&	
\Big|
\big\langle 
\Oincov_{(2)} 
\lord R^{\zeta_1}_{(n_1)}  R^{\zeta_2}_{(n_2)} \rord^\dagger
Y^{\zeta_1\zeta_2}_{\frak 2,  (n_1+n_2)} 
\big\rangle
\Big|^2
	=
\Big|	
	(A_{\zeta_1\zeta_2} + \tfrac{n_1}{n_2})
	(1 + \tfrac{n_1}{n_2})^{-2}
	n_1^{- \frac{2n_1}{n_1+n_2}}
	n_2^{-\frac{2n_1}{n_1+n_2}}
\Big|^2
\end{split}	
\end{align}\esub
and similarly for $G_{+-}$ and $G_{++}$.
We thus have two lists of products of structure constants, given in Appendix \ref{AppLists}.

\subsection{Equal twists $n_1 = n_2$}	\label{SectEqualtwsOPE}



When $n_1 = n_2 = n$, the results above must be revisited. 
The analytic structure of the map (\ref{uxm}) changes drastically
\be
u(x) = \left( \frac{x+ 1}{x-1} \right)^{2n} .
	\label{uxnn}
\ee
The coincidence limits (\ref{xfraka}) change as well. 
The limit $u \to 1$ now has the solutions $x = \infty$ and
$x =  0$, where we find the two inverses $x^1_{\frak a}(u)$ to be
\begin{align}
(u \to 1) \quad
	\begin{sqcases}
	x \to \infty , 
	&\quad x^1_{\frak1}(u) \approx - 4n (1-u)^{-1} 
\\
	x \to 0,
 	&\quad x^1_{\frak2}(u) 
		\approx 
				\tfrac{1}{4n}
				 (1 - u)
	\end{sqcases} \label{xaxp4nsCompM}
\end{align}
Meanwhile, for $u \to 0$ we now only have \emph{one} solution:
\be
(u \to 0) \quad
	\begin{sqcases}
x \to - 1 ,\quad 
	& x^0_{\frak2}(u) 
	\approx - 1
	+  2	\ u^{\frac{1}{2n}}
	+ \cdots
\end{sqcases}	\label{xto0minnuxCompM}
\ee
We denote this unique inverse by $x^0_{\frak2}(u)$ because this corresponds to the second channel in (\ref{xaxp4nsComp}) when $n_1 = n_2 = n$; the channel called $x^0_{\frak1}(u)$ \emph{disappears}.
This change in the coincidence limits reflects a change in the structure of the four-point functions (\ref{Gm21p})-(\ref{Gmmpp}), and in the fusion rules and OPEs.
We now have only three  different functions:
\begin{align}
\begin{split}	\label{g2n}
g_1(x) 
&= 
		C
		(x-1)^{1 + 2n}
		( x + 1 )^{ 1 - 2 n }
\\
&= 
G_{+-}(x) 
= G_{\dot1\pm}(x)
= G_{\dot1\dot2}(x)
\end{split}
\\
\begin{split}	\label{g3n}
g_2(x) 
&= 
	C x^{-2} (1 + x^2) 
	(x-1)^{1+2n}
	( x + 1 )^{1 - 2 n }
\\
&=  G_{\dot1\dot1}(x)
\end{split}
\\
\begin{split}	\label{g1n}
g_3(x) 
 	&= 
		 C
		x^{-2}
		(x-1)^{2+2n}
		( x + 1 )^{2 - 2n }
\\
&=  G_{++}(x)  
\end{split}
\end{align}

In the channel $x^1_{\frak 1}(x)$, all behave the same way:
$g_1(x) \approx g_2(x) \approx g_3(x) \approx C x^2$ when $x \to \infty$.
Therefore, as expected, we find again the usual identity channel with 
$C = 1/16n^2$.
Using (\ref{xaxp4nsCompM}),
in the channel $u\to 1$ with $x \to 0$ we have
\begin{align}
\begin{split}	\label{g2x0}
g_1(x^1_{\frak2}(u)) &= - \frac{1}{16n^2 } + \frac{x^1_{\frak2}}{4n} + \cdots
= - \frac{1}{16n^2 } + (1-u) + \cdots
\end{split}
\\
\begin{split}	\label{g3x0}
g_2(x^1_{\frak2}(u)) 
&= 
	\frac{-1}{(4n x^1_{\frak2})^2} + \frac{1}{4n x^1_{\frak2}} + \text{non-sing.} 
=  \frac{-1}{(1-u)^{2}} + \frac{1}{1-u} + \text{non-sing.}
\end{split}
\\
\begin{split}	\label{g1x0}
g_3(x^1_{\frak2}(u)) 
&= 
	\frac{1}{(4n x^1_{\frak2})^2} - \frac{1}{4n x^1_{\frak2}} + \text{non-sing.} 
=
	\frac{1}{(1-u)^{2}} - \frac{1}{1-u}	 + \text{non-sing.}
\end{split}
\end{align}
In the expansions (\ref{g1x0}) and (\ref{g3x0}) we can recognize again the identity channel.
The expansion (\ref{g2x0}) is \emph{not singular}, hence it does not correspond to an OPE channel.
In short, in the limit $u \to 1$ we only find the identity channel of the OPE 
$\Oint \times \Oint = \mathds1$.
The $\s_3$ channel of the fusion rule (\ref{OPEoints}) has disappeared.

The disappearance could be predicted by looking at the structure constants 
$
\langle 
\lord
R^{\zeta_1}_{(n_1)}
R^{\zeta_2}_{(n_2)}  
\rord^\dagger
\s_{(3)} 
\lord
R^{\zeta_1}_{(n_1)}
R^{\zeta_2}_{(n_2)}
\rord
\rangle 
$ in Eq.(\ref{strucconss3}) --- they all vanish when $n_1 = n_2$.
 This reflects the fact that Eq.(\ref{compto1}), which here assumes the form
\be
(n_1) (n_2) (3) (n_1)'(n_2)' = {\mathds1},
\ee
has no solutions satisfying the conditions: $(n_1)$ and $(n_2)$ are disjoint; 
$(n_1)'$ and $(n_2)'$ are disjoint;
and $(3)$ has one copy in $(n_1)$ and two copies in $(n_2)$ or vice-versa.

\bigskip

In the limit $u\to0$ we have seen that the channel $x^1_{\frak1}$, extant for $n_1 \neq n_2$, has also disappeared. In the remaining channel (\ref{xto0minnuxCompM}) we have the expansions
\begin{align}
g_1(x^0_{\frak2}(u))  		\label{g2xm1}
&= 	
	- \frac{1}{4n^2}
	\
	 u^{-1 + \frac{1}{2n}}
	 \left[ 1 -  (1+2n) \,  u^{\frac{1}{2n}} + \cdots \right]
\\
g_2(x^0_{\frak2}(u)) 		\label{g3xm1}
&= 	
	- \frac{1}{2n^2}
	\
	 u^{-1 + \frac{1}{2n}}
	 \left[ 1 +  (1-2n) \,  u^{\frac{1}{2n}} + \cdots \right]
\\
g_3(x^0_{\frak2}(u)) 		\label{g1xm1} 
&= 	
	\frac{1}{n^2}
	\
	 u^{-1 + \frac{1}{n}}
	 \left[ 1 +  2(1-n) \,  u^{\frac{1}{2n}} + \cdots \right]
\end{align}
and the fusion rules
\begin{align}	\label{OPEointRRM}
\Oint \times \lord R^{\zeta_1}_{[n]}  R^{\zeta_2}_{[n]} \rord
= 
 Y^{\zeta_1\zeta_2}_{\frak2, \ [2n]} 
\end{align}
which replaces (\ref{OPEointRR}).
The dimensions of the operators $Y^{\zeta_1\zeta_2}_{\frak2, \ [2n]}$, which have twist $2n$, are read from Eqs.(\ref{g3xm1})-(\ref{g2xm1}) to be
\begin{align}
\Delta^{+-}_{Y , \frak2} 
=
\Delta^{\dot1\pm}_{Y , \frak2} 
=
\Delta^{\dot1\dot2}_{Y , \frak2} 
=
\Delta^{\dot1\dot1}_{Y , \frak2} 
&=
\frac1n + n
\\
\Delta^{++}_{Y , \frak2} 
	&=  \frac{2}{n}  +  n 		\label{Depp2n}
\end{align}
and should be compared with the dimensions in Eqs.(\ref{DeltO2}) and (\ref{DeltO1}).
Inserting the OPEs back into the four-point functions, we find the structure constants
\begin{align}
\begin{split}
\Big|
\big\langle 
\Oincov_{(2)} 
\lord R^{+}_{(n)}  R^{-}_{(n)} \rord^\dagger
Y^{+-}_{{\frak2}(2n)} 
\big\rangle 
\Big|^2
&=
\Big|
\big\langle 
\Oincov_{(2)} 
\lord R^{\dot1}_{(n)}  R^{\pm}_{(n)} \rord^\dagger
Y^{\dot1\pm}_{{\frak2}(2n)} 
\big\rangle 
\Big|^2
\\
&=
\Big|
\big\langle 
\Oincov_{(2)} 
\lord R^{\dot1}_{(n)}  R^{\dot2}_{(n)} \rord^\dagger
Y^{\dot1\dot2}_{{\frak2}(2n)} 
\big\rangle 
\Big|^2
\\
&=
2^{-4}n^{-4}
\end{split}
\\
\Big|
\big\langle 
\Oincov_{(2)} 
\lord R^{\dot1}_{(n)}  R^{\dot1}_{(n)} \rord^\dagger
Y^{\dot1\dot1}_{{\frak2}(2n)} 
\big\rangle 
\Big|^2
&=
2^{-2} n^{-4}
\\
\Big|
\big\langle 
\Oincov_{(2)} 
\lord R^{+}_{(n)}  R^{+}_{(n)} \rord^\dagger
Y^{++}_{{\frak2}(2n)} 
\big\rangle 
\Big|^2
&=
n^{-4}		\label{Stucppnn}
\end{align}
One can compare these values with those obtained by taking $n_1 = n_2 = n$ in the list (\ref{List2}), and see that they agree.

\subsection{OPEs with full Ramond ground states}

The OPEs described above involve the product of $\Oint$ with a double-cycle Ramond field.
For a generic Ramond ground state,  several $Y$ operators appear in the OPE, according to the factorization of the Ramond field into pairs as in Eq.(\ref{Gufact}).
The interaction operator connects two components of the Ramond ground state to form the $Y$ operators, and leave the remaining components untouched. Hence we have the OPE%
	\footnote{%
	As usual, this is a schematic equation, without coefficients in front of operators.}
\begin{align}	\label{OPEgen}
\begin{split}
\Oint 
\times
\Big\lord
\prod_{i,\zeta_i}
( R^{\zeta_i}_{[n_i]})^{q^{\zeta_i}_i}
\Big\rord
&	= 
\sum_{i>j} \sum_{\zeta, \zeta' \neq \zeta}
\sum_{\frak a = \frak1,\frak2}
\Big\lord
Y^{\zeta \zeta'}_{\frak a,  [n_i+n_j]}
\!\!
\prod_{\substack{k \neq i,j \\ \zeta_k \neq \zeta,\zeta'}}
\!\!
( R^{\zeta_k}_{[n_k]})^{q^{\zeta_k}_k}
\Big\rord
\\
&
+
\sum_{i} \sum_{\zeta, \zeta' \neq \zeta}
\Big\lord
 Y^{\zeta \zeta'}_{\frak2,  [2 n_i]}
\!\!
\prod_{\substack{k \neq i \\ \zeta_k \neq \zeta,\zeta'}}
\!\!
( R^{\zeta_k}_{[n_k]})^{q^{\zeta_k}_k} 
\Big\rord
\\
&
+
\sum_{i,\zeta} 
\Big\lord
Y^{\zeta \zeta}_{\frak2,  [2 n_i]}
\!\!
\prod_{\substack{k \neq i \\ \zeta_k \neq \zeta}}
\!\!
( R^{\zeta_k}_{[n_k]})^{q^{\zeta_k}_k} 
\Big\rord
\end{split}
\end{align}
where the product of operators inside brackets, which now includes the $Y$, have ``normal-ordered cycles'' in the same sense as 
$
\lord \prod ( R^{\zeta_i}_{[n_i]})^{q^{\zeta_i}_i}\rord
$.
Note that, in the last two lines of Eq.(\ref{OPEgen}), there is only the operator  
$Y^{\zeta \zeta}_{\frak2,  [2 n_i]}$ in the ${\frak a} = \frak2$ channel, because this is the only one that exists when the twists of the double-cycle Ramond field are equal.

\section{Non-renormalization of the Ramond ground states}	\label{SectNonRenor}

We are now going to show explicitly that the Ramond ground states 
\be
\Big\lord \prod_i (R^{\zeta_i}_{[n_i]})^{q_i} \Big\rord , \quad \sum_i n_i q_i = N,
\ee
are protected, by computing the integral (\ref{JscrR}) using our formulae for the four-point function (\ref{Gu}), and showing that it vanishes.
We have shown in Sect.\ref{SectFactorization} that the four-point function (\ref{Gu}) factorizes into a sum of functions $G_{\zeta_1\zeta_2}(u,\bar u)$.
Hence (\ref{JscrR}) reduces to a sum of integrals
\be	\label{Jzeze}
\begin{split}
J_{\zeta_1\zeta_2} 
	&= \int\! d^2u \, G_{\zeta_1\zeta_2}(u,\bar u)
\\
	&= \int\! d^2x \, \big| u'(x) \,  G_{\zeta_1\zeta_2}(x) \big|^2 .
\end{split}
\ee
These can be computed analytically, and we are going to show that
\be
J_{\zeta_1\zeta_2} = 0 
\ee
for all $\zeta_i$. 
Therefore we will show explicitly that the composite Ramond fields are protected in the deformed theory, at order $\la^2$, and at order ${\bf g} = 0$ in the genus expansion.

\bigskip

\noindent
{\bfseries Different twists $n_1 \neq n_2$}

\noindent
Using the form in Eq.(\ref{GcompNeuGen}), and making the change of variables
\be
y = - (\tfrac{2n_2}{n_1+n_2})^{2}   (x-1)(x+ \tfrac{n_1}{n_2} ) ,
\ee
we have
\begin{align}	\label{Jzzy}
\begin{split}
J_{\zeta_1\zeta_2}
&= 
	\frac{1}{2^{10} n_1^4}
	\left( A_{\zeta_1\zeta_2} - \tfrac{2n_1}{n_2} \right) A_{\zeta_1\zeta_2}
	\int \! d^2y\, | 1 - y |^{-3} 
\\
&
	+
	\frac{1}{2^{10} n_1^4}
	\left( \frac{n_1+n_2}{2n_2} \right)^{4}
	\int \! d^2y\,  | 1 - y |^{-3} |y - w|^{2}
\\
&
	+
	\frac{1}{2^{10} n_1^4}
	\left( \frac{n_1+n_2}{2n_2} \right)^2
	A_{\zeta_1\zeta_2} 
	 \int\! d^2y\,  | 1 - y |^{-3} \big( y +\bar y  \big)
\end{split}	
\end{align}
The non-holomorphic integral in the last line vanishes:  $\int \! d^2y\, | 1 - y |^{-3}   \rm{Im} (y)$ cancels with $\int \! d^2y\, | 1 - y |^{-3} \rm{Im}(\bar y)$, and the integrand of $\int \! d^2y\, | 1 - y |^{-3}  2 \rm{Re}(y)$ is odd so the integral over the Real line vanishes.
%
At first sight, the remaining two integrals are not defined at $y = 1$, where the integrand diverges, but they can be defined via analytic continuation with the same method of deforming contours described in \cite{Lima:2020kek}. The analytic structure of the integrals in (\ref{Jzzy}) is however much simpler; their (unique) analytic continuation is given in detail in \cite{dotsenko1988lectures}, where one finds
\begin{align}
\begin{split}	\label{TheDFintn1}
 \int \! d^2y \, & |y|^{2a} | y-1 |^{2b}
		=  \sin( \pi b) 
			\frac{\Gamma(1+a) 
				\Gamma^2 (1+b) 
				\Gamma (-1-a-b)
				}{
				\Gamma(-a) 
				\Gamma(2+a+b)
				} .
\end{split}
\end{align}
The r.h.s.~is an analytic function of the parameters $a,b$. Thus we have
\begin{align}	
\begin{split}
 \int \! d^2y \,  | y-1 |^{-3} &=  \lim_{a \to 0} \frac{4 \pi \Gamma(1+a) \Gamma (\tfrac12 -a) }{ \Gamma(-a) \Gamma(\tfrac12+a) }
\\
&= \lim_{a \to 0} \frac{4 \pi }{ \Gamma(-a) } 		
\\
&= 0	.
\end{split}
\end{align}
Finally, the last remaining integral can be solved by Eq.(\ref{TheDFintn1}) with a further change of variables:
$u = (y-w)/(1+w)$,
\begin{align} \label{DF32}
\begin{split}
\int \! d^2y\,  | 1 - y |^{-3} |y - w|^{2}
&=
(1+w) \int \! d^2u \, |u|^2 |1-u|^{-3}
\\
&=  (1+w) \lim_{a \to 1} \frac{4 \pi \Gamma(1+a) \Gamma (\tfrac12 -a) }{ \Gamma(-a) \Gamma(\tfrac12+a) }
\\
&=  (1+w) \lim_{a \to 1} \frac{-16\pi}{ \Gamma(-a)}
\\
&= 0	.
\end{split}
\end{align}
Thus all integrals in the r.h.s.~of Eq.(\ref{Jzzy}) vanish.
As noted in Sect.\ref{SectOPEs}, Eq.(\ref{GcompNeuGen}) is not valid for the function $G_{++}(x)$ in (\ref{Gmmpp}), so the discussion above does not apply immediately; nevertheless, the integral $J_{++}$ was computed in Ref.\cite{Lima:2020nnx} --- it vanishes, and in fact it turns out to have the same form as in (\ref{DF32}).

\bigskip

\noindent
{\bfseries Equal twists $n_1 = n_2 = n$}

\noindent
When $n_1 = n_2 = n$, we have to integrate the functions (\ref{g2n})-(\ref{g1n}).
There are thus three integrals:
\begin{align}
\begin{split}
J_1 &= \int\! d^2x \big| u'(x) g_1(x) \big|^2
	= \frac{1}{16n^2} \int\! d^2x	
\end{split}
\\
\begin{split}
J_2 &= \int\! d^2x \big| u'(x) g_2(x) \big|^2
	= \frac{1}{16n^2} \int\! d^2x |x^{-2} (x^2+1) |^2	
\end{split}
\\
\begin{split}
J_3 &= \int\! d^2x \big| u'(x) g_3(x) \big|^2
	= \frac{1}{16n^2} \int\! d^2x |x^{-2} (x^2-1) |^2	
\end{split}
\end{align}
Again, they are divergent/undefined, but can be given a unique analytic continuation by the same method as before, after being put in the form (\ref{TheDFintn1}).
Making the change of variables $y = x^2$ in $J_1$ and $J_3$, and $y = - x^2$ in $J_2$, we get 
\begin{align}
\begin{split}
J_1 &= \frac{1}{2^6  n^2} \int\! d^2y \ | y |^{-1} 
	= \frac{1}{2^6  n^2} \times (-4 \sin 0 ) 	
	= 0
\end{split}
\\
\begin{split}
J_2 &= \frac{1}{2^6 n^2} \int\! d^2y \, |y|^{-3} |1-y|^2
	= \frac{1}{2^6 n^2} \times (16 \sin \pi)
	= 0
\end{split}
\\
\begin{split}
J_3 &= \frac{1}{2^6 n^2} \int\! d^2y \, |y|^{-3} |1-y|^2
	= \frac{1}{2^6 n^2} \times (16 \sin \pi)
	= 0
\end{split}
\end{align}
This concludes the demonstration that the integrals (\ref{Jzeze}) all vanish, for any values of $n_1,n_2$, and any double-cycle Ramond fields.

\section{Discussion and conclusions}		\label{SectConclusion}

The main result of the present work is the computation of the four-point function (\ref{GuIntro}). 
Its analysis illuminates the effects of how the deformation operator interacts with the Ramond ground states (\ref{CompleteRamondIntro}), as we move the CFT away from the free orbifold point. 
It does so by yielding two important pieces of information: fusion rules of the Ramond ground states with the interaction operator, and the non-renormalization of the conformal dimension of these states.
As a conclusion, we will now put these results in perspective, by discussing them in the context of previous literature.

\bigskip

\noindent
{\bfseries Protection of Ramond ground states}

\noindent
We have shown explicitly that the dimensions of the Ramond ground states in the D1-D5 CFT are protected at second order in the $\la$-expansion and at the level of genus-zero covering surfaces --- which is related to the large-$N$ expansion of the correlation functions but not exactly the same, as shown in \S\ref{SectNscaling}.

This was very expected from algebraic considerations: the Ramond ground states are related to BPS operators, the NS chiral ring, by spectral flow of the ${\cal N} = (4,4)$ super-conformal algebra with central charge $c = 6N$.

Recent work
\cite{Lima:2020boh,Lima:2020kek,Lima:2020nnx,Lima:2020urq}
has indicated that there are some subtleties with the use of spectral flow in the deformed theory.
The $n$-twisted sector of the free orbifold theory can be described as a ${\cal N}=(4,4)$ SCFT with central charge $c = 6n$, whose Ramond ground states  have conformal weight
$h= \frac14 n$, and
are mapped by the spectral flow of the $c = 6n$ algebra to the $n$-twisted extremal NS chiral fields.
Thus non-renormalization of the $n$-twisted NS chirals, proven in Ref.\cite{Pakman:2009mi}, could perhaps suggest that the $n$-twisted Ramond ground states would also be protected --- but they are not.
In the interacting theory, the twist-two $S_N$-invariant deformation operator $\Oint$ deforms the currents of the operator algebra, and the twisted sectors become mixed one with the other.
In other words, for $n < N$ the relation between the $n$-twisted Ramond ground states with $h= \frac14 n$ and the NS chiral ring is lost.

In light of such results of Refs.\cite{Lima:2020boh,Lima:2020kek,Lima:2020nnx,Lima:2020urq},
we see that the protection of the ``total'' Ramond ground states with $h = \frac14 N$ 
demonstrated here is not extended to the $n$-twisted components individually.
From another point of view, since the scalar modulus $\Oint$ preserves supersymmetry, the super-conformal algebra with central charge $c = 6N$ is preserved, even though the individual $n$-twisted algebras with $c = 6n  < 6N$ are not. Hence protection of the states with $h = \frac14 N$ follows from the protection of BPS NS chirals obtained by spectral flow of the algebra with $c = 6N$, but the same does not apply for each twisted sector individually.

\bigskip

\noindent
{\bfseries Symmetric states with same twists}

\noindent
Ramond ground states made by products of components with equal twists are dual to special, symmetric solutions of supergravity.
In particular, when all spins are aligned,
$\lord ( R^{+}_{[n]})^{N/n} \rord$ 
(assuming $N/n$ is an integer),
as well as descendants of this state, are dual to a family of axially symmetric SUGRA solutions \cite{Lunin:2001jy,Giusto:2004id,Giusto:2004ip,Giusto:2012yz}.
The resulting four-point functions are so simple that they are worth writing explicitly. 

The four-point function (\ref{Gufact}) is
\begin{align}	\label{GufactNncon}
\begin{split}
G(u,\bar u) 
	&= 
\Big\langle
\Big\lord
( R^{+}_{[n]})^{N/n}
\Big\rord^\dagger 
	(\infty,\bar \infty) 
\,	\Oint(1,\bar 1) 
\,	\Oint (u,\bar u)
\Big\lord
( R^{+}_{[n]})^{N/n}
\Big\rord
	(0,\bar 0)
\Big\rangle 
\\
&= 
{\scr P}^2(N/n)
\Big\langle
\big\lord
	R^{+}_{[n]} 
	R^{+}_{[n]}
\big\rord^\dagger
			(\infty,\bar \infty) 
\,	\Oint(1,\bar 1) 
\,	\Oint (u,\bar u)
\big\lord
	R^{+}_{[n]} 
	R^{+}_{[n]} 
\big\rord
			(0,\bar 0)
\Big\rangle 
\\
&=
\left( \frac{{\scr P}(\frac{N}{n})}{16n^2 {\scr S}_n(N) {\scr S}_2(N)} \right)^2
	\sum_{\frak a = \frak1}^{2 n}
	\Bigg|
	\frac{
		\left[ x_{\frak a}(u) -1\right]^{2+2n}
		\left[ x_{\frak a}(u) + 1 \right]^{2 - 2n }
		}{ x_{\frak a}^{2}(u)}
	\Bigg|^2	
\end{split}
\end{align}
where we have used Eq.(\ref{g1n}), along with the sum over the inverses of the map (\ref{uxnn}).
The OPE (\ref{OPEointRRM}), 
which has the very simple structure constant (\ref{Stucppnn}),
results in
\be
\Oint \times \big\lord ( R^{+}_{[n]})^{N/n} \big\rord
= 
\big\lord  Y^{++}_{\frak2, \ [2n]}  ( R^{+}_{[n]})^{\frac{N-2n}{n}} \big\rord
\ee
where the $2n$-twisted operator $Y^{++}_{\frak2, \ [2n]}$ has dimension 
$\Delta^{++}_{Y , \frak2} = n + 2/n$, cf.~Eq.(\ref{Depp2n}), R-charge $(j^3 , \tilde \jmath^3)  = (\frac12, \frac12)$ and SU(2)$_2$ charges $({\frak j}^3, \tilde{\frak j}^3) = (0, 0)$.
An operator with this dimension and the correct charges can easily be obtained by applying fractional modes of the R-current to a Ramond ground state:
$J^+_{\frac{2}{2n}} \tilde J^+_{\frac{2}{2n}}  R^{-}_{(2n)} (z,\bar z)$.
Recall that the R-current fractional mode $J^+_{k/M}$, for integer $k < M$, is well-defined in the $M$-twisted sector, raises the R-charge by one unit and the holomorphic dimension by $k/M$.

\bigskip

\noindent
{\bfseries Two OPE channels for different twists}

\noindent
The dynamical information obtained from the fusion rules (\ref{OPEointRRIntro}) reveals the existence of non-BPS operators 
$ Y^{\zeta_1\zeta_2}_{\frak a,  [n_1+n_2]}$
defining the conformal blocks in the OPE algebra of the Ramond ground states and the deformation operator. 
The two operators $Y^{\zeta_1\zeta_2}_{\frak a,  [n_1+n_2]}$,
distinguished by the label ${\frak a} = \frak1,\frak2$,
correspond to the two channels in the OPE (\ref{OPEointRR}). 
The dimensions of the operators in these two channels have a curious degenerate structure shown in Eqs.(\ref{DeltO1})-(\ref{DeltO2}).
The operators with at least one R-neutral index are all degenerate, with dimensions (\ref{DeltO1a}) in channel $\frak1$ and (\ref{DeltO2a}) in channel $\frak2$. 
The latter have dimensions
\be	\label{DimCon2}
\Delta^{\dot1+}_{\frak2}
=
\Delta^{\dot1\dot2}_{\frak2}
=
\Delta^{\dot1\dot1}_{\frak2}
=
\frac{2}{n_1+n_2}  + \frac{n_1 + n_2}{2}
\ee
and it is not hard to find possible realizations of these operators in terms of descendants of known fields. For example, for some constants $A, B$,
we can make the linear combination ansatz
\be	\label{ansfrac1}
Y^{\dot1+}_{\frak2,  (n_1+n_2)} (z,\bar z)
= 
A\, \psi^+_{\frac{1}{n_1+n_2}} \tilde \psi^+_{\frac{1}{n_1+n_2}}  R^{\dot1}_{(n_1 + n_2)} (z,\bar z) 
+
B\, {\psi}^{\dot1}_{\frac{1}{n_1+n_2}} \tilde {\psi}^{\dot1}_{\frac{1}{n_1+n_2}}  R^{+}_{(n_1 + n_2)} (z,\bar z) 
\ee
for $Y^{\dot1+}_{\frak2,  (n_1+n_2)}$, which has the same charges as $\lord R_{[n_1]}^{\dot1} R_{[n_2]}^{+} \rord$, namely
$j^3 = \frac12 = \tilde \jmath^3$ and ${\frak j}^3 = - \frac12 = \tilde{\frak j}^3$.
The r.h.s.~has the correct dimensions and charges, since the fermion fractional modes increase the conformal weight by $1/(n_1+n_2)$ and the SU(2) charges by $\frac12$.

The dimensions of the operators in channel $\frak1$,
\be	\label{DimCon1}
\Delta^{\dot1+}_{\frak1}
=
\Delta^{\dot1\dot2}_{\frak1}
=
\Delta^{\dot1\dot1}_{\frak1}
 =  \frac{2}{n_1-n_2}  + \frac{n_1 + n_2}{2} ,
\ee
are more curious. 
It is not possible to obtain the first term in the r.h.s.~by simply applying current fractional modes as in (\ref{ansfrac1}), because although the denominator is $n_1-n_2$, the operator is again in the $n_1 + n_2$ twisted sector.
Hence the operators in this sector do not seem to be simple descendants of primary fields.
The dimensions (\ref{DimCon1}) are singular when $n_1 = n_2$, but,
as we have shown in \S\ref{SectEqualtwsOPE}, this channel is not present when the twists are equal; only channel $\frak2$, with dimensions 
$\Delta^{\dot1+}_{\frak2}
=
\Delta^{\dot1\dot2}_{\frak2}
=
\Delta^{\dot1\dot1}_{\frak2}
= n + 1/n$
remains.
A similar thing happens for $\Delta^{+\pm}_{Y,\frak a}$.

The distinction between different OPE channels above is a very interesting feature of the orbifold theory. 
In this paper, the channels appear as a consequence of the existence of different solutions $x^{u_*}_{\frak a}(u)$ of the polynomial%
	\footnote{More precisely, $u(x) = u_*$ can be reduced to a polynomial equation since $u(x)$ in the form (\ref{uxm}) is a rational function.}
equation $u(x) = u_*$ which has, in general, ${\bf H} = 2 \max(n_1,n_2)$ solutions.
On the other hand, these solutions are related to the different classes of permutations solving Eq.(\ref{compto1}); $\bf H$ is a Hurwitz number, see \cite{Pakman:2009zz}.
In Appendix B of Ref.\cite{Lima:2020nnx}, we have shown how the $\bf H$ solutions of $u(x) = u_*$ correspond to ${\bf H}$ solutions of Eq.(\ref{compto1}). The discussion there makes it clear how there is an important qualitative change in the counting of different classes of permutations when $n_1 = n_2$, in synchronicity with the drastic changes in the structure of the equation $u(x) = u_*$. 
It would be interesting to examine this phenomenon more carefully from the point of view of spin chains/diagrams introduced in Refs.\cite{Pakman:2009mi,Pakman:2009zz}.

\bigskip

\noindent
{\bfseries Four-point functions of composite operators}

\noindent
The results of Sect.\ref{SectFactorization} for the factorization of the four-point function can be applied rather directly to similar functions where the Ramond ground states are replaced by other composite fields with the structure
$\big\lord \prod_i ( {\scr O}^i_{[n_i]})^{q_i} \big\rord$,  
$\sum_{i} n_i q_i = N$.
For example, we may consider powers of twisted NS chiral fields, etc.
We can also replace the interaction operators by other twist-two operators. One ends up with connected functions containing double-cycle operators, which can be computed with the covering map of \S\ref{SectCovMaps}.

There could be some additional complications. For example, depending on the operators involved, it may be that the three-point function factorization analogous to (\ref{3ptFact}) does not vanish. Note that these three-point function fall under the category analyzed in Ref.\cite{Tormo:2018fnt}; the authors computed correlators of NS chiral fields, but even for other fields their covering map can be used.

These different correlators --- along with the function (\ref{GuIntro}) computed here --- are examples of so-called `heavy-heavy-light-light' (HHLL) four-point functions: they contain two `heavy' operators, whose conformal weight is of the order of the central charge $c = 6N$, and two `light' operators, whose conformal weights remain finite in the limit of large $N$. 
This kind of four-point function is quite interesting for AdS$_3$/CFT$_2$ holography, as discussed in  \cite{Galliani:2016cai,Galliani:2017jlg}.
There, the authors consider HHLL correlators with Ramond ground states as heavy operators, while the light operators are untwisted NS fields with $h = \frac12$. It would be interesting to use our methods developed in the present paper to compute similar operators with \emph{twisted} light states, e.g.~the lowest-weight NS chiral $O^{(0,0)}_{[2]}$; we leave this for future work.

\vspace{10mm}

\noindent
{\bf Acknowledgements}

\noindent
The work of M.S.~is partially supported by the Bulgarian NSF grant KP-06-H28/5 and that of M.S.~and G.S.~by the Bulgarian NSF grant KP-06-H38/11.
 M.S.~is grateful for the kind hospitality of the Federal University of Esp\'irito Santo, Vit\'oria, Brazil, where part of his work was done.

\appendix

\section{Examples of factorizations and sums over orbits} 	\label{AppSumoveOrb1}

In this appendix we give detailed examples of the factorization of the four-point function (\ref{Gu}), so as to better illustrate the arguments of \S\ref{SectSumoveOrb1}

\bigskip

\bigskip

\noindent
{\bfseries Example 1}

\noindent
We start with a simple example: the four-point function involving the field
\be	\label{ex1Rq}
\lord (R^+_{[n]})^{q} \rord ,
 \qquad q n = N .
\ee
That is, we want to compute
\begin{align}	\label{GuConEx1}
\begin{split}
G = 
\Big\langle 
\Big\lord 
(R^{+\dagger}_{[n]_\infty})^{q} 
\Big\rord
	\Oincov_{(2)_1}
	\Oincov_{(2)_u} 
\Big\lord
(R^{+}_{[n]_0})^{q} 
\Big\rord
\Big\rangle
\end{split}
\end{align}
Very explicitly, Eq.(\ref{dobsumorbGen}) here reads
\begin{align*}
&
\Big\langle 
\sum_{h \in S_N} 
\Big[
R^{+\dagger}_{h (1,\cdots,n)_\infty h^{-1} }
R^{+\dagger}_{h (n+1,\cdots, 2n)_\infty h^{-1} }
R^{+\dagger}_{h (2n+1,\cdots, 3n)_\infty h^{-1} }
\cdots
\Big]
	\Oincov_{(2)_1}
\\
&\qquad\qquad
\times
	\Oincov_{(2)_u}
\sum_{g \in S_N}
\Big[
R^{+}_{g (n, \cdots,1)_0 g^{-1} }
R^{+}_{g (2n, \cdots, n+1)_0 g^{-1} }
R^{+}_{g (3n, \cdots, 2n+1)_0 g^{-1} }
\cdots
\Big]
\Big\rangle  
\end{align*}
with a product of $q$ Ramond fields on the left, and $q$ on the right.
In each term of this (double) sum, the copy indices entering the cycles $(2)_1$ will overlap with only two of the operators  inside the square brackets on the left;
the copies of the cycle $(2)_u$ will overlap with only two operators of the operators inside the brackets on the the right;
and the function will factorize into terms of the type shown in (\ref{DoubleCycle4ptSm}).
For example, we will have a term which pairs the first two components on the left with the first two components on the right:
\begin{align*}
&
\Big\langle 
\sum_{h \in S_N} 
\Big[
R^{+\dagger}_{h (1,\cdots,n)_\infty h^{-1} }
R^{+\dagger}_{h (n+1,\cdots, 2n)_\infty h^{-1} }
\Big]
	\Oincov_{(2)_1}
	\Oincov_{(2)_u}
\sum_{g \in S_N}
\Big[
R^{+}_{g (n, \cdots,1)_\infty g^{-1} }
R^{+}_{g (2n, \cdots, n+1)_0 g^{-1} }
\Big]
\Big\rangle  
\\
&
\times
\sum_{h,g \in S_N} 
\Big\langle 
R^{+\dagger}_{h (2n+1,\cdots, 3n)_\infty h^{-1} }
\Big|
R^{+}_{g (3n, \cdots, 2n+1)_0 g^{-1} }
\Big\rangle  
\cdots
\end{align*}
We will also have a term which pairs, again, the first two components on the left, but now with the second and third components on the right:
\begin{align*}
&
\Big\langle 
\sum_{h \in S_N} 
\Big[
R^{+\dagger}_{h (1,\cdots,n)_\infty h^{-1} }
R^{+\dagger}_{h (n+1,\cdots, 2n)_\infty h^{-1} }
\Big]
	\Oincov_{(2)_1}
	\Oincov_{(2)_u}
\sum_{g \in S_N}
\Big[
R^{+}_{g (n, \cdots,1)_\infty g^{-1} }
R^{+}_{g (3n, \cdots, 2n+1)_0 g^{-1} }
\Big]
\Big\rangle  
\\
&
\times
\sum_{h,g \in S_N} 
\Big\langle 
R^{+\dagger}_{h (2n+1,\cdots, 3n)_\infty h^{-1} }
\Big|
R^{+}_{g (2n, \cdots, n+1)_0 g^{-1} }
\Big\rangle  
\cdots
\end{align*}
and so on.%
	\footnote{%
	Note that the representatives $(n+1,\cdots,2n)$ and $(3n,\cdots,2n+1)$ appearing in this last four-point function are different, but this does not matter since we are summing over orbits independently on the right and on the left.}
In other words, we can pair the fields on the left and the fields on the right independently.	
All of these factorize four-point functions, of course, give the same numerical result. Hence we only need to compute 
\be	\label{DoubleCycle++Ap}
\Big\langle
\Big\lord
R^{+}_{[n]_\infty}
R^{+}_{[n]_\infty}
\Big\rord^\dagger
	\Oincov_{[2]_1}
	\Oincov_{[2]_u} 
\Big\lord
R^{+}_{[n]_0}
R^{+}_{[n]_0}
\Big\rord
\Big\rangle  
\equiv 
G^{n,n}_{+,+} ,
\ee
and multiply by the number of possible pairings.
Let ${\scr P}(q)$ denote the number of pairings of fields on the left; of course, there will also be ${\scr P}(q)$ pairs on the right. 
From what we saw above, we thus have
\be
G = {\scr P}^2(q) G^{n,n}_{+,+} .
\ee

Let us now find a formula for ${\scr P}(q)$.
If $q = 2p$ is even, the number of different pairings of $2p$ elements is
\be	\label{nmprs}
\text{Number of pairings of $2p$ objects}
=
\frac{ (2p)!}{p! 2^p}
\ee
This is well known, but we give a demonstration at the end of this Appendix.
If $q = 2p + 1$ is odd, we exclude one element, count the pairings of the remaining objects, then pair the excluded element with the $2p$ non-excluded objects. This gives
\be
\text{Number of pairings of $(2p+1)$ objects}
=
\frac{ (2p)!}{p! 2^p} + 2p .
\ee
One way of writing these number in a unified fashion is
\be
{\scr P}(q) 
\equiv
\left[
\parbox{10em}{\begin{center}Number of pairings of $q$ objects \\ between themselves \end{center}}
\right]
=
\frac{ (2 \lfloor \tfrac12 q \rfloor)!}{( \lfloor \frac12 q \rfloor)! \ 2^{ \lfloor \frac12 q \rfloor} } +2 (\tfrac12 q - \lfloor \tfrac12 q \rfloor) \lfloor \tfrac12 q \rfloor ,
\ee
where$\lfloor q \rfloor$ is the floor function.

\bigskip

\noindent
{\bfseries Example 2}

\noindent
Now consider the field
\be
\lord (R^+_{[n_1]})^{q_1} (R^{\dot1}_{[n_2]})^{q_2} \rord ,
 \qquad q_1 n_1 + q_2 n_2 = N ,
\ee
and its function
\begin{align}	\label{GuConEx1}
\begin{split}
G =
\Big\langle 
\Big\lord 
(R^{+\dagger}_{[n_1]_\infty})^{q_1} 
(R^{\dot1 \dagger}_{[n_2]_\infty})^{q_2}
\Big\rord
	\Oincov_{(2)_1}
	\Oincov_{(2)_u} 
\Big\lord
(R^{+}_{[n_1]_0})^{q_1} 
(R^{\dot1}_{[n_2]_0})^{q_2}
\Big\rord
\Big\rangle
\end{split}
\end{align}
Now we will have pairings between components with the same charge, as before; but we will also have pairings between the fields with different charges.
We have
\be	\label{Pairingq1q2}
\left[
\parbox{15em}{\begin{center}Number of pairs of $q_1$ objects \\ with $q_2$ different objects \end{center}}
\right]
= q_1 \times q_2
\ee
since we can choose one among the $q_1$ objects of the first group and pair will all $q_2$ objects of the second group; then choose the next object of the first group and do the same, and so on.
Thus we will have
\be
q_1 \times q_2 \quad \text{different pairs of the type} \quad 
R^{+\dagger}_{[n_1]_\infty} R^{\dot1 \dagger}_{[n_2]_\infty}
\ee
and the same number $q_1q_2$ of pairs of the type 
$R^{+}_{[n_1]_\infty} R^{\dot1}_{[n_2]_\infty}$.
Thus we have found
\begin{align}	\label{GuConEx1Final}
\begin{split}
G &=
(q_1 q_2)^2
\Big\langle 
\lord
R^{+}_{[n_1]_\infty} R^{\dot1}_{[n_2]_\infty}
\rord^\dagger
	\Oincov_{[2]_1}
	\Oincov_{[2]_u}
\lord
R^{+}_{[n_1]_0}  R^{\dot1}_{[n_2]_0}
\rord
\Big\rangle  
\\
&
+
{\scr P}^2(q_1)
\Big\langle 
\lord
R^{+}_{[n_1]_\infty} R^{+ }_{[n_1]_\infty}
\rord^\dagger
	\Oincov_{[2]_1}
	\Oincov_{[2]_u}
\lord
R^{+\dagger}_{[n_1]_0} R^{+ \dagger}_{[n_1]_0}
\rord
\Big\rangle  
\\
&
+
{\scr P}^2(q_2)
\Big\langle 
\lord
R^{\dot1}_{[n_2]_\infty} R^{\dot1}_{[n_2]_\infty}
\rord^\dagger
	\Oincov_{[2]_1}
	\Oincov_{[2]_u}
\lord
R^{\dot1\dagger}_{[n_2]_0} R^{\dot1 \dagger}_{[n_2]_0}
\rord
\Big\rangle  
\end{split}
\end{align}

It may be instructive to give an explicit example of our counting factors.
Take
$n_1 = 3$, $q_1 = 2$; $n_2=1$, $q_2 = 3$; $N = 9$.
A representative of the Ramond field 
$(R^{+}_{(n_1)})^{q_1} (R^{\dot1}_{(n_2)})^{q_2}$
is
\begin{align*}
(R^{+}_{(n_1)})^{q_1} (R^{\dot1}_{(n_2)})^{q_2}
	=
(R^{+}_{(3)})^{2} (R^{\dot1}_{(1)})^{3}
	=
R^{+\dagger}_{(1,2,3)} \
R^{+\dagger}_{(4,5,6)} \
R^{\dot1\dagger}_{(7)} \
R^{\dot1\dagger}_{(8)} \
R^{\dot1\dagger}_{(9)}
\end{align*}
We have the following pairings:
\begin{align*}
&
\text{$q_1 q_2 = 6$ pairs of different cycles} : \
	\begin{cases}
	R^{+\dagger}_{(1,2,3)}  R^{\dot1\dagger}_{(7)} ,
	\quad
	R^{+\dagger}_{(1,2,3)}  R^{\dot1\dagger}_{(8)} ,
	\quad
	R^{+\dagger}_{(1,2,3)}  R^{\dot1\dagger}_{(9)} ,
	\\
	R^{+\dagger}_{(4,5,6)}  R^{\dot1\dagger}_{(7)} ,
	\quad
	R^{+\dagger}_{(4,5,6)}  R^{\dot1\dagger}_{(8)} ,
	\quad
	R^{+\dagger}_{(4,5,6)}  R^{\dot1\dagger}_{(9)}
	\end{cases}
\\
&
\text{
${\scr P}(2) =
\frac{2!}{1! 2^1} = 1
$
pair of $n_1$ cycles} : \
	\begin{cases}
	R^{+\dagger}_{(1,2,3)} R^{+\dagger}_{(4,5,6)}
	\end{cases}
\\
&
\text{
${\scr P}(3) =
\frac{2!}{1! 2^1} + 2 = 3
$
pairs of $n_2$ cycles} : \
	\begin{cases}
	R^{\dot1\dagger}_{(7)} R^{\dot1\dagger}_{(8)} ,
	\
	R^{\dot1\dagger}_{(7)} R^{\dot1\dagger}_{(9)} ,
	\
	R^{\dot1\dagger}_{(8)} R^{\dot1\dagger}_{(9)} .
	\end{cases}
\end{align*}

\bigskip

\noindent
{\bfseries Example 3 (crossed terms)}

\noindent
Finally, we have the following delicate possibility:
\be	\label{crossexfl}
(R^+_{[n_1]})^{q_1} 
(R^{\dot1}_{[n_1]})^{s_1}
(R^{-}_{[n_2]})^{q_2} 
(R^{\dot2}_{[n_2]})^{s_2} 
\qquad 
(q_1+s_1) n_1 + (q_2+s_2) n_2 = N.
\ee
At first glance, it seems that the factorization of (\ref{crossexfl}) would include the function (note the charges of the Ramond fields)
\be	\label{crosstr}
\Big\langle 
	R^{+\dagger}_{[n_1]_\infty} 
	R^{- \dagger}_{[n_2]_\infty}
	\Oincov_{[2]_1}
	\Oincov_{[2]_u}
	R^{\dot1}_{[n_1]_0}  
	R^{\dot2}_{[n_2]_0}
\Big\rangle  .
\ee
Indeed, this correlator has non-trivial solutions of the permutation equation (\ref{compto1}), and it does not vanish automatically because both double-cycle Ramond fields at $\infty$ and at $0$ have precisely the same SU(2) charges, see Table \ref{TabQuNaRamComp}.
This factorization \emph{does not occur}, however.
If we look at how to obtain the ``crossed function''  (\ref{crosstr}) from the full field (\ref{crossexfl}), we actually need individual factorizations with the structure
\be
\Big\langle 
	R^{+\dagger}_{(n_1)_\infty} 
	R^{- \dagger}_{(n_2)_\infty}
	\Oincov_{(2)_1}
	\Oincov_{(2)_u}
	R^{\dot1}_{(n_1)_0}  
	R^{\dot2}_{(n_2)_0}
\Big\rangle  
\;
\Big\langle
	R^{\dot1\dagger}_{(n_1)_\infty} 
	R^+_{(n_1)_0}
\Big\rangle	
\;
\Big\langle
	R^{\dot2\dagger}_{(n_2)_\infty} 
	R^-_{(n_2)_0}
\Big\rangle
\;
\times 
\cdots	
\ee
and although connected the four-point function does not vanish, the factorized two-point functions do.

The same applies to combination such as
$(R^+_{[n_1]})^{q_1} 
(R^{-}_{[n_1]})^{s_1}
(R^{-}_{[n_2]})^{q_2} 
(R^{+}_{[n_2]})^{s_2} 
$
or
$
(R^{\dot1}_{[n_1]})^{q_1} 
(R^{\dot2}_{[n_1]})^{s_1}
(R^{\dot2}_{[n_2]})^{q_2} 
(R^{\dot1}_{[n_2]})^{s_2} 
$,
with
$(q_1+s_1) n_1 + (q_2+s_2) n_2 = N$,
which at first sight could have contributions like (in the former case) the ``antisymmetric'' function
\be
\Big\langle 
	R^{+\dagger}_{(n_1)_\infty} 
	R^{- \dagger}_{(n_2)_\infty}
	\Oincov_{(2)_1}
	\Oincov_{(2)_u}
	R^{-}_{(n_1)_0}  
	R^{+}_{(n_2)_0}
\Big\rangle  
\;
\Big\langle
	R^{-\dagger}_{(n_1)_\infty} 
	R^+_{(n_1)_0}
\Big\rangle	
\Big\langle
	R^{+\dagger}_{(n_2)_\infty} 
	R^-_{(n_2)_0}
\Big\rangle
\;
\times 
\cdots	
\ee
Again, the four-point function does not vanish, but the two-point functions are zero.

\bigskip
\begin{center}
***
\end{center}

We can derive Eq.(\ref{nmprs}) as follows.
	The number of ways of choosing a first pair out of the $2p$ elements is ${2p \choose 2}$. Then there are $2p-2$ remaining objects, and the number of ways of forming a second pair is therefore ${2p-2 \choose 2}$, and so on, leading to a product of $p$ binomial coefficients
\begin{align*}
& 
{2p \choose 2} \times {2(p-1) \choose 2} \times {2(p-2) \choose 2} 
\times \cdots \times
{2\times2 \choose 2} \times {2 \choose 2}
\\
& =	
\frac{[2p]!}{2! \textcolor{magenta}{[2p-2]!}}
\frac{\textcolor{magenta}{[2(p-1)]!}}{2! \textcolor{green}{[2(p-1)-2]!}} 
\frac{\textcolor{green}{[2(p-2)]!}}{2! \textcolor{cyan}{[2(p-2)-2]!}} 
\cdots 
\frac{\textcolor{blue}{[2\times2]!}}{2! \textcolor{red}{2!}}
\frac{\textcolor{red}{2!}}{2!}
=
\frac{[2p]!}{(2!)^p}
 =	
\frac{[2p]!}{2^p}
\end{align*}	
In the second line,  same-color factors cancel.
By construction, the number $(2p)!/2^p$ counts 
\emph{the number of ways of choosing $p$ pairs in a fixed order}. That is, we have counted the number of ways of forming
$
\{\text{Pair \#1 ; Pair \#2 ; $\cdots$ ; Pair \#$p$}\}
$
If we just want the total number of pairs, i.e. if the configurations
$$
\{\text{Pair \#1 ; Pair \#2 ; $\cdots$ ; Pair \#$p$}\}
\cong
\{\text{Pair \#5 ; Pair \#3 ; $\cdots$ ; Pair \#$7$}\}
\cong
\cdots
$$
are all equivalent, we must divide by the total number of permutations of $p$ objects, i.e.~by $p!$. Thus we get (\ref{nmprs}).

\section{General formula for the double-cycle four-point function}	\label{General4pt}

In this appendix we derive the four-point function (\ref{Gcfull}) for the generic double-cycle operator $R^{\zeta_1}_{[n_1]} R^{\zeta_2}_{[n_2]}$.
In order to do so, we define the following operator
\begin{align}	\label{Rn1n2Gen}
\begin{split}
&\scr R^{\{\hat \sigma , \hat \varrho \}}_{[n_1]} 
	\scr R^{\{ \check \sigma , \check \varrho \}}_{[n_2]}  (z) 
	\equiv 
\\
&	\frac{1}{\scr C_{n_1n_2}}
	\sum_{h\in S_N} 
	\exp \Bigg[ 
	\frac{i}{2n_1}
	\sum_{I=1}^{n_1} 
	\big( \hat \sigma \phi_{1,h(I)} + \hat \varrho \phi_{2,h(I)} \big) 
	+
	{i\over 2n_2}
	\sum_{I=n_1+1}^{n_1+n_2}
	 \big( 
	 \check \sigma \phi_{1,h(I)} + \check \varrho \phi_{2,h(I)} 
	\big)
	\Bigg]
\\	&\qquad\qquad\qquad
	\times
	\sigma_{h^{-1}(1\cdots n_1)h} 
	\sigma_{h^{-1}(n_1+1\cdots n_1+n_2)h} .
\end{split}
\end{align} 
Choosing combinations of the parameters $\hat\sigma, \hat\varrho,\check\sigma,\check\varrho = \pm 1$ we can obtain all the possible composite Ramond fields, see Table \ref{ZNSR}.
%

\begin{table}
\begin{center}
\begin{tabular}{r || c c c c c c }
$\scr R^{\{\hat \sigma , \hat \varrho \}}_{[n_1]} 
\scr R^{\{ \check \sigma , \check \varrho \}}_{[n_2]}$
&$R^{\dot1}_{[n_1]} R^{+}_{[n_2]}$
&$R^{\dot1}_{[n_1]} R^{-}_{[n_2]}$
&$R^{\dot1}_{[n_1]} R^{\dot2}_{[n_2]}$
&$R^{\dot1}_{[n_1]} R^{\dot1}_{[n_2]}$
&$R^{+}_{[n_1]} R^{-}_{[n_2]}$
&$R^{+}_{[n_1]} R^{+}_{[n_2]}$
\\
\hline
$\{\hat \sigma , \hat \varrho \}$
& {\footnotesize $\{-1 , -1 \}$}
& {\footnotesize $\{-1 , -1 \}$}
& {\footnotesize $\{-1 , -1 \}$}
& {\footnotesize $\{-1 , -1 \}$}
& {\footnotesize $\{+1 , -1 \}$} 
& {\footnotesize $\{+1 , -1 \}$}
\\
$\{ \check \sigma , \check \varrho \}$ 
& {\footnotesize $\{ +1 , -1 \}$}
& {\footnotesize $\{ - 1 , +1 \}$}
& {\footnotesize $\{ +1 , +1 \}$} 
& {\footnotesize $\{ - 1 , -1 \}$}
& {\footnotesize $\{ - 1 , +1 \}$}
& {\footnotesize $\{  +1 , -1 \}$}
\end{tabular}
\caption{Translating $\scr R^{\{\hat \sigma , \hat \varrho \}}_{[n_1]} \scr R^{\{ \check \sigma , \check \varrho \}}_{[n_2]}$ to the Ramond ground states.}
\label{ZNSR}
\end{center}
\end{table}

\subsection{LM technique}

We need to find the correlator of the fields lifted to the covering surface appearing in Eq.(\ref{CoverBaseG}), namely
\be	\label{Gcoverx}
\begin{split}
&G_{\scr R}^{\rm{(cover)}}(x,\bar x)
\\
&	= 
	\Big\langle
	\scr R^{\{\hat \sigma , \hat \varrho \} \dagger}(\infty,\bar \infty) 
	\scr R^{\{ \check \sigma , \check \varrho \}\dagger} (t_\infty, \bar t_\infty) 
	\Oincov (t_1,\bar t_1) 
	\Oincov (x, \bar x)
	\scr R^{\{ \check \sigma , \check \varrho \}} (t_0,\bar t_0) 
	\scr R^{\{\hat \sigma , \hat \varrho \}}(0,\bar 0) 
	\Big\rangle	,
\end{split}
\ee
Recall we drop twist indices on the covering. 
Fermionic exponentials inserted at the ramification points, lift to the covering surface as \cite{Lunin:2001pw}
\bsub	\label{rim}
\begin{align}
\exp \big( i p \big[ \phi_{1,I}(z_*) \pm \phi_{2,I}(z_*)  \big] \big)
	&\mapsfrom b_*^{- p^2 / n_* }
	\exp \big( i p \big[ \phi_{1}(t_*) \pm \phi_{2}(t_*)  \big] \big)
\\
\exp \big( i p \big[ \phi_{1,I}(\infty) \pm \phi_{2,I}(\infty)  \big] \big)
	&\mapsfrom b_{t_\infty}^{+ p^2 / n_{t_\infty} }
	\exp \big( i p \big[ \phi_{1}(\infty) \pm \phi_{2}(\infty)  \big] \big)
\\
\exp \big( i p \big[ \phi_{1,I}(\infty) \pm \phi_{2,I}(\infty)  \big] \big)
	&\mapsfrom b_\infty^{+ p^2 / n_\infty }
	\exp \big( i p \big[ \phi_{1}(t_\infty) \pm \phi_{2}(t_\infty)  \big] \big)
\end{align}
\esub
hence 
\begin{align}
\scr R^{\{\hat \sigma , \hat \varrho \}\dagger}(\infty) 
\scr R^{\{ \check \sigma , \check \varrho \}\dagger} (t_\infty) 
&=
	b_\infty^{\frac{1}{4n_1}}
	b_{t_\infty}^{\frac{1}{4n_2}}
	e^{
	- \frac{i}{2}
	[ \hat \sigma \phi_{1} (\infty) + \hat \varrho \phi_{2} (\infty) ] 
	}
	\
	e^{
	- \frac{i}{2}
	[ \check \sigma \phi_{1}(t_\infty) + \check \varrho \phi_{2}(t_\infty) ]
	} ,
\\
\scr R^{\{\hat \sigma , \hat \varrho \}}(0) 
\scr R^{\{ \check \sigma , \check \varrho \}} (t_0) 
&=
	b_0^{-\frac{1}{4n_1}}
	b_{t_0}^{-\frac{1}{4n_2}}
	e^{
	\frac{i}{2}
	[ \hat \sigma \phi_{1} (0) + \hat \varrho \phi_{2} (0) ] 
	}
	\
	e^{
	\frac{i}{2}
	[ \check \sigma \phi_{1}(t_0) + \check \varrho \phi_{2}(t_0) ]
	} .
\end{align}
As explained in \S\ref{SectStressTensor},  several terms appearing in the product $\Oint \Oint$ in (\ref{Gcoverx}) cancel because they have factorized bosonic two-point functions $\langle \pa X^{\dot A}(t_1) \pa X^{\dot B B}(x)\rangle$ which vanish unless $ \pa X^{\dot B B} = (\pa X^{\dot A A})^\dagger$, in which case they are given by Eq.(\ref{twopntboconj}).
Note that the bosonic contribution only depends on the two interaction operators --- it is the same whatever Ramond fields enter the four-point function.
Hence we have the same structure as in Eq.(\ref{Gbosn}),
\be	\label{GbosnAp}
\begin{split}
G_{\scr R}^{\rm{(cover)}} 
	&= 
	4 
	\left|
	b_\infty^{\frac{1}{4n_1}}	
	b_{t_\infty}^{\frac{1}{4n_2}}
	b_{t_1}^{- \frac{5}{8}}
	b_{x}^{- \frac{5}{8}}
	b_0^{-\frac{1}{4n_1}}	
	b_{t_0}^{-\frac{1}{4n_2} }
	\right|^2
	 \left| \frac{2}{ (t_1 - x)^{2} } \right|^2 
	\times 
		  G^F_{\scr R} ,
\end{split}	
\ee
but with a new fermionic contribution
\be	\label{GFIII}
\begin{split}
G^F_{\scr R} = 
\Big\langle
&	e^{ - \frac{i}{2} [ \hat \sigma \phi_{1}  + \hat \varrho \phi_{2}  ] } (\infty)
	e^{ - \frac{i}{2} [ \hat \sigma \bphi_{1}  + \hat \varrho \bphi_{2}  ] } (\bar\infty) 
	e^{- \frac{i}{2} [ \check \sigma \phi_{1} + \check \varrho \phi_{2} ]} (t_\infty)
	e^{- \frac{i}{2} [ \check \sigma \bphi_{1} + \check \varrho \bphi_{2} ]} (\bar t_\infty)
\\
&\quad
\times
\Big(I + II \Big)
	e^{\frac{i}{2} [ \hat \sigma \phi_{1} + \hat \varrho \phi_{2}  ] } (0)
	e^{\frac{i}{2} [ \hat \sigma \bphi_{1} + \hat \varrho \bphi_{2}  ] } (\bar0)
	e^{\frac{i}{2}[ \check \sigma \phi_{1} + \check \varrho \phi_{2}]} (t_0)
	e^{\frac{i}{2}[ \check \sigma \bphi_{1} + \check \varrho \bphi_{2}]} (\bar t_0)
\Big\rangle
\end{split}
\ee
where $I$ and $II$ are given by (\ref{ints}).
Let us start with the term $I$; it has a holomorphic factor 
\begin{align}
\begin{split}
%
%
%
&
\Big\langle
	e^{- \frac{i}{2} \hat \sigma \phi_1 } (\infty)
	e^{- \frac{i}{2} \check \sigma \phi_1 } (t_\infty)
	e^{- \frac{i}{2} \phi_1 } (t_1)
	e^{\frac{i}{2} \phi_1 } (x)
	e^{\frac{i}{2} \check \sigma \phi_1 } (t_0)
	e^{\frac{i}{2} \hat \sigma \phi_1 } (0)
\\
&\qquad\qquad
\times
	e^{- \frac{i}{2} \hat \varrho \phi_2} (\infty)
	e^{- \frac{i}{2}  \check \varrho \phi_2} (t_\infty)
	e^{- \frac{i}{2}  \phi_2} (t_1)
	e^{\frac{i}{2}  \phi_2} (x)
	e^{\frac{i}{2}  \check \varrho \phi_2} (t_0)
	e^{\frac{i}{2}  \hat \varrho \phi_2} (0)
\Big\rangle	
\\
&
= 
	\left( 
	\frac{ 
	(t_\infty - t_1) (t_0 - x) 
	}{
	(t_\infty - x) (t_0 - t_1) 
	} 
	\right)^{\frac{ \check \sigma + \check \varrho }{4} }
	\left( \frac{x}{t_1} \right)^{ \frac{\hat \sigma + \hat \varrho}{4}}
	( t_\infty - t_0 )^{- \frac12 }
	(t_1 - x )^{-\frac12}
	\left( \frac{t_0}{t_\infty} 
	\right)^{
	\frac{\check \sigma \hat \sigma + \hat \varrho \check \varrho }{4} 
			}
\end{split}
\end{align}	
and an anti-holomorphic factor
\begin{align}
\begin{split}
&
\Big\langle
	e^{- \frac{i}{2} \hat \sigma \bphi_1 } (\infty)
	e^{- \frac{i}{2} \check \sigma \bphi_1 } (t_\infty)
	e^{\frac{i}{2} \bphi_1 } (t_1)
	e^{-\frac{i}{2} \bphi_1 } (x)
	e^{\frac{i}{2} \check \sigma \bphi_1 } (t_0)
	e^{\frac{i}{2} \hat \sigma \bphi_1 } (0)
\\
&\qquad\qquad
\times
	e^{- \frac{i}{2} \hat \varrho \bphi_2} (\infty)
	e^{- \frac{i}{2}  \check \varrho \bphi_2} (t_\infty)
	e^{\frac{i}{2}  \bphi_2} (t_1)
	e^{- \frac{i}{2}  \bphi_2} (x)
	e^{\frac{i}{2}  \check \varrho \bphi_2} (t_0)
	e^{\frac{i}{2}  \hat \varrho \bphi_2} (0)
\Big\rangle	
\\
&
+
\Big\langle
	e^{- \frac{i}{2} \hat \sigma \bphi_1 } (\infty)
	e^{- \frac{i}{2} \check \sigma \bphi_1 } (t_\infty)
	e^{-\frac{i}{2} \bphi_1 } (t_1)
	e^{\frac{i}{2} \bphi_1 } (x)
	e^{\frac{i}{2} \check \sigma \bphi_1 } (t_0)
	e^{\frac{i}{2} \hat \sigma \bphi_1 } (0)
\\
&\qquad\qquad
\times
	e^{- \frac{i}{2} \hat \varrho \bphi_2} (\infty)
	e^{- \frac{i}{2}  \check \varrho \bphi_2} (t_\infty)
	e^{-\frac{i}{2}  \bphi_2} (t_1)
	e^{ \frac{i}{2}  \bphi_2} (x)
	e^{\frac{i}{2}  \check \varrho \bphi_2} (t_0)
	e^{\frac{i}{2}  \hat \varrho \bphi_2} (0)
\Big\rangle	
\\
&
= 
	\left( \frac{\bar t_0}{\bar  t_\infty} 
	\right)^{
	\frac{\check \sigma \hat \sigma + \hat \varrho \check \varrho }{4} 
			}
	( \bar t_\infty - \bar t_0 )^{ - \frac12 }
	(\bar t_1 - \bar x)^{- \frac{1}{2} }
\\
&
\times
\Bigg[
	\left( 
	\frac{ 
	(\bar t_\infty - \bar t_1) (\bar t_0 - \bar x)
	}{ 
	(\bar t_\infty - \bar x)(\bar t_0 - \bar t_1) } 
	\right)^{\frac{ \check \sigma + \check \varrho }{4} }
	\left( \frac{\bar x}{\bar t_1} \right)^{ \frac{\hat \sigma + \hat \varrho}{4}}
+
	\left( 
	\frac{ 
	(\bar t_\infty - \bar t_1) (\bar t_0 - \bar x)
	}{ 
	(\bar t_\infty - \bar x)(\bar t_0 - \bar t_1) } 
	\right)^{- \frac{ \check \sigma + \check \varrho }{4} }
	\left( \frac{\bar x}{\bar t_1} \right)^{- \frac{\hat \sigma + \hat \varrho}{4}}
\Bigg]
\end{split}
\end{align}	
The term $II$ gives similar contributions, but with some crucial differences: the holomorphic factor is
\begin{align}
\begin{split}
&
\Big\langle
	e^{- \frac{i}{2} \hat \sigma \phi_1 } (\infty)
	e^{- \frac{i}{2} \check \sigma \phi_1 } (t_\infty)
	e^{\frac{i}{2} \phi_1 } (t_1)
	e^{-\frac{i}{2} \phi_1 } (x)
	e^{\frac{i}{2} \check \sigma \phi_1 } (t_0)
	e^{\frac{i}{2} \hat \sigma \phi_1 } (0)
\\
&\qquad\qquad
\times
	e^{- \frac{i}{2} \hat \varrho \phi_2} (\infty)
	e^{- \frac{i}{2}  \check \varrho \phi_2} (t_\infty)
	e^{\frac{i}{2}  \phi_2} (t_1)
	e^{- \frac{i}{2}  \phi_2} (x)
	e^{\frac{i}{2}  \check \varrho \phi_2} (t_0)
	e^{\frac{i}{2}  \hat \varrho \phi_2} (0)
\Big\rangle	
\\
&
= 
	\left( \frac{(t_\infty - t_1) (t_0 - x)}{(t_\infty - x) (t_0 - t_1)} \right)^{- \frac{ \check \sigma + \check \varrho }{4} }
	\left( \frac{x}{t_1} \right)^{- \frac{\hat\sigma + \hat \varrho}{4} }
	( t_\infty - t_0 )^{- \frac12 }
	(t_1 - x )^{-\frac12}
	\left( \frac{t_0}{t_\infty} 
	\right)^{
	\frac{\check \sigma \hat \sigma + \hat \varrho \check \varrho }{4} 
			}
\end{split}
\end{align}	
and the anti-holomorphic factor is
\begin{align}
\begin{split}
&
\Big\langle
	e^{- \frac{i}{2} \hat \sigma \bphi_1 } (\infty)
	e^{- \frac{i}{2} \check \sigma \bphi_1 } (t_\infty)
	e^{- \frac{i}{2} \bphi_1 } (t_1)
	e^{\frac{i}{2} \bphi_1 } (x)
	e^{\frac{i}{2} \check \sigma \bphi_1 } (t_0)
	e^{\frac{i}{2} \hat \sigma \bphi_1 } (0)
\\
&\qquad\qquad
\times
	e^{- \frac{i}{2} \hat \varrho \bphi_2} (\infty)
	e^{- \frac{i}{2}  \check \varrho \bphi_2} (t_\infty)
	e^{-\frac{i}{2}  \bphi_2} (t_1)
	e^{ \frac{i}{2}  \bphi_2} (x)
	e^{\frac{i}{2}  \check \varrho \bphi_2} (t_0)
	e^{\frac{i}{2}  \hat \varrho \bphi_2} (0)
\Big\rangle	
\\
&
+
\Big\langle
	e^{- \frac{i}{2} \hat \sigma \bphi_1 } (\infty)
	e^{- \frac{i}{2} \check \sigma \bphi_1 } (t_\infty)
	e^{\frac{i}{2} \bphi_1 } (t_1)
	e^{-\frac{i}{2} \bphi_1 } (x)
	e^{\frac{i}{2} \check \sigma \bphi_1 } (t_0)
	e^{\frac{i}{2} \hat \sigma \bphi_1 } (0)
\\
&\qquad\qquad
\times
	e^{- \frac{i}{2} \hat \varrho \bphi_2} (\infty)
	e^{- \frac{i}{2}  \check \varrho \bphi_2} (t_\infty)
	e^{\frac{i}{2}  \bphi_2} (t_1)
	e^{- \frac{i}{2}  \bphi_2} (x)
	e^{\frac{i}{2}  \check \varrho \bphi_2} (t_0)
	e^{\frac{i}{2}  \hat \varrho \bphi_2} (0)
\Big\rangle	
\\
&
= 
	\left( \frac{\bar t_0}{\bar  t_\infty} 
	\right)^{
	\frac{\check \sigma \hat \sigma + \hat \varrho \check \varrho }{4} 
			}
	( \bar t_\infty - \bar t_0 )^{ - \frac12 }
	(\bar t_1 - \bar x)^{- \frac{1}{2} }
\\
&
\times
\Bigg[
	\left( 
	\frac{ 
	(\bar t_\infty - \bar t_1) (\bar t_0 - \bar x)
	}{ 
	(\bar t_\infty - \bar x)(\bar t_0 - \bar t_1) } 
	\right)^{- \frac{ \check \sigma + \check \varrho }{4} }
	\left( \frac{\bar x}{\bar t_1} \right)^{- \frac{\hat \sigma + \hat \varrho}{4}}
+
	\left( 
	\frac{ 
	(\bar t_\infty - \bar t_1) (\bar t_0 - \bar x)
	}{ 
	(\bar t_\infty - \bar x)(\bar t_0 - \bar t_1) } 
	\right)^{\frac{ \check \sigma + \hat \varrho }{4} }
	\left( \frac{\bar x}{\bar t_1} \right)^{ \frac{\hat \sigma + \hat \varrho}{4}}
\Bigg]
\end{split}
\end{align}

Combining everything, we find that the entire expression can be written as a square-modulus, as expected: 
\begin{align}	\label{GFfinal}
\begin{split}
&G^F_{\scr R}
=
	2
	\Bigg|
	\left(\frac{ t_0}{ t_\infty} \right)^{
	\frac{\check \sigma \hat \sigma + \hat \varrho \check \varrho }{4} 
			}
	(  t_\infty -  t_0 )^{ - \frac12 }
	( t_1 -  x)^{- \frac{1}{2} }
\\
&
\times
\Bigg[
	\left( 
	\frac{ 
	( t_\infty -  t_1) ( t_0 -  x) 
	}{ 
	( t_\infty -  x)( t_0 -  t_1) } 
	\right)^{- \frac{ \check \sigma + \check \varrho }{4} }
	\left( \frac{x}{t_1} \right)^{- \frac{\hat\sigma + \hat \varrho}{4} }
+
	\left( 
	\frac{ 
	( t_\infty -  t_1) ( t_0 -  x) 
	}{ 
	( t_\infty -  x)( t_0 -  t_1)  } 
	\right)^{\frac{ \check \sigma + \check \varrho }{4} }
	\left( \frac{x}{t_1} \right)^{ \frac{\hat\sigma + \hat \varrho}{4} }
\Bigg]	
\Bigg|^2
\end{split}
\end{align}	
Now using Eqs.(\ref{tim}), Eqs.(\ref{bofx}), and combining with the Liouville factor (\ref{SLofx}), we finally arrive at 
\begin{align}	\label{Gcofx}
\begin{split}
G_{\scr R} &(x,\bar x) 
 =
 	\Bigg|
	C \
	x^{ n_2 - n_1 
	+ \frac{6 
		- \check\sigma 
		- \check\varrho 
		- \hat\sigma 
		- \hat\varrho 
		- \check\sigma \hat\sigma
		- \check\varrho \hat\varrho 
		}{4}
	}
	(x-1)^{ n_2 + n_1 
	+ \frac{6 
		- \check\sigma 
		- \check\varrho 
		- \hat\sigma 
		- \hat\varrho 
		+ \hat \sigma \check \sigma
		+ \hat \varrho \check \varrho 
		}{4}
	}
\\
&
\times
	(x + \tfrac{n_1}{n_2})^{- n_2 - n_1 
	+ \frac{6 
		- \check\sigma 
		- \check\varrho 
		- \hat\sigma 
		- \hat\varrho 
		+ \hat \sigma \check \sigma
		+ \hat \varrho \check \varrho 
		}{4}
	}
	(x + \tfrac{n_1}{n_2} - 1)^{- n_2 + n_1 
	+ \frac{6
		- \check\sigma 
		- \check\varrho 
		- \hat\sigma 
		- \hat\varrho 
		- \check\sigma \hat\sigma
		- \check\varrho \hat\varrho 
		}{4}
	}
\\
& 
\times
	(x + \tfrac{n_1-n_2}{2n_2} )^{ -4}
\Bigg(
	\Big[ ( x+ \tfrac{n_1}{n_2} ) x \Big]^{\frac{\hat\sigma + \hat\varrho}{2}}
	\Big[  ( x+ \tfrac{n_1}{n_2} ) ( x+ \tfrac{n_1}{n_2} -1) \Big]^{\frac{\check\sigma + \check\varrho}{2}}
\\
&\qquad\qquad\qquad\qquad
+	
	\Big[ (x-1) ( x+ \tfrac{n_1}{n_2} -1) \Big]^{\frac{\hat\sigma + \hat\varrho}{2}}
	\Big[  ( x-1) x \Big]^{\frac{\check\sigma + \check\varrho}{2}}
\Bigg)
\Bigg|^2
\end{split}
\end{align}
which is the final, general expression for the double-cycle four-point function parameterized by $x$.
The constant $C$ is introduced to take into account the arbitrariness of normalization of the twists.

\subsection{The stress-tensor method}	\label{AppStressTensor}

Let us give here the results of the stress-tensor for the general field (\ref{Rn1n2Gen}).
As explained in the main text, to find $G_{\scr R}(x,\bar x)$ we want to solve the differential equation
\be	\label{difeqGsr}
\begin{split}
\pa_x \log G_{\scr R}(x) 
&
= 
2 \left( \frac{du}{dx} \right)
\underset{z = u}{ \rm{Res} } 
	\Bigg[
	\{t,z\} + \left( \frac{dt}{dz} \right)^2 F^{\{\hat\sigma, \hat\varrho; \check\sigma, \check\varrho \}}_{\rm{cover}} (t(z))
	\Bigg]
\end{split}
\ee
so we must only compute the function
\be\label{methodcovChComp}
\begin{split}
&
F^{\{\hat\sigma, \hat\varrho; \check\sigma, \check\varrho \}}_{\rm{cover}} (t)
\\
&
=
	 \frac{
\big\langle
T(t)
\scr R^{\{\hat \sigma , \hat \varrho \}\dagger}(\infty) 
\scr R^{\{ \check \sigma , \check \varrho \}\dagger} (t_\infty) 
\Oincov (t_1,\bar t_1) 
\Oincov (x,\bar x)
\scr R^{\{\hat \sigma , \hat \varrho \}}(0) 
\scr R^{\{ \check \sigma , \check \varrho \}} (t_0) 
\big\rangle
	}{
\big\langle
\scr R^{\{\hat \sigma , \hat \varrho \}\dagger}(\infty) 
\scr R^{\{ \check \sigma , \check \varrho \}\dagger} (t_\infty) 
\Oincov (t_1,\bar t_1) 
\Oincov (x,\bar x)
\scr R^{\{\hat \sigma , \hat \varrho \}}(0) 
\scr R^{\{ \check \sigma , \check \varrho \}} (t_0) 
\big\rangle
		} 
\\
&
=
	 \frac{
\big\langle
T(t)
\scr R^{\{\hat \sigma , \hat \varrho \}\dagger}(\infty) 
\scr R^{\{ \check \sigma , \check \varrho \}\dagger} (t_\infty) 
\Oincov (t_1,\bar t_1) 
\Oincov (x,\bar x)
\scr R^{\{\hat \sigma , \hat \varrho \}}(0) 
\scr R^{\{ \check \sigma , \check \varrho \}} (t_0) 
\big\rangle
	}{
G_{\scr R}^{\rm{(cover)}}
	} ,
\end{split}
\ee
and then follow the same procedure described in \cite{Lima:2020nnx} to obtain the residue and integrate Eq.(\ref{difeqGsr}).

To compute 
$F^{\{\hat\sigma, \hat\varrho; \check\sigma, \check\varrho \}}_{\rm{cover}} (t)$, first let us introduce the notation
\bsub
\begin{flalign}
&& \Oincov (t , \bar t) &\equiv  V_- (t , \bar t) + V_+ (t , \bar t) , &&
\\
\text{where}
&&
V_+(t,\bar t) &= \Big[ 
		\big( a.h.)_{\dot 1 1} \pa X^{\dot 11} 
		-  
		\big( a.h.)_{\dot 1 2} \pa X^{\dot 12} 
		\Big] \lor e^{+ \frac{i}{2} ( \phi_1 +  \phi_2 )} \ror
&&
\\
&&V_- (t,\bar t) &= \Big[ 
		\big( {a.h.})_{\dot 1 1\dagger} (\pa X^{\dot 11})^\dagger 
		+  
		\big( {a.h.})_{\dot 1 2\dagger} (\pa X^{\dot 12})^\dagger 
		\Big] \lor e^{- \frac{i}{2} ( \phi_1 +  \phi_2 )} \ror
&&		
\end{flalign}\label{DefofVpm}\esub
the $(a.h)$s being combinations of anti-holomorphic fields which can be read from (\ref{InteraOpera}).
The point here is that, since only the bosonic two-point functions (\ref{twopntboconj}) are not zero, we have
$
\langle \cdots V_\pm (t_1,\bar t_1) V_\pm(x, \bar x) \cdots \rangle = 0
$
where the ellipses indicate any pure-fermionic operators that are inserted in the correlation.
This leads to the useful identities
\bsub\label{TrickForm}
\begin{align}
& 
\big\langle \cdots V_+(t_1,\bar t_1) V_-(x, \bar x) \cdots \big\rangle 
+ \big\langle \cdots V_-(t_1,\bar t_1) V_+(x, \bar x) \cdots \big\rangle 
= G_{\scr R}^{\rm{(cover)}}
\\
&
\big\langle \cdots \Oincov (t_1) \big[ V_-(x) - V_+(x) \big] 
		\cdots \big\rangle 
	=	
	G_{\scr R}^{\rm{(cover)}}  
	- 2 \big\langle \cdots V_+(t_1) V_-(x ) \cdots  \big\rangle	 
	\label{Eqwineer}
\\	
&\big\langle \cdots \big[ V_-(t_1) - V_+(t_1) \big]  \Oincov(x) \cdots \big\rangle 
	=	
	G_{\scr R}^{\rm{(cover)}}
	-2 \big\langle \cdots V_-(t_1) V_+ (x)  \cdots \big\rangle	 
	\label{Eqwineer2}
\end{align}\esub

Now let us go back to Eq.(\ref{methodcovChComp}).
Contraction of $T(t)$ with the bosons in $\Oincov$ is precisely the same as in the four-point functions with other (fermionic) fields discussed in \cite{Lima:2020boh,Lima:2020kek,Lima:2020nnx,Lima:2020urq}.
Contraction of $T(t)$ with the fermions is more complicated than in \cite{Lima:2020boh,Lima:2020kek,Lima:2020nnx,Lima:2020urq}, but it follows the same lines, for example the contractions of the fermions $\pa \phi_1 \pa\phi_1(t)$ in $T(t)$, after normal-ordering, give
\begin{align}	\label{corrphphT}
\begin{split}
&
\big\langle
\pa \phi_1 \pa\phi_1(t)
\scr R^{\{\hat\sigma , \hat\varrho \}\dagger}(\infty) 
\scr R^{\{ \check\sigma , \check\varrho \}\dagger} (t_\infty) 
\Oincov (t_1,\bar t_1) 
\Oincov (x,\bar x)
\scr R^{\{\hat\sigma , \hat\varrho \}}(0) 
\scr R^{\{ \check\sigma , \check\varrho \}} (t_0) 
	   \big\rangle 
\\
&
=
- \frac{1}{4}
\Bigg[
	\left( \frac{1}{t-t_1} - \frac{1}{t-x} \right)^2
	+
	\left(
 	\frac{\check\sigma}{t-t_\infty}
	-
	\frac{\check\sigma}{t-t_0}
	-
	\frac{\hat\sigma}{t} 
	\right)^2
\\
&
+
	2
	\left(
 	\frac{\check\sigma}{t-t_\infty}
	-
	\frac{\check\sigma}{t-t_0}
	-
	\frac{\hat\sigma}{t} 
	\right)
\\
&
\times
	\Bigg( 
	\frac{1}{t-t_1} + \frac{1}{t-x} 
	 -
	 \frac{2}{t-t_1} 
	\frac{\langle \cdots V_+  V_-  \cdots \rangle}{G_{\scr R}^{\rm{(cover)}}}
	-
	\frac{2}{t-x} 
	\frac{\langle \cdots V_-  V_+  \cdots \rangle}{G_{\scr R}^{\rm{(cover)}}}
	\Bigg)
	\Bigg] G_{\scr R}^{\rm{(cover)}}
\end{split}
\end{align}	
We have made use of the identities (\ref{TrickForm}) to arrive at the form above.
The equivalent contractions with $\pa\phi_2\pa\phi_2(t)$ give the same result with $\hat\sigma \mapsto \hat\varrho$ and $\check\sigma \mapsto \check\varrho$.
In the end, we get
\bsub\label{FCompCover}
\begin{align}
\begin{split}
&F^{\{\hat\sigma, \hat\varrho; \check\sigma, \check\varrho \}}_{\rm{cover}} 
=
	\frac{(t_1 - x)^2}{(t-t_1)^2 (t - x)^2} 
\\
&+
\frac{1}{4}
\Bigg\{
	\left( \frac{1}{t-t_1} - \frac{1}{t-x} \right)^2
	+
	\frac{1}{2}
	\left(
 	\frac{\check\sigma}{t-t_\infty}
	-
	\frac{\check\sigma}{t-t_0}
	-
	\frac{\hat\sigma}{t} 
	\right)^2
	+
	\frac{1}{2}
	\left(
	\frac{\check\varrho}{t-t_0}
	-
 	\frac{\check\varrho}{t-t_\infty}
	-
	\frac{\hat\varrho}{t} 
	\right)^2
\\
&\qquad
+
	\left(
 	\frac{\check\sigma + \check\varrho}{t-t_\infty}
	-
	\frac{\check\sigma + \check\varrho}{t-t_0}
	-
	\frac{\hat\sigma + \hat\varrho}{t} 
	\right)
	\Bigg(
	\frac{1-2\frak w_+}{t-t_1} + \frac{1-2\frak w_-}{t-x} 
	\Bigg)
	\Bigg\}
\end{split}
\end{align}
where
\begin{align}
\frak w_{-}
	&=	
	\frac{ [(x - t_0)(t_1 - t_\infty)]^{\frac{\check\sigma + \check\varrho}{2} }
		 x^{ \frac{\hat\sigma + \hat\varrho}{2} }
		}{
		[(x - t_0)(t_1 - t_\infty)]^{\frac{\check\sigma + \check\varrho}{2} } 
		 x^{\frac{\hat\sigma + \hat\varrho}{2} }
		+
		[(t_1 - t_0)(x - t_\infty)]^{\frac{\check\sigma + \check\varrho}{2} } 
		 t_1^{ \frac{\hat\sigma + \hat\varrho}{2} }
		}
	\label{FCompCoverWm}	
\\
\frak w_{+}
	&=	
	\frac{ 
		[(t_1 - t_0)(x - t_\infty)]^{\frac{\check\sigma + \check\varrho}{2} } 
		 t_1^{ \frac{\hat\sigma + \hat\varrho}{2} }
		}{
		[(x - t_0)(t_1 - t_\infty)]^{\frac{\check\sigma + \check\varrho}{2} } 
		 x^{\frac{\hat\sigma + \hat\varrho}{2} }
		+
		[(t_1 - t_0)(x - t_\infty)]^{\frac{\check\sigma + \check\varrho}{2} } 
		 t_1^{ \frac{\hat\sigma + \hat\varrho}{2} }
		}
	\label{FCompCoverWp}	
\end{align}
\esub

Let us give some detail about the how to obtain formulae (\ref{FCompCover}). 
The functions $\frak w_\pm$ are initially defined as the following expressions, appearing in contractions such as (\ref{corrphphT}),
\begin{align}
\begin{split}
\frak w_{\pm} 
&\equiv \frac{	
	\big\langle 
	\cdots
	V_\pm (t_1, \bar t_1) V_\mp (x , \bar x) 
	\cdots
	\big\rangle
	}{G_{\scr R}^{\rm{(cover)}}}
\qquad\quad	
\frak y_\pm \equiv 
	\big\langle \cdots V_\pm(t,\bar t_1) V_\mp (x,\bar x) \cdots \big\rangle
\\
&= 
	\frac{ \frak y_\pm}{\frak y_+ + \frak y_-} 
\end{split}	
\label{frakwpmDef}	
\end{align}
The functions $\frak y_\pm$ differ by a swap of the points $t_1$ and $x$.
They have non-trivial holomorphic and anti-holomorphic parts, but the anti-holomorphic parts happen to be the same for both functions and therefore cancel in $\frak w_\pm$.
To see this, we can show that there is symmetry under $\bar t_1 \leftrightarrow \bar x$. Consider $\frak y_+$,
\begin{align*}
&\big\langle \cdots V_+(t_1,\bar t_1) V_-(x, \bar x)  \cdots \big\rangle
\\
&	=
	\frac{2}{(t_1 - x)^2}
	\Big\langle 
	\cdots 
	\Big[
	(a.h.)_{\dot11} (a.h.)_{\dot11\dagger} 
	+
	(a.h.)_{\dot12} (a.h.)_{\dot12\dagger} 
		\Big]
	S^{\dot2}(t_1)
	S^{\dot1}(x)
	\cdots
	\Big\rangle		
\end{align*}
with $2 /(t_1 - x)^2$ the bosonic two-point function.
The	anti-holomorphic factors are
\begin{align*}
\big\langle
	\cdots 
 (a.h.)_{\dot11} (a.h.)_{\dot11\dagger} 
 	\cdots 
\big\rangle
&	=
	\big\langle
	\cdots 
		\big[
		\bar\pa X^{\dot12}(\bar t_1) \tilde S^{\dot 2}(\bar t_1) 
		-
		\bar\pa X^{\dot11\dagger}(\bar t_1) \tilde S^{\dot1}(\bar t_1) 
		\big]
\\		
&\quad
		\times
		\big[
		\bar\pa X^{\dot12\dagger}(\bar x) \tilde S^{\dot 1}(\bar x) 
		+
		\bar\pa X^{\dot11}(\bar x) \tilde S^{\dot2}(\bar x) 
		\big]
		\cdots
		\big\rangle	
\\
&	=
	\frac{-2}{(\bar t_1 - \bar x)^2}
	\big\langle
	\cdots 
		\big[
		S^{\dot2}(\bar t_1) \tilde S^{\dot 1}(\bar x) 
		+
		S^{\dot1}(\bar t_1)	\tilde S^{\dot 2}(\bar x) 
		\big]
		\cdots
		\big\rangle	
\\
\big\langle 
	\cdots 
(a.h.)_{\dot12} (a.h.)_{\dot12\dagger} 
	\cdots 
\big\rangle
&	=
	\big\langle
	\cdots 
		\big[
		\bar\pa X^{\dot12\dagger}(\bar t_1) \tilde S^{\dot 1}(\bar t_1) 
		+
		\bar\pa X^{\dot11}(\bar t_1) \tilde S^{\dot2}(\bar t_1) 
		\big]
\\		
&\quad
		\times
		\big[
		\bar\pa X^{\dot12}(\bar x) \tilde S^{\dot 2}(\bar x) 
		-
		\bar\pa X^{\dot11\dagger}(\bar x) \tilde S^{\dot1}(\bar x) 
		\big]
		\cdots
		\big\rangle	
\\
&	=
	\frac{-2}{(\bar t_1 - \bar x)^2}
	\big\langle
	\cdots 
		\big[
		\tilde S^{\dot1}(\bar t_1) \tilde S^{\dot2}(\bar x) 
		+
		\tilde S^{\dot2}(\bar t_1) \tilde S^{\dot1}(\bar x) 
		\big]
		\cdots
		\big\rangle	
\end{align*}
which is clearly symmetric under $\bar t_1 \leftrightarrow \bar x$.
We therefore only have to compute the holomorphic parts of the correlators $\frak y_\pm$, that is e.g.
\begin{align*}
\frak y_{-}
&=\Big\langle 
		\scr R^{\{ \hat\sigma, \hat\varrho\}\dagger} (\infty)
		\scr R^{\{ \check\sigma, \check\varrho\}\dagger}(t_\infty)
		S^{\dot1}(t_1)
		S^{\dot2}(x)
		\scr R^{\{ \check\sigma, \check\varrho\}}(t_0)
		\scr R^{\{ \hat\sigma, \hat\varrho\}}(0)
\Big\rangle		
\\
&=
e^{i \theta}
\Big\langle
	e^{	-\frac{i\hat\sigma}{2} \phi_1(\infty)  }
	e^{	-\frac{i\check\sigma}{2} \phi_1(t_\infty)  }
	e^{	- \frac{i}{2}  \phi_1(t_1)  }
	e^{	+ \frac{i}{2} \phi_1(x)  }
	e^{ 	+ \frac{i \check\sigma}{2} \phi_1(t_0)  }	
	e^{ 	+ \frac{i \hat\sigma}{2} \phi_1(0)  }	
\Big\rangle
\\
&\quad
\times	
\Big\langle
	e^{	- \frac{i\hat\varrho}{2} \phi_2(\infty) }
	e^{	- \frac{i \check\varrho}{2} \phi_2(t_\infty) }
	e^{	- \frac{i}{2}   \phi_2(t_1) }
	e^{	+ \frac{i}{2} \phi_2(x)   }
	e^{ 	+ \frac{i\check\varrho}{2} \phi_2(t_0) }	
	e^{ 	+ \frac{i \hat\varrho}{2} \phi_2(0) }	
\Big\rangle	
\end{align*}
Computation of $\frak y_{+}$ is similar, with $S^{\dot1}$ and $S^{\dot2}$ swapped. 
The phase $e^{i\theta}$ comes from possible cocycles, but has to be the same in $\frak y_{-}$ and $\frak y_{+}$, so it cancels in $\frak w_\pm$ and can be ignored. Direct computation results in
\begin{align*}
\begin{split}
&\frak y_{-}
=
	\left( 
	\frac{x - t_0 }{t_1 - t_0}
	\,
	\frac{t_1-t_\infty }{x-t_\infty}
	\right)^{ \frac{\check\sigma + \check\varrho }{4} }
	\left( \frac{x}{t_1} \right)^{ \frac{\hat\sigma + \hat\varrho}{4}  }
	\left( t_\infty - t_0 \right)^{- \frac{\check\sigma^2 + \check\varrho^2}{4} }
	\left( \frac{t_0}{t_\infty} \right)^{   \frac{\check\sigma \hat\sigma + \check\varrho \hat\varrho }{4} }
	(t_1 - x)^{- \frac{1}{2} } \,  
\end{split}
\\
\begin{split}
&\frak y_{+}
=
	\left( 
	\frac{x - t_0 }{t_1 - t_0}
	\,
	\frac{t_1-t_\infty }{x-t_\infty}
	\right)^{- \frac{\check\sigma + \check\varrho }{4} }
	\left( \frac{x}{t_1} \right)^{- \frac{\hat\sigma + \hat\varrho}{4}  }
	\left( t_\infty - t_0 \right)^{- \frac{\check\sigma^2 + \check\varrho^2}{4} }
	\left( \frac{t_0}{t_\infty} \right)^{   \frac{\check\sigma \hat\sigma + \check\varrho \hat\varrho }{4} }
	(t_1 - x)^{- \frac{1}{2} } \,  
\end{split}
\end{align*}
When we build up the $\frak w$, only the ratio of $\frak y$s matter, so all but the two first factors in each expression cancel and, after some manipulation, we obtain the Formulae (\ref{FCompCover}).

\section{Lists of products of structure constants}	\label{AppLists}

Eqs.(\ref{stcconsOsO}) and (\ref{OsORsR}) combined give the following list of structure constants:
\bsub	\label{strucconss3}
\begin{align}
\begin{split}
&
	\big\langle 
	\lord
	R^{\dot1}_{(n_1)} R^+_{(n_2)}
	\rord^\dagger  
	\s_{(3)}
	\lord
	R^{\dot1}_{(n_1)} R^+_{(n_2)}
	\rord
	\big\rangle 
	=
	 2^{-\frac{25}{3}} 3^{-\frac{16}{3}} 
	(n_1^2-n_2^2)^{\frac23} 
	n_1^{-\frac43} 
	n_2^{-\frac43} 
\end{split}	
\\
\begin{split}
&
	\big\langle 
	\lord
	R^{\dot1}_{(n_1)} R^{\dot2}_{(n_2)}
	\rord^\dagger  
	\s_{(3)}
	\lord
	R^{\dot1}_{(n_1)} R^{\dot2}_{(n_2)}
	\rord
	\big\rangle 
	=
	 2^{-\frac{25}{3}} 3^{-\frac{16}{3}} 
	(n_1^2-n_2^2)^{\frac83} 
	(n_1+n_2)^{-\frac43}
	n_1^{-\frac43} 
	n_2^{-\frac43} 
\end{split}	
\\
\begin{split}
&
	\big\langle 
	\lord
	R^{\dot1}_{(n_1)} R^{\dot1}_{(n_2)}
	\rord^\dagger  
	\s_{(3)}
	\lord
	R^{\dot1}_{(n_1)} R^{\dot1}_{(n_2)}
	\rord
	\big\rangle 
	=
	 2^{-\frac{25}{3}} 3^{-\frac{16}{3}} 
	(n_1^2-n_2^2)^{-\frac43} 
	(n_1+n_2)^{\frac83}
	n_1^{-\frac43} 
	n_2^{-\frac43} 
\end{split}	
\\
\begin{split}
&
	\big\langle 
	\lord
	R^{+}_{(n_1)} R^{-}_{(n_2)}
	\rord^\dagger  
	\s_{(3)}
	\lord
	R^{+}_{(n_1)} R^{-}_{(n_2)}
	\rord
	\big\rangle 
	=
	 2^{-\frac{25}{3}} 3^{-\frac{16}{3}} 
	(n_1-n_2)^{\frac83} 
	(n_1+n_2)^{-\frac43}
	n_1^{-\frac43} 
	n_2^{-\frac43} \ 
\end{split}	
\\
\begin{split}
&
	\big\langle 
	\lord
	R^{+}_{(n_1)} R^{+}_{(n_2)}
	\rord^\dagger  
	\s_{(3)}
	\lord
	R^{+}_{(n_1)} R^{+}_{(n_2)}
	\rord
	\big\rangle 
	=
	 2^{-\frac{25}{3}} 3^{-\frac{16}{3}} 
	(n_1-n_2)^{-\frac43} 
	(n_1+n_2)^{\frac83}
	n_1^{-\frac43} 
	n_2^{-\frac43} \ 
\end{split}	
\end{align}
\esub

Now we give the explicit structure constants (\ref{ProdRO}).
In the channel 
$$
\Oint \times \lord R^{\zeta_1}_{[n_1]}  R^{\zeta_2}_{[n_2]} \rord 
	=  Y^{\zeta_1\zeta_2}_{\frak1,  [n_1+n_2]}  
$$
we find the following list from Eq.(\ref{ProdRO1})
\bsub \label{List1}
\begin{align}
\begin{split}
&	
\Big|
\big\langle 
\Oincov_{(2)} 
\lord R^{\dot2}_{(n_1)}  R^{-}_{(n_2)} \rord
Y^{\dot1+}_{\frak 1,  (n_1+n_2)} 
\big\rangle
\Big|^2
	=
	 2^{-2}
	(1 - \tfrac{n_1}{n_2})^{-2}
	n_1^{-\frac{4n_1}{n_1-n_2} } 
	n_2^{- \frac{4n_2}{n_2-n_1} } 
\end{split}	
\\
\begin{split}
&
\Big|
\big\langle 
\Oincov_{(2)} 
\lord R^{\dot2}_{(n_1)}  R^{\dot1}_{(n_2)} \rord
Y^{\dot1\dot2}_{\frak 1,  (n_1+n_2)} 
\big\rangle
\Big|^2
	=
	 2^{-2}
	n_1^{-\frac{4n_1}{n_1-n_2} } 
	n_2^{- \frac{4n_2}{n_2-n_1} } 
\end{split}	
\\
\begin{split}
&
\Big|
\big\langle 
\Oincov_{(2)} 
\lord R^{\dot2}_{(n_1)}  R^{\dot2}_{(n_2)} \rord
Y^{\dot1\dot1}_{\frak 1,  (n_1+n_2)} 
\big\rangle
\Big|^2
	=
	 2^{-2}
	(n_1 - n_2)^{-4}
	(n_1^2 + n_2^2)^2 
	n_1^{-\frac{4n_1}{n_1-n_2} } 
	n_2^{- \frac{4n_2}{n_2-n_1} } 
\end{split}	
\\
\begin{split}
&
\Big|
\big\langle 
\Oincov_{(2)} 
\lord R^{-}_{(n_1)}  R^{+}_{(n_2)} \rord
Y^{+-}_{\frak 1,  (n_1+n_2)} 
\big\rangle
\Big|^2
	=
	n_1^{- \frac{3n_1+n_2}{n_1-n_2}}
	n_2^{ \frac{n_1+3n_2}{n_1-n_2}}
\end{split}	
\\
\begin{split}
&
\Big|
\big\langle 
\Oincov_{(2)} 
\lord R^{-}_{(n_1)}  R^{-}_{(n_2)} \rord
Y^{++}_{\frak 1,  (n_1+n_2)} 
\big\rangle
\Big|^2
	=
	(n_1-n_2)^{-4}
	n_1^{- 2 \frac{n_1+n_2}{n_1-n_2}}
	n_2^{ 2 \frac{n_1+n_2}{n_1-n_2}}
\end{split}	
\end{align}
\esub
Here we have used that
$
\lord R^{\dot1}_{(n_1)}  R^{\dot2}_{(n_2)} \rord^\dagger
=
\lord R^{\dot2}_{(n_1)}  R^{\dot1}_{(n_2)} \rord
$,
$
\lord R^{+}_{(n_1)}  R^{-}_{(n_2)} \rord^\dagger
=
\lord R^{-}_{(n_1)}  R^{+}_{(n_2)} \rord
$,
etc.

In the channel 
$$
\Oint \times \lord R^{\zeta_1}_{[n_1]}  R^{\zeta_2}_{[n_2]} \rord 
	=  Y^{\zeta_1\zeta_2}_{\frak2,  [n_1+n_2]} 
$$
we find the following list of structure constants from Eq.(\ref{ProdRO2})
\bsub 	\label{List2}
\begin{align}
\begin{split}
&	
\Big|
\big\langle 
\Oincov_{(2)} 
\lord R^{\dot2}_{(n_1)}  R^{-}_{(n_2)} \rord
Y^{\dot1+}_{\frak 2,  (n_1+n_2)} 
\big\rangle
\Big|^2
	=
	 2^{-2}
	(1 + \tfrac{n_1}{n_2})^{-2}
	n_1^{-\frac{4n_1}{n_1+n_2} } 
	n_2^{- \frac{4n_2}{n_2+n_1} } 
\end{split}	
\\
\begin{split}
&
\Big|
\big\langle 
\Oincov_{(2)} 
\lord R^{\dot2}_{(n_1)}  R^{\dot1}_{(n_2)} \rord
Y^{\dot1\dot2}_{\frak 2,  (n_1+n_2)} 
\big\rangle
\Big|^2
	=
	 2^{-2}
	 (n_1+n_2)^{-4}
	 (n_1^2 + n_2^2)^2
	n_1^{-\frac{4n_1}{n_1+n_2} } 
	n_2^{- \frac{4n_2}{n_2+n_1} } 
\end{split}	
\\
\begin{split}
&
\Big|
\big\langle 
\Oincov_{(2)} 
\lord R^{\dot2}_{(n_1)}  R^{\dot2}_{(n_2)} \rord
Y^{\dot1\dot1}_{\frak 2,  (n_1+n_2)} 
\big\rangle
\Big|^2
	=
	 2^{-2}
	n_1^{-\frac{4n_1}{n_1+n_2} } 
	n_2^{- \frac{4n_2}{n_2+n_1} } 
\end{split}	
\\
\begin{split}
&
\Big|
\big\langle 
\Oincov_{(2)} 
\lord R^{-}_{(n_1)}  R^{+}_{(n_2)} \rord
Y^{+-}_{\frak 2,  (n_1+n_2)} 
\big\rangle
\Big|^2
	=
	(n_1+n_2)^{-4}
	n_1^{- \frac{n_1-n_2}{n_1+n_2}}
	n_2^{\frac{n_1-n_2}{n_1+n_2}}
\end{split}	
\\
\begin{split}
&
\Big|
\big\langle 
\Oincov_{(2)} 
\lord R^{-}_{(n_1)}  R^{-}_{(n_2)} \rord
Y^{++}_{\frak 2,  (n_1+n_2)} 
\big\rangle
\Big|^2
	=
	n_1^{-2 \frac{3n_1-n_2}{n_1+n_2}}
	n_2^{ 2 \frac{n_1-3n_2}{n_1+n_2}}
\end{split}	
\end{align}
\esub

\bibliographystyle{utphys}

\bibliography{LongMultiCycleReferencesV9} 

\end{document}